\documentclass[%
 reprint,amsmath, amssymb, aps, superscriptaddress,pra]{revtex4-2}
\usepackage{graphicx}% Include figure files
\usepackage{multirow}
\usepackage{dcolumn}% Align table columns on decimal point
\usepackage{bm}% bold math
\usepackage{hyperref}% add hypertext capabilities
\usepackage[mathlines]{lineno}% Enable numbering of text and display math
\usepackage{physics}
\usepackage{diagbox}
\usepackage{booktabs}
\usepackage[normalem]{ulem}
\usepackage{resizegather}
\usepackage[mathscr]{euscript}
\usepackage{placeins}

\hypersetup{
    colorlinks=true,
    linkcolor=red,
    filecolor=magenta,
    citecolor=blue,
    urlcolor=blue,
}

\newcommand{\hide}[1]{}

\newcommand{\IITB}{Department of Physics, Indian Institute of Technology Bombay, Powai, Mumbai, Maharashtra 400076, India}
\newcommand{\Quicst}{Centre of Excellence in Quantum Information, Computation, Science and Technology,
Indian Institute of Technology Bombay, Powai, Mumbai, Maharashtra 400076, India}
\newcommand{\SFU}{Department of Physics,
Simon Fraser University, Burnaby, British Columbia, Canada V5A 1S6}

\begin{document}

\title{Quantum error correction for an unresolvable spin ensemble}

\author{Harsh Sharma}
\email{harsh.sharma@iitb.ac.in} 
\affiliation{\IITB}

\author{Himadri Shekhar Dhar}
\email{himadri.dhar@iitb.ac.in}
\affiliation{\IITB}\affiliation{\Quicst}

\author{Hoi-Kwan Lau}
\email{kero\_lau@sfu.ca} 
\affiliation{\SFU}


\begin{abstract}
Spin ensembles are promising quantum technological platforms, but their utility relies on the ability to perform quantum error correction (QEC) for decoherences in these systems. Typical QEC for ensembles requires addressing {individually resolved} qubits, but this is practically challenging in most realistic architectures. Here, we propose QEC schemes for unresolvable spin ensembles. By using degenerate superpositions of excited states, which are fundamentally mixed, we {find} codes that can protect against both individual and collective errors, including dephasing, decay, and pumping. We show how information recovery can be achieved with only collective measurement and control, and illustrate its applications in extending memory lifetime and loss-tolerant sensing.
\end{abstract}

\maketitle

\section{Introduction\label{Sec:Intro}}

Atomic {and} solid-state spin ensembles have recently become promising platforms for quantum technologies~\cite{nielsen_chuang_2019,Tordrup2008,Wesenberg2009,Cox2022,Afzelius2009,lvovsky2009,Sangouard2011}. 
{Systems such as} atomic cloud~\cite{Muschik2008,Saffman2010,Heller2020}, defect centers~\cite{Morello2020,Nemoto2014,Pezzagna2021} and ions in Penning traps~\cite{Gilmore2021,Polloreno2022}, generally exhibit superior coherence times~\cite{Longdell2005,Dudin2013,Bar-Gill2013} and collectively enhanced coupling~\cite{Kubo2010,Schuster2010}, while {remaining} highly controllable and readily {adaptable} in hybrid architectures~\cite{Xiang2013,Kurizki2015}. 
These properties make spin ensembles favorable for implementing quantum {computing}~\cite{Tordrup2008,Wesenberg2009,Cox2022}, repeaters~\cite{Afzelius2009,lvovsky2009,Sangouard2011}, simulators~\cite{bloch2012,Endres2016} and {sensors~\cite{Pezze2018,Mu2023,Ouyang2026a}}.

Nevertheless, the practicality of ensemble-based platforms is hindered by two main issues: decoherence and limited resolvability. Typically, quantum information suffers from decoherence when the encoding system interacts with the {environment~\cite{Wineland1998,Chase2008,Biercuk2009,Saffman2010,Choi2017,Kirton2017,Shammah2018}}. Decoherence in spin ensembles tend to be complicated as the spins can both couple individually to local environments and collectively to a common environment, and even lose spins. Various quantum error correction (QEC) schemes have been proposed to protect information in {spin ensembles~\cite{Ruskai2000,Pollatsek2004,Brion2008,Ouyang2014,Ouyang2026b,Omanakuttan2023a,Omanakuttan2023b}}. 
{However, complete decoding in these schemes typically require addressing individual spins to identify and correct the  {errors~\cite{Ruskai2000,Pollatsek2004,Ouyang2014,Ouyang2026b,Omanakuttan2023a}}. 
In schemes without any spin resolution, only errors such as collective noise~\cite{Omanakuttan2023b} or {spin loss~\cite{Ouyang2026b}}, can be corrected, limiting the scope of error protection in unresolved ensembles.}

Unfortunately, in most ensemble-based platforms individual spin resolvability is practically challenging, if not impossible. For example, resolving an atom in a cloud is infeasible as the wavelength of the controlling optical photon $(\sim 10^3 \textup{\AA})$ is much longer than the inter-atomic spacing $(\sim 1-10 \textup{\AA})$ \cite{Saffman2010,Biron2007}.
{Moreover, addressing the rapidly moving ions in Penning traps also requires demanding setups and configuration that have been partly realized only very recently \cite{Pezze2018,Polloreno2022,McMahon2024}.}
By contrast, sophisticated collective spin Hamiltonians have been realized for over a decade~\cite{Gilmore2021,Bohnet2016}, enabling the generation of large-scale entanglement and quantum-enhanced sensing. This gap in technological maturity suggests that collective global control is far more accessible than individual spin addressing in large ensembles. However, without individual spin addressing, the possibility of {correcting the} dominant decoherence errors in an ensemble remains unknown.

In this work, we answer {this} question {affirmatively} by proposing a systematic construction of QEC codes for unresolvable spin ensembles. In analogy to QEC codes for bosonic systems~\cite{Albert2018,Joshi_2021}, large spins~\cite{Gross2021}, and molecules~\cite{Albert2020,Jain2024,Furey2024}, we leverage the abundantly available {quantum levels} in an ensemble as {the} resource. {Our} code is effective against individual and collective errors, as well as spin loss {in the ensemble.}
Every process of the {proposed} QEC can be implemented with only global control and measurement, {without any spin resolution.}
{Moreover, the efficacy of our protocol can be extended to achieve enhanced} quantum memory lifetime and protecting quantum sensing against decay.

The remainder of the paper is organized as follows. Section~\ref{Sec:II} introduces the general representation of states and dynamics in unresolvable spin ensembles and shows how logical information can be encoded in these states. Building on this framework, Sec.~\ref{Sec:III} presents our qubit encoding and demonstrates that the resulting memory is protected against collective, individual, and spin-loss errors. Section~\ref{Sec:IV} details the practical implementation of the different steps of the protocol, while Sec.~\ref{Sec:V} illustrates an additional application of the scheme to enhance quantum sensing. Possible physical realizations in several experimental platforms are discussed in Sec.~\ref{Sec:VI}. We conclude with a discussion and outlook in Sec.~\ref{Sec:VII}.

%---------------------------------------------------------------------------
\section{Permutationally invariant states\label{Sec:II}}
The state of an ensemble of $N$ spin-1/2 particles can always be expressed as a linear combination of the $2^N$ basis states $\{\ket{J,M,i_J,N}\}$. 
Here $J$ and $M$ are the quantum numbers corresponding to the total angular momentum $\hat{J}^2$ and magnetization $\hat{J}_z$, where $\hat{J}^2 = \hat{J}_x^2 + \hat{J}_y^2 + \hat{J}_z^2$ and $\hat{J}_{\alpha} = \frac{1}{2}\sum_{n=1}^N \hat{\sigma}_{\alpha,n}$ is the sum of all single-spin Pauli operators $\hat{\sigma}_{\alpha}$ with $\alpha = \{x,y,z\}$~\cite{Varshalovich1988}. The degeneracy index {$i_J$} denotes the states that share the same $J$ and $M$~\cite{Chase2008}. 

{In unresolvable spin ensembles, the hardware cannot distinguish individual spins, so all accessible controls and readouts are built from these collective spin operators $\hat{J}_{\alpha}$, and any Hamiltonian $\mathcal{H}$ made from them are invariant under exchange or permutations ($\Pi_{ij}$) of any two spins, i.e., $\Pi_{ij} \mathcal{H} \Pi_{ij} = \mathcal{H}$. In contrast, any Hamiltonian $\mathcal{H}_i$ that targets a particular spin $i$ cannot be permutationally invariant (PI), because swapping spin $i$ with $j$ would turn it into a control for $j$ instead, $\Pi_{ij} \mathcal{H}_i \Pi_{ij} = \mathcal{H}_j \neq \mathcal{H}_i$.} 
{Due to this symmetry of the system,} the description of the state of an unresolvable spin ensemble {must also be} PI~\cite{Chase2008,Shammah2018}. 
{An immediate} consequence is that only $N^2/4+\mathcal{O}(N)$ states {can be distinguished} by their different $J$ and $M$, while degenerate states are indistinguishable.
The definition of PI state thus implicitly requires every degenerate state is equally probable. The most general PI state is {then} given by
\begin{equation}
    \hat{\rho}_c = \sum_{J=J_{min}}^{N/2} \sum_{M,M'=-J}^J \rho_{J,M;J,M'} \overline{\ket{J,M}\bra{J,M'}},
    \label{Eq:genSt}
\end{equation}
where $J_{min} = 0 ~(1/2)$ for even (odd) $N$, and 
\begin{equation}
    \overline{\ket{J,M}\bra{J,M'}} \equiv \frac{1}{d_N^J} \sum_{i_J=1}^{d_N^J} \ket{J,M,i_J}\bra{J,M',i_J}. 
    \label{Eq:CollStDef}
\end{equation}
Here $d_N^J$ is the total number of degenerate states with the same $J$~\cite{Chase2008}. For clarity, we drop index $N$ whenever the number of spins is preserved. 

We note that Eq.~\eqref{Eq:genSt} implies that PI symmetry forbids any coherence between states with different $J$.
{While no collective operation can transform states from one $J$ subspace to another, it is possible for individual processes to change $J$.}
Moreover, PI states with maximum $J = N/2$ are the well-known Dicke states \cite{Dicke1954}. It is the only PI subspace where the basis states are non-degenerate and thus pure, i.e., $d^J_N =1$, while any PI state with $J< N/2$ {is inherently} mixed in the computational basis or collective spin basis $\{\ket{J,M,i_J,N}\}$.
\begin{figure}[t]
    \centering
    \includegraphics[width=0.95\columnwidth]{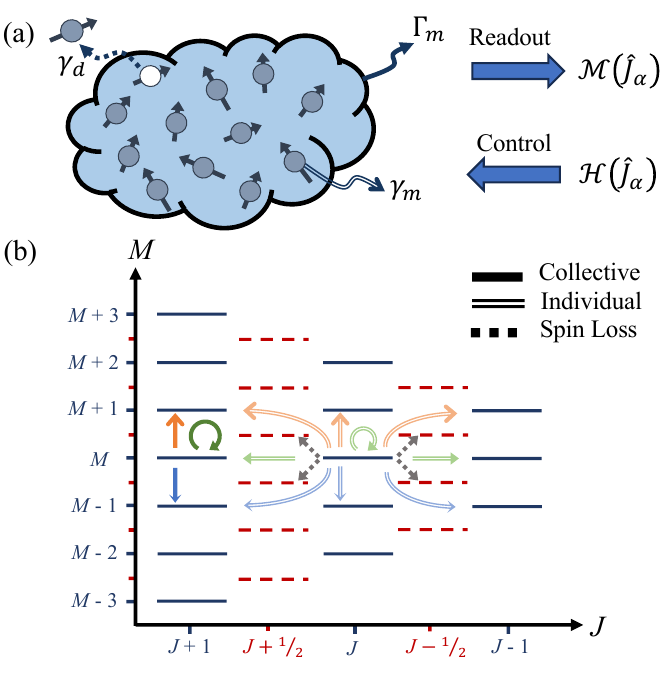}
    \caption{\textbf{Illustration of our system:} (a) An unresolvable spin ensemble that can be manipulated and measured only collectively, but the spins can suffer from both individual and collective decoherence. (b) Level diagram of PI states. Collective decoherences do not change the total angular momentum $J$ (solid arrows) while individual decoherence can change $J$ by $\{0,\pm1\}$ (hollow arrows). Spin loss will shift both $J$ and $M$ by a half-integer (dotted arrows).}
    \label{fig:ErrorMaps}
\end{figure}

%---------------------------------------------------------------------------
\subsection{System dynamics\label{Sec:IIA}}
We consider a spin ensemble that can only be controlled and measured collectively, i.e., both the control Hamiltonian 
$\mathscr{H}$ and measurement operators ${\mathscr{M}}$ are {only} functions of $\hat{J}_\alpha$ (see Fig.~\ref{fig:ErrorMaps}(a)). 
Upon interaction with the environment, the ensemble suffers from decoherence, {the effect of which} depends on the nature of the environment: whether all spins are \textit{collectively} interacting with the same environment, or each spin is \textit{individually} interacting with its own environment. 
The decoherence dynamics of the system follows the Lindblad master equation~\cite{Breuer2007},
\begin{equation}
    \dot{\hat{\rho}}_c = \sum_{m =-1}^1 \left( \frac{\Gamma_{m}}{2} \hat{\mathcal{L}}_{\hat{J}_{m}}[\hat{\rho}_c] + \frac{\gamma_{m}}{2} \sum_{n=1}^{N} \mathcal{L}_{\hat{\sigma}_{m,n}}[\hat{\rho}_c] \right),
    \label{Eq:errModel}
\end{equation}
where $\mathcal{L}_{\hat{A}}[\hat{\rho}] \equiv 2 \hat{A} \hat{\rho} \hat{A}^{\dagger} - \{\hat{A}^{\dagger}\hat{A},\hat{\rho}\}$; $\Gamma_{m}$ and $\gamma_{m}$ respectively represent the decoherence rates for collective and individual processes. The magnetization change due to decoherence, $m = -1, 0 ~\text{and}~ +1$, respectively denote the decay, dephasing and pumping processes. The corresponding jump operators are given by $\hat{J}_{\pm1} = \hat{J}_x \pm i \hat{J}_y$, $\hat{J}_0 = \hat{J}_z$, $\hat{\sigma}_{\pm1,n} = (\hat{\sigma}_{x,n} \pm i \hat{\sigma}_{y,n})/2$ and  {$\hat{\sigma}_{0,n} = \hat{\sigma}_{z,n}/2$}.

Since both collective and {individual decoherence preserves permutation symmetry~\footnote{The individual decoherence rates of all spins are assumed to be same, which preserves permutational invariance of the system. This condition is naturally satisfied when the spins are identical and couple identically to the environment.}}~\cite{Chase2008}, their effects are characterized by the transformation of PI state elements (see Fig.~\ref{fig:ErrorMaps}(b)). For short evolution time, i.e., $\gamma_m dt, \Gamma_m dt \ll 1$, decoherence can be approximated by quantum jumps \cite{Dalibard1992}. For collective decoherence, we have
\begin{multline}
    \overline{\ket{J,M}\bra{J,M'}} \xrightarrow[]{\Gamma_m} X_{m}^{J,M} X_{m}^{J,M'} \\ \times \overline{\ket{J,M+m}\bra{J,M'+m}},
    \label{Eq:ChaseMapColl}
\end{multline}
and for individual processes,
\begin{multline}
    \overline{\ket{J,M}\bra{J,M'}} \xrightarrow[]{\gamma_m} \sum_{j=-1}^1 \chi_{j,m}^{N,J,M} \chi_{j,m}^{N,J,M'} \\ \times  \overline{\ket{J+j,M+m}\bra{J+j,M'+m}}.
    \label{Eq:ChaseMap}
\end{multline}
The expressions of the coefficients $X_{m}^{J,M}$ and $\chi_{j,m}^{N,J,M}$ are {discussed in Appendix~\ref{App:A1}.}
{The above transformations show that collective decoherence preserves $J$, but individual processes can change $J$ by $0,\pm1$.} 

%-----------------------------------------------------------------------------
\subsection{Logical Mapping\label{sec:logicmap}}

Physical state of an ensemble that is prepared in the Dicke subspace $(J=N/2)$ would eventually become mixed when the ensemble is subjected to individual errors.  Without spin resolvability to reveal the degeneracy index $i_J$, the error state will remain mixed, so existing QEC schemes cannot be used to protect the state afterwards.  Our scheme embraces the mixedness of the state after errors, so that the information can be continuously protected by re-encoding into the physically mixed state in $J<N/2$ subspaces. Encoding the pure logical information in physically mixed states might seem paradoxical, but it has been discussed in detail in the literature~\cite{Grace2006,Lau2017}. Mathematically, this is possible because the information is encoded in a collective degree of freedom while the mixedness of the state affects only other degrees of freedom. Physically, every components in the mixture share the same superposition amplitudes, so the quantum information (as superposition amplitudes) can be represented even though the identity of the component is not known.

Using the general description of the state in Eq.~\eqref{Eq:genSt}, we now discuss how quantum information can be represented in an unresolvable spin ensemble. Since no superposition is allowed between different $J$ subspaces, we encode information in a state with a specific $J$. A logical qudit $\hat{\rho}_L$, where information is represented by a density matrix in the logical basis $\{\ket{M_L}\}$, can be encoded by a PI state $\hat{\rho}_c$ that shares the same matrix elements:
\begin{multline}
    \hat{\rho}_c = \sum_{M,M'=-J}^J \rho_{M,M'} \overline{\ket{J,M}\bra{J,M'}} \\ \longleftrightarrow \sum_{M,M'=-d/2}^{d/2} \rho_{M,M'} {\ket{M_L}{\bra{M'_L}}} = \hat{\rho}_L
    \label{Eq:LogMap}
\end{multline}
for any $J$ subspace that contains sufficient levels, i.e., $J \geq d/2$. 

\begin{figure*}
    \centering
    \includegraphics[width = 0.95\textwidth]{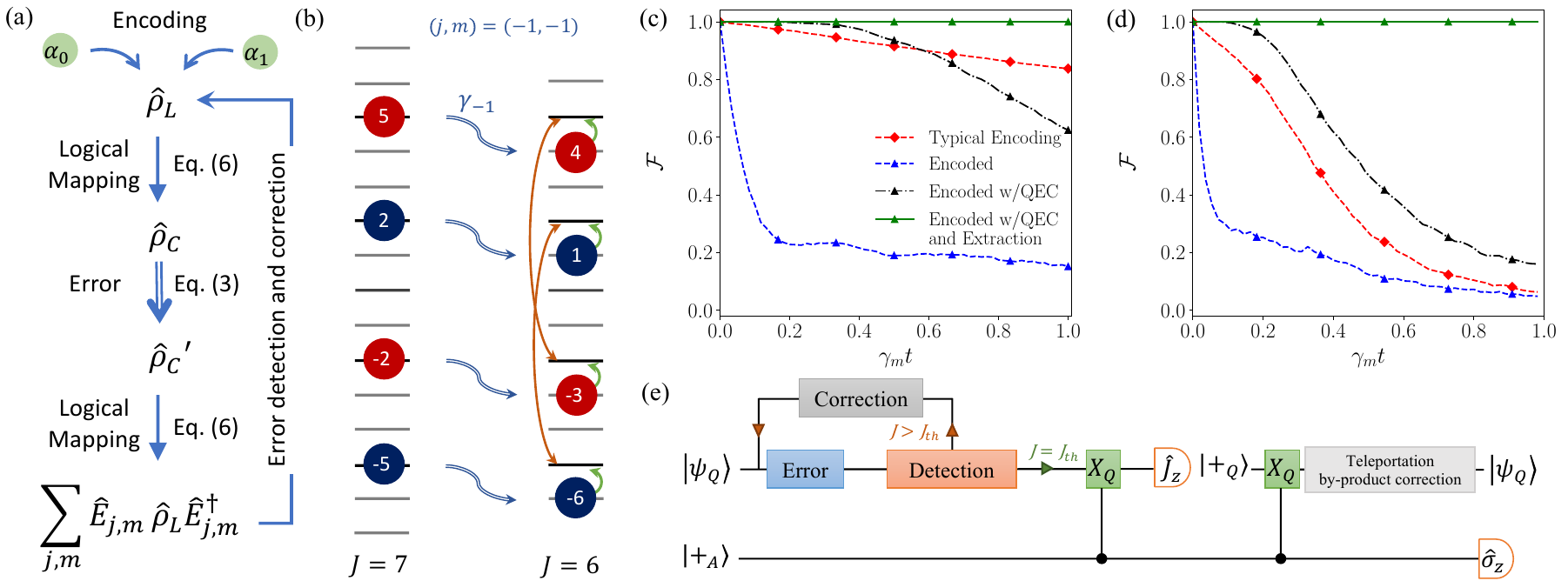}
    \caption{\textbf{Procedure for spin ensemble QEC.} (a) Qubit information $(\alpha_0, \alpha_1)$ is encoded in PI states. After decoherence, error is detected and ensemble is then restored to the code state. (b) Level diagram of QEC code in Eq.~\eqref{Eq:AssyCode}. Red and blue balls represent the levels occupied by $\ket{0_Q}$ and $\ket{1_Q}$. After identifying the angular momentum change $j$ and spin excitation change $m$ due to decoherence, collective unitary is applied to restore the code state. Fidelity $\mathcal{F}$ of a qubit encoded with lowest excitation states (diamond) and QEC code in Eq.~\eqref{Eq:AssyCode} (triangle) in presence of (c) individual decay ($\gamma_{-1} = 0.5$) and (d) all individual decoherence processes ($\gamma_0 = \gamma_{\pm1} = 1$). Initial state is taken to be {$\ket{+_Q} = (\ket{0_Q}+\ket{1_Q})/\sqrt{2}$} in $J=N/2$ subspace for both encodings. Error correction (dot-dashed and solid) shows clear improvement over bare evolution (dashed). Dynamics is obtained by averaging over $10^3$ quantum trajectories~\cite{Dalibard1992}, for an ensemble of 20 spins with $M_1 = 5$ and $M_2 = 2$. (e) Quantum circuit for the teleportation process when $J$ reaches threshold $J_{th}$. It involves two one-bit teleportation~\cite{Zhou2000}, first from the ensemble to ancilla, then from ancilla back to the ensemble that is prepared with a larger $J$. The final teleportation by-product is a logical bit-flip $X_Q$ conditional on measurement outcomes.}
    \label{fig:Encoding}
\end{figure*}

{Importantly, this allows us the flexibility to encode information in the larger PI space and also identify} the resultant logical information after the physical states suffer from decoherence (see Fig.~\ref{fig:Encoding}(a)). 
A collective jump introduces to the logical state a decoherence characterized by the logical Kraus operator 
\begin{equation}
    \hat{E}^c_{0,m} = \sqrt{\Gamma_m dt} \sum_{M} X_{m}^{J,M} \ket{(M+m)_L}\bra{M_L}.
    \label{Eq:KrausC}
\end{equation}
Similarly, after determining the total angular momentum of the physical state (e.g. by non-destructive $\hat{J}^2$ measurement), the logical Kraus operators introduced by individual processes are 
\begin{equation}
    \hat{E}^l_{j,m} = \sqrt{\gamma_m dt} \sum_{M} \chi_{j,m}^{N,J,M} \ket{(M+m)_L}\bra{M_L}.
    \label{Eq:KrausI}
\end{equation} 
Finally, the Kraus operator without any jumps is {$\hat{E}_{\varnothing,\varnothing} = \sqrt{\hat{\mathbb{I}}-\sum_{m=-1}^1(\hat{E}^{c\dagger}_{0,m}\hat{E}^c_{0,m}+\sum_{j= -1}^1\hat{E}^{l\dagger}_{j,m}\hat{E}^l_{j,m})}$}, which also changes the collective state. However, the second term in no-jump case is $\mathcal{O}(dt)$, which for small $dt$ is negligible compared to terms in Eqs.~\eqref{Eq:KrausC} and \eqref{Eq:KrausI}. 
Therefore, we can write $\hat{E}_{\varnothing,\varnothing}\approx \hat{\mathbb{I}}$. Alternatively, we can still keep this term and perform approximate QEC~\cite{Leung1997}.%

%-----------------------------------------------------------------------------

\section{Error correcting code\label{Sec:III}}

\subsection{Qubit encoding\label{sec:encoding}}
The abundant states available in spin ensembles can be divided into separate subspaces for encoding information and error syndromes.
{Let us consider the simplest case of encoding information for a qubit state $\alpha_0 \ket{0} + \alpha_1 \ket{1}$.} 
Two orthogonal logical qudit states are first chosen to represent the logical values 0 and 1:
\begin{equation}
\ket{0_Q} \equiv \sum_{M} c_{M}^{(0)} \ket{M_L}~~,~~\ket{1_Q} \equiv \sum_{M} c_{M}^{(1)} \ket{M_L}.
\end{equation}
For simplicity, we consider them occupying different levels, i.e.,  $c_M^{(0)}=0$ if $c_M^{(1)} \neq 0$ and vice versa. The ensemble is then prepared according to Eq.\eqref{Eq:LogMap} with $\hat{\rho}_L = \left( \alpha_0 \ket{0_Q} + \alpha_1 \ket{1_Q} \right) \left( \alpha_0^\ast \bra{0_Q} + \alpha_1^\ast \bra{1_Q} \right)$. %

The physical state of the spin ensemble will generally be corrupted by decoherence and become a state with a mixture of different $J$. The first step of QEC is to learn $J$ through non-destructive measurement, e.g., via phase estimation with an ancilla \cite{Wang2021}. The resultant ensemble state will represent a logical qudit via Eq.~\eqref{Eq:LogMap}, which can be related to the initial qudit via the Kraus operators.  To protect the encoded qubit information, we can construct QEC encodings that satisfy the Knill-Laflamme (KL) conditions~\cite{Knill1997,nielsen_chuang_2019},
\begin{equation}
     \bra{k'_Q} \hat{E}_{j',m'}^\dagger \hat{E}_{j,m} \ket{k_Q} = K_{j,m;j',m'}  \delta_{kk'},
    \label{Eq:KL}
\end{equation}
where $K_{j,m;j',m'}$ are the elements of a Hermitian positive semi-definite matrix. We have dropped the superscripts from the Kraus operators because $\chi_{0,m}^{N,J,M} \propto X_{m}^{J,M}$, i.e., the state after collective decoherence is the same as that after individual decoherence if $J$ is unchanged. 

The systematic procedure to construct such encodings is discussed in Appendix~\ref{App:B}. 
Here, we discuss the results for a specific example where both encoding states occupies only two levels (see Fig.~\ref{fig:Encoding}(b)). Consider $\ket{0_Q}$ involves levels $-M_1$ and $M_2$ and $\ket{1_Q}$ involves levels $M_1$ and $-M_2$. We find that this code can protect against one round of individual or collective dephasing, decay, and pumping errors {if the 
amplitudes satisfy} 
\begin{equation}
    c_{-M_1}^{(0)} = c_{M_1}^{(1)} = \sqrt{\frac{M_2 }{ M_1 + M_2}},
    \label{Eq:AssyCode}
\end{equation}
and $c_{\pm M_2}^{(k)} = \sqrt{1-\left| c_{\mp M_1}^{(k)}\right|^2}$, for $M_2 \geq 3/2 $ and $M_1 - M_2 \geq 3$. The latter two conditions guarantee that the code distance is sufficient to distinguish between decay and pumping errors. Satisfying these conditions require a subspace with $J\geq 9/2$, which can be implemented in any ensemble with $N\geq9$ spins. Remarkably, this encoding applies to both odd and even number of spins $N$, where $J$ is respectively half and full integers. 
%%%
{Let us look at the specific}
{state in Fig.~\ref{fig:Encoding}(b), where the information is encoded in an ensemble of $N=20$ spins, with qubits encoded in levels $M_1=5$ and $M_2=2$, and $J\geq 5$. The logical qubits are,
\begin{subequations}
\begin{equation}
\ket{0_Q}=\sqrt{\frac{2}{7}}\ket{-5_L}+\sqrt{\frac{5}{7}}\ket{2_L}~\textrm{and}
\end{equation}
\begin{equation}
\ket{1_Q}=\sqrt{\frac{5}{7}}\ket{-2_L}+\sqrt{\frac{2}{7}}\ket{5_L},
\end{equation}
\end{subequations}
where the subscript $L$ means that the $M$ levels belong to the logical qudit and depending on the $J$ subspace in which information is being encoded,
{the dimension} of Hilbert space of logical states is $2J+1$. 
Then, using the logical map in Eq.~\eqref{Eq:LogMap}, the logical state $\ket{\psi_Q}=\alpha_0\ket{0_Q}+\alpha_1\ket{1_Q}$ or equivalently the logical density matrix $\hat{\rho}_L = \sum_{k,k' = 0 }^1 \alpha_k^{} \alpha^*_{k'} \ket{k^{}_Q}\bra{k'_Q}$ can be mapped to the collective physical state as,
\begin{multline}
    \hat{\rho}_c = \frac{1}{d_{N}^J}\sum_{i_J = 1}^{d_{N}^J}\sum_{k,k'=0}^1\sum_{l,l'=1}^2 \alpha_k^{} \alpha_{k'}^* c^{}_{M_{l^{}}}c^*_{M_{l'}} \\
    \big|J,(-1)^{k+l}M^{}_l,i_J\big\rangle \big\langle J,(-1)^{k'+l'}M_{l'},i_J \big|,
\end{multline}
where $c_{M_1} = \sqrt{2/7}$ and $c_{M_2} = \sqrt{5/7}$. In Appendix~\ref{App:D0}, we look at the encoding of state $\ket{+_Q}=\frac{1}{\sqrt{2}}(\ket{0_Q}+\ket{1_Q})$ and its transformation under individual errors.}

%---------------------------------------------------------------------------
\subsection{Error detection and correction\label{sec:detect}}

After identifying $J$ of the ensemble (and thus the change of total angular momentum $j$), the
next step is to determine the change of magnetization $m$, which can also be obtained non-destructively, {for example, using} phase-estimation~\cite{Wang2021,Ouyang2026b} or {via} quantum signal processing~\cite{Motlagh2024,Fong2026}. 
{The decay and pumping errors correspond to $m = -1, +1$.}  Getting $j\neq 0$ but $m=0$ signifies individual dephasing. However, getting $j=m=0$ could mean either no jump or dephasing. 
Distinguishing between them requires addressing the phase between the encoding levels. This can be achieved by engineering a Hamiltonian to map the code and dephased states to different levels, then distinguish them by non-destructive $M$ measurement. See Sec.~\ref{Sec:IVB} and \ref{Sec:IVC} for more details on control Hamiltonian and syndrome extraction.

After determining the type of error, the corrupted state should be restored to an encoding state for subsequent QEC. Unlike typical QEC schemes that aim at recovering the original code state, for unresolvable spin ensembles no method is yet known to change total angular momentum while preserving the superposition which encodes quantum information. Fortunately, since QEC code is available whenever $J\geq 9/2$, the corrupted state could instead be restored to the code state with the new angular momentum, $J+j$. 

The correction transformation can be implemented by applying collective control, i.e., the control Hamiltonian is a function of $\{\hat{J}_x, \hat{J}_y, \hat{J}_z\}$ only, and an arbitrary  {transformation} can be engineered by applying a sequence of universal gates \cite{Gutman2023}. As an explicit example, we propose an implementation that requires only linear drive and one-axis-twisting (OAT) spin squeezing~\cite{Kitagawa1993}, which has been proposed (if not routinely implemented) in spin ensemble systems such as atomic clouds~\cite{Keating2016}, defect centers~\cite{Bennett2013,Teissier2014,Pradana2021} and Penning trap ions~\cite{Bohnet2016}.  
This process along with linear drive $\hat{J}_x$ can be used to arbitrarily adjust the superposition in any $J$ subspace.
The recovery process is illustrated with an example in 
Fig.~\ref{fig:Encoding}(b), {where} sequential transformations are first applied to restore the corrupted state to the encoding levels {(green arrows)}, and then the superposition amplitudes are adjusted {(orange arrows)} to that of the QEC code.

In Figs.~\ref{fig:Encoding}(c)-(d), we simulate the performance of an ensemble being used as a memory for qubit. We compare the logical fidelity of our QEC code with the typical encoding where the logical values are represented by the two lowest excited Dicke states, i.e., $\ket{N/2,-N/2}~\text{and}~\ket{N/2,-N/2+1}$. 
When there is only individual decay (Fig.~\ref{fig:Encoding}(c)), without QEC our encoding state decays much faster, because the typical encoding is the least prone to decay {as it has the} lowest excitation number.  However, QEC greatly improves the logical fidelity of our code {in comparison to} the typical counterpart. The advantage of QEC is more prominent when the system is subjected to all individual errors (Fig.~\ref{fig:Encoding}(d)). 

The logical fidelity in Figs.~\ref{fig:Encoding}(c)-(d)  eventually drops despite QEC, because repeated individual errors can reduce $J$ to below the threshold $9/2$ {after several rounds}. This problem can be resolved if we retrieve the information from the ensemble when $J$ approaches the threshold, and transfer it to {another} ensemble initialized at a higher $J$. {The} teleportation circuit {in Fig.~\ref{fig:Encoding}(e) illustrates} such information retrieval and transfer.
This requires an ancilla qubit that dispersively couples to the ensemble and the ability to perform spin excitation number measurement. {In Fig.~\ref{fig:Encoding}(c) and (d), this process ensures that} the encoded information is essentially undepleted over the simulated period. We discuss the recovery process and teleportation protocol in more detail in Sec.~\ref{Sec:IVC} and Sec.~\ref{Sec:IVD}, respectively.

%===========================================================================================================
\subsection{Spin loss\label{spinloss}}
In addition to the decoherence that preserves $N$, the ensemble can also suffer from spin loss, e.g., atoms escape from the trap or conversion of defect centers~\cite{Tsyganok2023, Choi2017}. Such error cannot be described by the master equation in Eq.~\eqref{Eq:errModel} which preserves $N$. Since each spin is equally likely to be lost, the state of the remaining $N-1$ spins will remain PI. After a single spin loss, we find that the PI state transforms as 
\begin{multline}
    \overline{\ket{J,M,N}\bra{J,M',N}} \xrightarrow[]{\gamma_d} \sum_{j,m=-1/2}^{1/2} \xi_{j,m}^{N,J,M} \xi_{j,m}^{N,J,M'}\\ \times \overline{\ket{J+{j},M+{m},N-1}\bra{J+{j},M'+{m},N-1}}.
    \label{Eq:DeletionErr}
\end{multline}
Explicit expressions for the coefficients $\xi_{j,m}^{N,J,M}$ 
{are provided in Appendix~\ref{App:D}}. 
%To the best of our knowledge, the effect of spin loss on general PI states {has not been previously reported.}

During QEC, spin loss error is identified by an half-integer change of $J$ and $M$. The logical Kraus operators are 
$
    \hat{E}^d_{j,m} = \sqrt{\gamma_d dt} \sum_{M}\xi_{j,m}^{N,J,M} {\ket{(M+m)_L}}\bra{M_L}
$, 
where $\gamma_d$ is the spin loss rate. A code for spin loss can be constructed using our formalism.  Somewhat surprisingly, we find that the code in Eq.~\eqref{Eq:AssyCode} can already tackle this error. 
To elaborate, consider {that} a spin polarized as $\sigma_z=-1/2~(1/2)$ is added to the ensemble if {an increase (decrease) in $M$ is detected} after a spin loss. The overall effect is equivalent to individual dephasing, which is protected by {the} code. We note that adding spin is not necessary for recovery, as one can restore the ensemble to the code states of $N-1$ spins. {Recovery follows the same procedure as in  $N$-preserving decoherence, and requires only collective control.}

%---------------------------------------------------------------------------

\section{Implementation of QEC protocol\label{Sec:IV}}
\subsection{Initial state preparation\label{Sec:IVA}}
{The first step in implementing the scheme is to prepare an initial state that encodes the desired information. This can be achieved by first initializing the ensemble in the all-spin-down ground state $\ket{\downarrow}^{\otimes N}$, for instance via cooling and/or optical pumping. The information is then encoded as a superposition within the Dicke ($J=N/2$) subspace spanned by the collective ground and first-excited states. Finally, the target encoded state superpositions specified in Eq.~\eqref{Eq:AssyCode} are prepared by sequential application of a universal control unitary generated by a one-axis-twisting interaction together with a linear drive (to be discussed in the next subsection), which maps the logical amplitudes onto the required physical levels.} {See Appendix~\ref{App:D0} for more details.}

However, the possibility to encode in the $J$ subspaces other than the pure Dicke subspace prompts us to ask the question: if we can prepare the low $J$ states, is there any advantage in protecting quantum information?  Here we argue that if the ensemble is subjected to both collective and individual decoherence, then a lower $J$ encoding state can reduce the error rate.
Explicitly, the rates of collective processes generally scale as $J^2$ due to superradiance \cite{Dicke1954}, while individual decoherence rate is insensitive to $J$. Therefore, encoding in lower $J$ subspaces can typically reduce the decoherence rate~\cite{Guerin2016}. See Appendix~\ref{App:A2} for more details on decoherence rates, and Fig.~\ref{Apfig:ErrRate} and Table~\ref{ApTab:ErrRate} therein. 

Engineered dissipation protocols~\cite{Plenio1999,Reiter2016} offer a robust route to preparing states in low $J$ subspaces of collective spin systems. By tailoring system environment couplings, population can be selectively removed from higher $J$ manifolds, stabilizing target states such as singlets~\cite{Kastoryano2011,Lin2013,Behbood2014,Ilo2022} and W states~\cite{Cole2021}. Such protocols can further be used for deterministically initializing arbitrary states within low $J$ subspaces~\cite{Tullu2023}.
{Additionally, probabilistic approaches can also be used to prepare states in the low $J$ subspaces. For example, one may prepare an ensemble with mixed spins such as by working at non-zero temperature or by subjecting a Dicke state to individual dephasing, which leads to support in lower $J$ subspaces. A subsequent phase-estimation measurement of the total spin angular momentum $J$ can then post-selectively identify the desired subspace.}

%---------------------------------------------------------------------------
\subsection{Universal control\label{Sec:IVB}}
As mentioned earlier, all encoding and recovery processes can be performed by using only one-axis-twist (OAT) squeezing, for which the control Hamiltonian $\mathscr{H} \propto \hat{J}_z^2$~\cite{Kitagawa1993} and linear drive, for which $\mathscr{H} \propto \hat{J}_x$. The key idea in this method is that $\hat{J}_z^2$ introduces nonlinear energy shifts such that each $M \leftrightarrow M-1$ transition has a unique frequency splitting $\Delta_M$. By applying the linear drive resonant to each frequency splitting, we can selectively Rabi oscillate between any two neighboring levels without affecting others. Sequential application of these operations then allows us to carry out the encoding and decoding processes.

Explicitly, the control Hamiltonian is given by,
\begin{equation}
     {\mathscr{H}}_{M} = \omega_0 \hat{J}_z + \chi \hat{J}_z^2 + 2 \Omega \cos{(\omega_M t + \phi)} \hat{J}_x,
     \label{AppEq:HamUC}
\end{equation}
where $\omega_0$ is the splitting between the levels, $\chi$ is the OAT strength, $\Omega$, $\omega_M$, and $\phi$ are respectively the strength, frequency, and phase of the linear drive. First thing to notice is that, the time independent part of the Hamiltonian creates a frequency gap between the levels $M$ and $M-1$ by $\Delta_M = \omega_0 + \chi (2M-1)$. For $\chi \leq \omega_0/N$, $\Delta_M$ is guaranteed to be different for all $M$. Secondly, moving to the frame rotating with the frequency of drive $\omega_M = \Delta_M$, in the regime $\omega_0,\chi \gg \Omega$ only the levels $M$ and $M-1$ are driven. We can effectively write the control Hamiltonian under Rotating Wave Approximation (RWA) as,
\begin{multline}
    {\mathscr{H}}_{M}^{\text{RWA}} = \sum_J \frac{g^J_M}{ d_N^J} \sum_{i_J=1}^{d_N^J}  e^{-i \phi} \ket{J,M,i_J}\bra{J,M-1,i_J} \\+ e^{i \phi} \ket{J,M-1,i_J}\bra{J,M,i_J},
\end{multline}
where $g^{J}_{M} = \Omega \sqrt{(J-(M-1))(J+(M-1)+1)}/2$ is the frequency of the Rabi oscillations. 

\begin{figure*}[t!]
    \centering
    \includegraphics[width=0.78\linewidth]{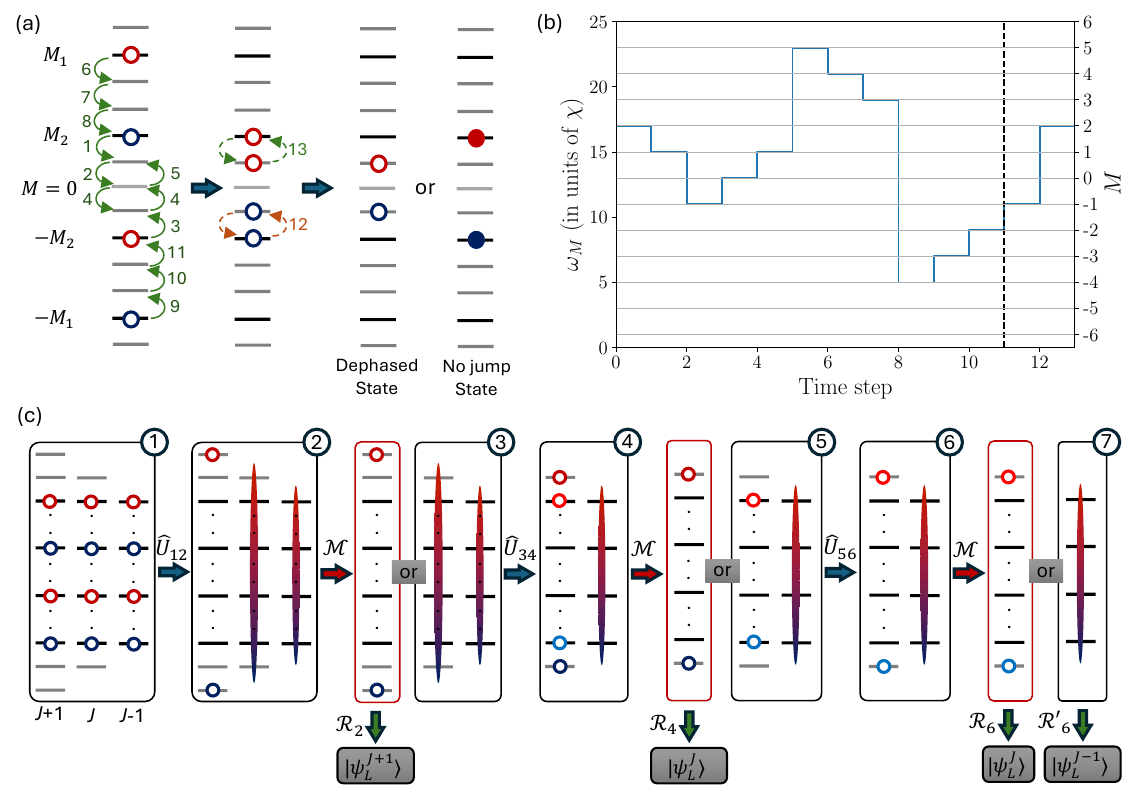}
    \caption{\textbf{Schematic of distinguishing states using collective control}. (a) Outline of the process to distinguish dephased and no jump state, with time in steps (1-11) being $\pi/2g_M^J$ and in steps (12) and (13) being $\cos^{-1}(c_{M_1}/g_{-1}^J)$ and $\pi + \cos^{-1}(c_{M_1}/g_{2}^J)$. (b) Frequency $\omega_M$ of drive tone during each time step of the process in (a). (c) Schematic for QND $\hat{J}^2$ measurement. At first a process similar to (a) is used to transfer populations to levels $\pm (J+1)$, but this changes the populations in other $J$ subspaces. Then generalised parity measurement $\mathscr{M}$ is performed using dispersive coupling to distinguish the states in $J+1$ and other subspaces. Depending on the outcome on ancilla, the ensemble state is either in $J+1$ subspace (red box) or not (black box numbered 3). If the state after measurement is in $J+1$ subspace, the information is re-encoded ($\mathscr{R}_2$) to get the new encoding state $\ket{\psi_L^{J+1}}$ in $J+1$ subspace. If the state is not in $J+1$ subspace, we repeat all the steps again to distinguish remaining states and/or subspaces.}
    \label{Apfig:QND}
\end{figure*}

The unitary generated from the control Hamiltonian, $\hat{U}_{M}(t,\phi) = \text{Exp}\left[-i {\mathscr{H}_{M }^{\text{RWA}} }t  \right]$, transforms the state $\ket{\psi} = \alpha \ket{J,M,i_J} + \beta \ket{J,M-1,i_J}$ as,
\begin{multline}
    \hat{U}_{M}(t,\phi) \ket{\psi} = \left( \alpha \cos{(g_{M}^Jt)} -ie^{-i\phi}\beta \sin{(g_{M}^Jt)} \right) \ket{J,M,i_J} \\
    + \left( \beta \cos{(g_{M}^Jt)} -ie^{i\phi}\alpha \sin{(g_{M}^Jt)} \right) \ket{J,M-1,i_J} .
    \label{AppEq:RabiOsc}
\end{multline}
Here, the direction of the rotation in $x-y$ plane can be controlled by changing the initial phase $\phi$ in the linear drive. Thus by choosing specific values of $t$ and initial phase $\phi$, the levels $M$ and $M-1$ can be manipulated as per requirements. Sequentially applying this control between different levels we can then perform both encoding and recovery. We have outlined one such recovery process via sequential application of control unitary in Fig~\ref{Apfig:QND}.

%-------------------------------------===========================

\subsection{Quantum non-demolition measurement for error syndrome detection\label{Sec:IVC}}
To implement error syndrome measurement, we need to measure two key observables: $\hat{J}_z$ and $\hat{J}^2$. While not strictly necessary, this can be done using an auxiliary qubit with a suitable coupling. Here we outline an exemplary syndrome measurement process with the widely used dispersive coupling,
\begin{equation}
    \mathcal{H}^{\text{disp}}_{int} = \hat{\sigma}_z \hat{J}_z.
\end{equation}
This interaction introduces a phase shift on the auxiliary qubit that depends on the spin excitation number $M$. Together with auxiliary qubit rotation, we can use quantum signal processing (QSP)~\cite{Motlagh2024,Fong2026} to engineer arbitrary $M$ dependence on the   phase shift, i.e. $\ket{0} \rightarrow \ket{0}$ and $\ket{1} \rightarrow e^{i \Phi_M} \ket{1}$, for any $\Phi_M$. For $M$ belonging to encoding levels we pick $\Phi_M = \pi$, and $\Phi_M = 0$ otherwise. Then one can use ancilla $X$ measurement to distinguish between the no--jump/dephasing case $(m=0)$ and the spin gain/loss case $(m \neq 0)$. Similarly, one can further distinguish between spin gain $(m=1)$ and loss $(m=-1)$ cases. This method is also known as the generalized parity measurement.

After that, we can distinguish states corresponding to different total angular momentum $J$ by employing dispersive coupling and individual level transitions. Without loss of generality, we illustrate the process when $m=0$ is obtained in the generalized parity measurement. We first distinguish the states within the $J+1$ subspace from those belonging to other subspaces $(J,J-1)$. To achieve this, we map the corrupted logical state $|0_L\rangle$ in the $J+1$ subspace to $|M = -(J + 1)\rangle$, and similarly that of $|1_L\rangle$ to $|M = J + 1\rangle$. We then perform a generalized parity measurement that differentiates the states $|M = \pm (J + 1)\rangle$ from all other $M$ states. A positive measurement outcome indicates that the state resides within the $J+1$ subspace. Otherwise, the state must belong to either $J$ or $J-1$ subspace because $M=\pm(J+1)$ states are not available. Sequentially, we map the uncorrupted code states to $M=\pm J$ levels. A generalized parity measurement is then performed to differentiate the states $|M = \pm J \rangle$ from all other $M$ states, thus identifying the uncorrupted states. We note that although the dephased state can reside also in the $J$ subspace, they will not occupy the $M=\pm J$ state after the aforementioned operation.
This happens because the code is designed such that the dephased state is orthogonal to the uncorrupted state. The unitary which takes the no jump state to levels $M=\pm J$, will only take the orthogonal dephased state to levels other than $M=\pm J$ (see Fig.~\ref{Apfig:QND} (a)-(b)). 
Finally, we map the dephased states to levels $M=\pm J$ again and perform the generalized parity measurement. This measurement distinguishes the dephased state from the state in $J-1$ subspace. This whole process is schematically represented in Fig.~\ref{Apfig:QND}(c). See Fig.~\ref{Apfig:collcont} in Appendix~\ref{App:C} for an alternate recovery process, which can be performed after both $j$ and $m$ have been determined using syndrome measurements.

%-------------------------------------===========================
\subsection{Information extraction and re-encoding\label{Sec:IVD}}
\begin{figure}
    \centering
    \includegraphics[width = \linewidth]{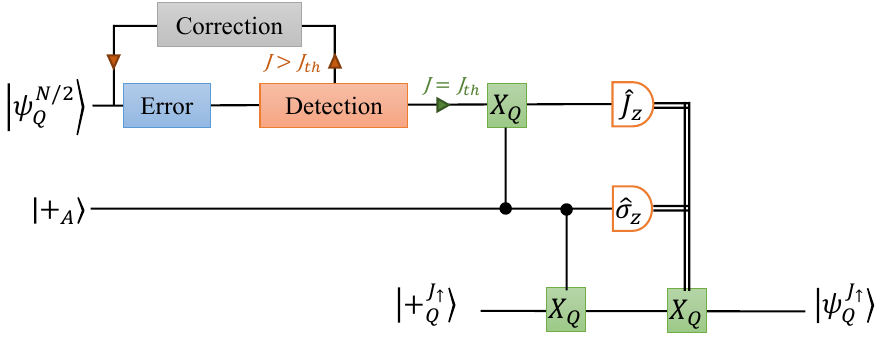}
    \caption{\textbf{Quantum circuit for teleportation.} This protocol is implemented when $J$ reaches threshold $J_{th}$. If measurement outcome for operators $(\hat{J}_z,\hat{\sigma}_z)$ is any one of $\{(M_1,0), (-M_2,0), (M_2,1), (-M_1,1)\}$, then we perform teleportation by-product correction by applying $X_Q$ gate in the end.}
    \label{Apfig:TeleCir}
\end{figure}

As described earlier, once the state decoheres to a state below threshold $J\leq J_{th}$ subspace, it is not possible to encode information any more. In such cases, we need to perform information retrieval followed by transfer to a fresh ensemble initialized in $J_{\uparrow} > J_{th}$ subspace. The circuit which performs this procedure is given in Fig.~\ref{Apfig:TeleCir}. The first part involves usual syndrome extraction and recovery process outlined in previous section as long as $J>J_{th}$. As soon as $J = J_{th}$, we start extraction and re-encoding process and at this point the joint state of the logical qubit and ancilla is given by $\ket{\Psi} = \ket{\psi_Q^{J_{th}}}\ket{+_A}$, where the superscript in logical state denotes the $J$ subspace in which the information was encoded. We perform conditional rotation on this joint state, which gives us the state
\begin{multline}
    \hat{C}_{AL} \ket{\Psi} = \frac{\ket{0_Q^{J_{th}}}}{\sqrt{2}} \left(\alpha_0 \ket{0_A} + \alpha_1 \ket{1_A}\right) \\+ \frac{\ket{1_Q^{J_{th}}}}{\sqrt{2}} \left(\alpha_0 \ket{1_A} + \alpha_1 \ket{0_A}\right).
\end{multline}

Note that this operation requires dispersive coupling between ancilla and ensemble along with global rotation, which effectively implements the unitary $\hat{U}_{AL} = \text{Exp}[-i \pi \hat{J}_x \otimes \hat{\sigma}_z]$ on the joint physical state. As evident from the equation above, measurement of the $\hat{J}_z$ operator transfers the encoded information to the ancilla. Since measurement of $\hat{J}_z$ operator collapses the logical state to one of the levels in encoding space, we consider the outcome 0 (or 1) if the level after measurement corresponds to encoding space of logical qubit 0 (or 1).   Next step is to re-encode this information into a fresh ensemble, for which we first initialize the ensemble in $\ket{+_Q^{J_{\uparrow}}}$ state. Then we perform the conditional rotation on the new ensemble and ancilla joint state $\ket{\Phi}$,
\begin{multline}
    \hat{C}_{AL} \ket{\Phi} =  \left(\alpha_{l} \ket{0_Q^{J_{\uparrow}}} + \alpha_{\overline{l}} \ket{1_Q^{J_{\uparrow}}}\right) \frac{\ket{0_A}}{\sqrt{2}}\\+  \left(\alpha_{l} \ket{1_Q^{J_{\uparrow}}} + \alpha_{\overline{l}} \ket{0_Q^{J_{\uparrow}}}\right)\frac{\ket{1_A}}{\sqrt{2}},
\end{multline}
where $l$ is $0$ if measurement outcome of $\hat{J}_z$ is $-M_1$ or $M_2$ and it is $1$ otherwise. Finally we perform measurement of $\hat{\sigma}_z$ operator on ancilla which re-encodes the information in the new ensemble. The final logical state at this step is given by 
\begin{equation}
    \ket{\psi_Q} = \alpha_k \ket{0_Q^{J_{\uparrow}}} + \alpha_{\overline{k}} \ket{1_Q^{J_{\uparrow}}},
\end{equation}
where $k$ is 0 if measurement of $(\hat{J}_z,\hat{\sigma}_z)$ operators have outcomes $\{(-M_1,0), (M_2,0), (-M_2,1), (M_1,1)\}$ and it is 1 otherwise. If $k$ is 1, we apply another logical $\hat{X}_Q$ operator on the logical qubit to get the original encoding in larger $J$ subspace back. This step is required if and only if one measures $\ket{1}$ state only once, i.e., either on logical qubit or on ancilla. We call it the \textit{teleportation by-product} correction step.
Note that implementing a logical $\hat{X}_Q$ gate is equivalent to flipping all the spins of  the encoding state and it can be done via unitary evolution $\hat{U}_{X} = \text{Exp}[-i \pi \hat{J}_x]$. However, this is not necessarily required as technically we can redefine the logical qubits and keep on updating the definition if we find $k=1$.

%----------------------------------------------------------------------------------
\section{Application in quantum sensing\label{Sec:V}}

{Additionally, our} QEC scheme can also protect spin-ensemble quantum sensors. {Before presenting the scheme, we note that sensing requires the state to respond to (i.e., accumulate) the signal, which constitutes an additional design constraint on the code states. Accordingly, the quantum sensing code will, in general, differ from the codes in the preceding sections, which are optimized only for protecting quantum information.} We follow the standard Ramsey interferometry \cite{Giovannetti2006, Degen2017}, where the ensemble is initialized as $(\ket{0_c}+\ket{1_c})/\sqrt{2}$ and evolves under the signal Hamiltonian $\mathscr{H}_\textrm{sig} =  \omega \hat{J}_z$. After time $t$, a phase $\varphi$ {is} accumulated between the logical basis
\begin{equation}
    \varphi = \omega t \left(\bra{0_c} \hat{J}_z \ket{0_c} - \bra{1_c} \hat{J}_z \ket{1_c} \right).
    \label{Eq:QSPhase}
\end{equation}
Measuring the ensemble then allows us to estimate the phase and hence the unknown parameter $\omega$. Though entanglement in the ensemble can enhance sensitivity, such enhancement is volatile against decoherence~\cite{Huelga1997}.

{While it is tempting} to suppress the decoherence {using} QEC, its efficacy crucially depends on the {structure of decoherence} and {the signal Hamiltonian \cite{Zhou2018}. 
Explicitly,} the KL condition requires identical decoherence rates for the logical basis states, i.e.,
\begin{multline}
    \bra{0_c}\hat{J}_z\ket{0_c} = \text{Tr} \left[ \sum_n \hat{\sigma}_{-,n} \ket{0_c}\bra{0_c} \hat{\sigma}_{+,n} \right]\\= \text{Tr} \left[\sum_n \hat{\sigma}_{-,n} \ket{1_c}\bra{1_c} \hat{\sigma}_{+,n}\right] =  \bra{1_c}\hat{J}_z\ket{1_c}.
\end{multline}
This implies the accumulated phase in Eq.~\eqref{Eq:QSPhase} {vanishes}, which implies that we cannot find any encoded probe state that can protect sensing against individual decay {or pumping} errors. See Appendix~\ref{App:E} for detailed calculations.

However, it is possible to construct QEC codes for other errors, such as collective decay. Since the collective decoherence rate is modified by superradiance{~\cite{Dicke1954,Gross1982}}, the KL condition becomes
\begin{equation}
    \bra{0_c} \hat{J}_z^2-\hat{J}_z \ket{0_c} = \bra{1_c}\hat{J}_z^2-\hat{J}_z\ket{1_c},
\end{equation}
which can be satisfied while the phase in Eq.~\eqref{Eq:QSPhase} remains non-zero. An example of such code involves only one Dicke state in each logical basis: $\ket{0_c} = \ket{N/2,-N/2+1}$ and $\ket{1_c} = \ket{N/2,N/2}$.
\begin{figure}[t]
    \centering
    \includegraphics[width=0.9\columnwidth]{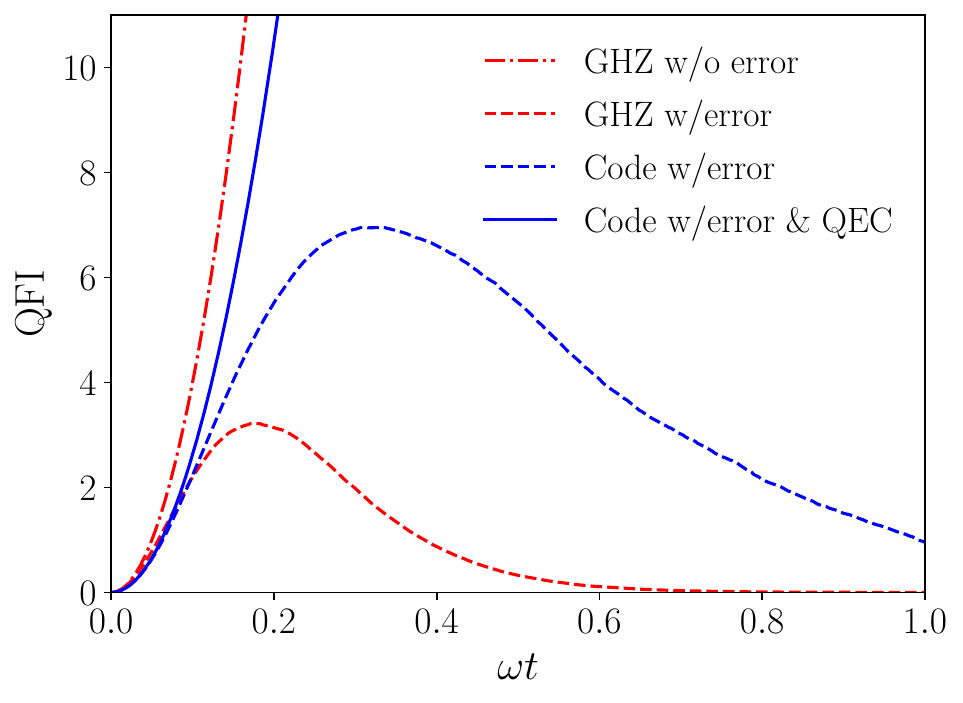}
    \caption{\textbf{Quantum Fisher information under collective decay.} The plot shows QFI of a GHZ (red) and encoded (blue) probe state under collective decay. It is characterized by Bures distance between the probe state evolved with and without signal \cite{Braunstein1994}. QEC (solid) significantly improves the QFI over bare evolution (dashed), and approaches the maximal lossless QFI (dot-dashed). Dynamics is obtained by averaging over $10^4$ quantum trajectories \cite{Dalibard1992}, for an ensemble of 20 spins with collective decay $\Gamma_{-1} = 0.5$ and signal strength $\omega= 0.1$.}
    \label{fig:QS_QEC}
\end{figure}

In Fig.~\ref{fig:QS_QEC}, we compare the quantum {Fisher} information (QFI) in our QEC sensing states with the most sensitive probe state when there is no loss: the GHZ state $(\ket{N/2,N/2}+\ket{N/2,-N/2})/\sqrt{2}$. Without QEC, the QFI of both states is suppressed because collective decay will eventually destroy the superposition. However, QEC can restore the QFI of our probe state to be comparable with the lossless GHZ state.

%----------------------------------------------------------------------------
\section{Potential Physical platforms for implementation\label{Sec:VI}}

Our protocol can be {adequately} implemented by an one-axis twisting  interaction~\cite{Kitagawa1993} for universal control and a dispersive coupling for syndrome measurements. Both interactions can be engineered in various ensemble-based platforms. Here, we discuss two such platforms in detail: 
{ensemble of ions in a Penning trap and nitrogen-vacancy (NV) centers in diamond.} 
 
\subsection{One-axis twisting interaction\label{sec:oat}}

In Penning trap systems, OAT interaction can be generated by applying two detuned laser drives. This creates a spin-dependent optical dipole force that couples the internal states of ions to their collective, center-of-mass motion. After tracing out the motion that is barely excited, the effective Hamiltonian of the internal states takes the form of OAT \cite{Bohnet2016}. 

Alternatively, in the hybrid setup where NV center ensemble is coupled to a superconducting resonator, spin squeezing can be achieved by operating the system in dispersive regime~\cite{Kubo2010,Pradana2021}. In this regime, the resonator mode is far detuned from the ensemble transition energy and can be adiabatically eliminated. The effective Hamiltonian thus obtained in the spin ensemble subspace contains the $\hat{J}_z^2$ term and provides the mechanism from OAT squeezing. Interestingly, the same interaction can also be achieved using a mechanical resonator~\cite{Bennett2013,Teissier2014}. 

\subsection{Dispersive coupling\label{sec:dispersive}}
In Penning traps, it was recently demonstrated that the ions near the center of the crystal are generally stationary or slowly moving, so they can be addressed by focused lasers~\cite{McMahon2024}. Therefore, they can be used as auxiliary qubits for dispersive coupling. In such a setup, we can initialize the central ion into an auxiliary state distinct from the computational basis. By employing suitable sideband interactions, we create a hybrid quantum system where both the collective ion ensemble states and the auxiliary state couple coherently to the ion crystal's center-of-mass vibrational mode via a spin-dependent force~\cite{Sawyer2012,Safvi2018}. The effective interaction Hamiltonian for this hybrid system {is}
\begin{equation}
\mathcal{H}_{{int}} = \Delta \hat a^{\dagger} \hat a + \chi \hat\sigma_z (\hat a + \hat a^{\dagger}) + g (\hat J_- \hat a^{\dagger} + \hat J_+ \hat a),
\end{equation}
where $\hat \sigma_z$ corresponds to the auxiliary state operator, $\hat a$ and $\hat a^{\dagger}$ denote the vibrational mode's annihilation and creation operators, $\hat J_+$ and $\hat J_-$ represent collective spin raising and lowering operators of the ion ensemble, and $\Delta$ indicates the detuning parameter. In the regime of large detuning (large $\Delta$), we can adiabatically eliminate the motional mode, effectively obtaining a dispersive coupling (upto a global rotation) between the auxiliary ion state and the ensemble of ions.

In the NV center platform, the dispersive coupling can be implemented in a hybrid setup that involves superconducting qubits. For example, as shown in Ref.~\cite{Zhu2011}, a diamond crystal containing a dense layer of NV centers is glued directly on top of the superconducting circuit (flux qubit) at a separation of less than one micrometer. The current flowing in the flux qubit produces a magnetic field, which couples to the NV$^-$ spins in the diamond crystal and gives rise to the dispersive coupling between the two {quantum} systems. 

%-------------------------------------------------------------------------------------
\section{Discussion\label{Sec:VII}}
{In summary, we {have} developed a general scheme {to implement} quantum error correction {in} spin ensemble systems that cannot be resolved, which is robust  against all prominent errors, including individual and collective errors {arising from} decay, dephasing and pumping, and also spin loss.}
{The versatility of our approach suggests several promising directions.} {First,} the framework can be {adequately} adapted to {several} {experimental platforms such as solid-state defects, atomic ensembles, or trapped-ion crystals, where collective degrees of freedom are accessible but single particle} {resolvability is practically not accessible}. 
Secondly, the ability of our QEC code to encode pure logical information in mixed physical states may also facilitate robust quantum memories in noisy systems. And finally, {implementing a powerful quantum error correction protocol in applications} {where spin ensembles serve as resource for quantum metrology, communication, or distributed computing architectures, could significantly broaden the practicality of spin ensemble based quantum technologies.}

% \section*{Data availability}
% All relevant data and codes related to the simulations and figures are available online~\footnote{All relevant data and codes related to the simulations and figures are available in GitHub repository and can be accessed via this link \url{https://github.com/qid-iitb/SpinEnsembleQEC}.}.

\begin{acknowledgments}
The authors thank R. Tullu, P. Iyer and S. Vinjanampathy for useful discussions. 
H.S. acknowledges financial support from the Prime Minister’s Research Fellowship (ID: 1302055), Govt. of India and Mitacs Globalink Research Award (Ref: IT34115), Canada. 
H.S.D. acknowledges funding from SERB-DST, India under Core-Research Grant (CRG/2021/008918) and IRCC, IIT Bombay (RD/0521-IRCCSH0-001 and RD/0523-IOE00I0-082). H.-K.L. acknowledges support from NSERC Discovery (RGPIN-2021-02637) and Canada Research Chairs (CRC-2020-00134).
\end{acknowledgments}

\appendix
\section{Details on System dynamics}\label{App:A}
%------------------------------------============================
\begin{table*}[t]
    \caption{\textbf{Values of $X_{j,m}^{J,M}$ and $C_j^{N,J}$.}\label{tab:decoherences}}
    \centering
    \begin{ruledtabular}
    \begin{tabular}{cccccc} 
     &\multicolumn{3}{c}{$X_{j,m}^{J,M}$}&$C_j^{N,J}$\\
         \backslashbox{$j$}{$m$}&-1&0& +1&--\\ \hline
         -1&     $-\sqrt{(J+M)(J+M-1)}$&$\sqrt{(J+M)(J-M)}$& $\sqrt{(J-M)(J-M-1)}$&${1\over 2J}\left({{N/2}+J+1 \over 2J+1} \right)$\\ 
         0&     $\sqrt{(J+M)(J-M+1)}$&$M$& $\sqrt{(J-M)(J+M+1)}$&${1\over 2J}\left({{N/2}+1 \over J+1} \right)$\\
         +1&     $\sqrt{(J-M+1)(J-M+2)}$&$\sqrt{(J+M+1)(J-M+1)}$& $-\sqrt{(J+M+1)(J+M+2)}$&${1\over 2(J+1)}\left({{N/2}-J \over 2J+1} \right)$\\
    \end{tabular}
    \end{ruledtabular}
\end{table*}
\subsection{Error processes\label{App:A1}}
In order to correct the errors that distort the quantum information, we need to know how the encoding state is transformed under these errors. We consider broadly two kinds of errors, one {that preserves the} number of spins $N$ in the ensemble, i.e., collective and individual spin decay, dephasing and pumping, while the other changes $N$, i.e., spin loss errors. In this section we will discuss the former and discuss spin loss separately {in Appendix~\ref{App:D}}.

Consider the master equation introduced in the main text,
\begin{equation}
    \dot{\hat{\rho}}_c = \sum_{m =-1}^1 \left( {\Gamma_{m} \over 2} \hat{\mathcal{L}}_{\hat{J}_{m}}[\hat{\rho}_c] + \sum_{n=1}^{N} {\gamma_{m} \over 2} \hat{\mathcal{L}}_{\hat{\sigma}_{m,n}}[\hat{\rho}_c] \right),
    \label{AppEq:errModel}
\end{equation}
where $\mathcal{L}_{\hat{A}}[\hat{\rho}] = 2 \hat{A} \hat{\rho} \hat{A}^{\dagger} - \{\hat{A}^{\dagger}\hat{A},\hat{\rho}\}$ is the Lindblad superoperator and $\Gamma_{m}$ and $\gamma_{m}$ represents the decoherence rates for collective and individual processes respectively. This equation can be rewritten as
\begin{multline}
    \dot{\hat{\rho}}_c = -{i \over \hbar} [\mathcal{H}_{eff},\hat{\rho}_c] + \sum_{m \in \{0,\pm 1\}} \left( {\Gamma_{m}} {\hat{J}_{m}}\hat{\rho}_c {\hat{J}_{m}^\dagger}\right.
    \\ \left. + {\gamma_{m}}  \sum_{n=1}^{N}  {\hat{\sigma}_{m,n}}\hat{\rho}_c{\hat{\sigma}_{m,n}^\dagger} \right),
    \label{AppEq:ErrMod}
\end{multline}
where $\mathcal{H}_{eff}$ is the effective non--hermitian Hamiltonian 
\begin{equation}
    \mathcal{H}_{eff} = - i \sum_{m \in \{0,\pm 1\}} \left({\Gamma_{m} \over 2} \hat{J}_{m}^\dagger \hat{J}_{m}  + {\gamma_{m} \over 2} \sum_{n=1}^{N} \hat{\sigma}_{m,n}^\dagger\hat{\sigma}_{m,n} \right).
\end{equation}
Without quantum jump, the state evolves under this non--hermitian Hamiltonian (i.e. first term in Eq.~\eqref{AppEq:ErrMod}). The evolution is not unitary, so both the state and the norm of $\hat{\rho}_c$ will change. The second term gives the information about how the state is transformed when a jump happens. For collective decoherence processes, the second term changes the collective basis state as follows~\cite{Shammah2018},
\begin{multline}
   \hat{J}_m \overline{\ket{J,M}\bra{J,M'}} \hat{J}^\dagger_m 
    \\= X_{0,m}^{J,M} \overline{\ket{J,M+m}\bra{J,M'+m}} X_{0,m}^{J,M'},
\end{multline}
where $m = \{-1,0,+1\}$ corresponds to a decay, dephasing and pumping process respectively and values of coefficient $X_{0,m}^{J,M/M'}$ {are} mentioned in Table~\ref{tab:decoherences}. Note that there is an extra subscript `0' in the coefficient `$X$' in comparison to the one mentioned in the main text, which is introduced here to clarify the differences and similarities between individual and collective processes in the discussion later on.

Similarly, the dynamics due to individual decoherence processes in Eq.~\eqref{AppEq:ErrMod} transforms the collective basis states as \cite{Chase2008},
\begin{align}
    &\sum_{n=1}^N \hat{\sigma}_{m,n} \overline{\ket{J,M}\bra{J,M'}} \hat{\sigma}_{m,n}^{\dagger} \nonumber\\
    &= \sum_{j} C_{j}^{N,J} X_{j,m}^{J,M}X_{j,m}^{J,M'} \overline{\ket{J+j,M+m}\bra{J+j,M'+m}},
    \label{Eq:ChaseIdentity}
\end{align}
where the coefficients take the value from the Table~\ref{tab:decoherences}. Comparing the equation above with Eq.~\eqref{Eq:ChaseMap}, it can be seen that $\chi_{j,m}^{N,J,M} = \sqrt{C_{j}^{N,J}} X_{j,m}^{J,M}$. It is evident from this expression that for the case of $j=0$, up to some normalization that determines the probability of the jump, both individual and collective errors are the same.

For the state at time $t$ given by,
\begin{equation}
    \hat{\rho}_c(t) = \sum_{M,M'} \rho_{M,M'} \overline{\ket{J,M}\bra{J,M'}},
\end{equation}
the evolved state at time $t+dt$ for $dt\ll1$ can be found by expanding left side of Eq.~\eqref{AppEq:ErrMod} as,
\begin{widetext}
\begin{align}
    \hat{\rho}_c(t+dt) &\approx (\hat{\mathbb{I}}-i\mathcal{H}_{eff}dt) \hat{\rho}_c(t) (\hat{\mathbb{I}}+i\mathcal{H}_{eff}^\dagger dt) + dt \sum_{m \in \{0,\pm 1\}} \left( {\Gamma_{m}} {\hat{J}_{m}}\hat{\rho}_c(t) {\hat{J}_{m}^\dagger}+ {\gamma_{m}}  \sum_{n=1}^{N} {\hat{\sigma}_{m,n}}\hat{\rho}_c(t) {\hat{\sigma}_{m,n}^\dagger} \right) \nonumber\\
    &= \hat{{\rho}}_{\varnothing,\varnothing}(t) + \sum_{m \in \{0,\pm 1\}} \left( \Gamma_m dt \sum_{M,M'} \rho_{M,M'} X_{0,m}^{J,M} \overline{\ket{J,M+m}\bra{J,M'+m}} X_{0,m}^{J,M'} \right) \nonumber \\
    & \qquad \qquad + \sum_{j,m \in \{0,\pm 1\}} \left( \gamma_m dt \sum_{M,M'} \rho_{M,M'} C_j^{N,J} X_{j,m}^{J,M} \overline{\ket{J,M+m}\bra{J+j,M'+m}} X_{j,m}^{J,M'} \right)  \nonumber \\
    &= \hat{{\rho}}_{\varnothing,\varnothing}(t) + \sum_{m \in \{0,\pm 1\}} \hat{\rho}_{0,m}^c(t)  + \sum_{j,m \in \{0,\pm 1\}} \hat{\rho}_{j,m}^l(t),
    \label{AppEq:StDyn}
\end{align} 
\end{widetext}
where $\hat{\rho}_{j,m}^c$ and $\hat{\rho}_{j,m}^l$ represents the error state in $J+j$ subspace after a collective and individual error process $m$ respectively, and  $\hat{{\rho}}_{\varnothing,\varnothing}$ represents the state if no jump occurs. 

As discussed earlier, the non--Hermitian term in the master equation that corresponds to the no--jump case also causes an error.
If error correction is performed frequently such that $dt \ll 1$, we can effectively redefine the logical qubits as $\ket{\Tilde{k}_Q} \propto (\hat{\mathbb{I}}-i\mathcal{H}_{eff}dt)\ket{k_Q}$, which includes the errors introduced by the effective Hamiltonian. The Kraus operator for this error is discussed in the \emph{Logical Mapping} section of the main text. Approximate QEC can be performed with the redefined logical qubits, where for sufficient level spacing, the non--trivial (diagonal) KL condition is required to hold only at the linear order of $dt$:
\begin{equation}
    \bra{\tilde{0}_Q} \hat{E}_{j,m}^\dagger \hat{E}_{j,m} \ket{\tilde{0}_Q} = \bra{\tilde{1}_Q} \hat{E}_{j,m}^\dagger \hat{E}_{j,m} \ket{\tilde{1}_Q} + \mathcal{O}(dt^{3/2}).
    \label{AppEq:KL_approx}
\end{equation}

Since we have exact information about how the state evolves in no jump case, we can use universal collective control~\cite{Gutman2023} to recover encoding states with high fidelity. However, exact QEC is still possible if the no--jump term is included in the KL conditions.

%-----------------------------------------------------
\subsection{Error rates\label{App:A2}}
For small enough $dt$ in Eq.~\eqref{AppEq:StDyn}, at any given time $t$, trace of the states after error tells the probability of a particular decoherence process occurring in the system. These probabilities can be interpreted as the error rates, which are defined for collective and individual processes as $r^{c/l}_{j,m}(t) := \text{Tr}[\hat{\rho}_{j,m}^{c/l} (t)]/dt$. If a logical qubit occupies the levels $\{M_i\}$, then the error rates for an individual error process parameterized by $(j,m)$ is
\begin{align}
    r^{l}_{j,m} &= \gamma_m C_j^{N,J} \sum_{i} \rho_{M_i,M_i}\left( X_{j,m}^{J,M_i} \right)^2, 
\end{align}
and the same for a collective process is 
\begin{align}
    r^{c}_{m} &= \Gamma_m \sum_{i} \rho_{M_i,M_i}\left( X_{0,m}^{J,M_i} \right)^2. 
\end{align}
Since individual process can map state to different $J$ subspace, we define total individual error rate as $r^l_m = \sum_{j=-1}^1 r^l_{j,m}$. It is easy to show that for individual decay and pumping errors, 
\begin{align}
    r^l_{\pm1} &= \gamma_{\pm1} \text{Tr}\left[\sum_{n=1}^N \hat{\sigma}_{\pm,n} \hat{\rho}_c \hat{\sigma}_{\pm,n}^{\dagger}\right]\nonumber = \gamma_{\pm1} \text{Tr}\left[\sum_{n=1}^N {\mathbb{I} \mp \hat{\sigma}_{0,n}\over2} \hat{\rho}_c \right]\nonumber \nonumber\\
    &= \gamma_{\pm1} \left( {N\over 2} \mp \text{Tr}\left[ \hat{J}_z \hat{\rho}_c\right] \right).
   \label{AppEq:locErr}
\end{align}
Since Eq.~\eqref{Eq:KL} (also see Eq.~\eqref{Eq:Appdeperr}) requires $\text{Tr}\left[ \hat{J}_z \hat{\rho}_c\right] = 0$, $r^l_\pm = \gamma_{\pm1} N/2$. Similarly, for individual dephasing errors {$r^l_0 = \gamma_{0}\text{Tr}\left[\sum_{i=1}^N \mathbb{I}~\hat{\rho}_c /4\right] = \gamma_{0}N/4$}.
\begin{table}[h]
    \caption{\textbf{Error rates for an arbitrary encoding state $\ket{\psi_Q}$.}\label{ApTab:ErrRate}}
    \centering
    \begin{ruledtabular}
    \begin{tabular}{ccccc} 
         \backslashbox{$l/c$}{$m$}&-1&0& +1\\ \hline 
         $r^c_m/\Gamma_m$&     $J(J+1)-M_1M_2$&$M_1M_2$& $J(J+1)-M_1M_2$\\
         $r^l_m/\gamma_m$&     $N/2$&$N/4$& $N/2$\\
    \end{tabular}
    \end{ruledtabular}
\end{table}

\begin{figure}[t!]
    \centering
    \includegraphics[width=\linewidth]{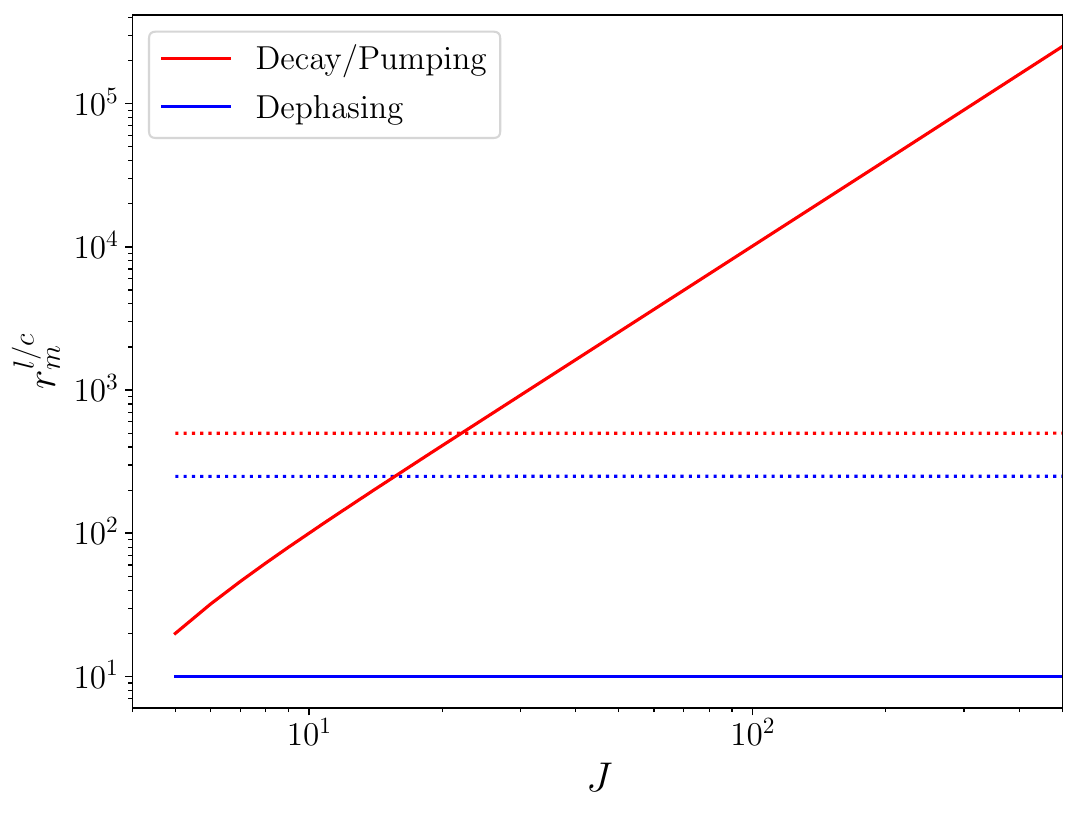}
    \caption{\label{Apfig:ErrRate} \textbf{Error rates for different decoherence processes.} The plot shows the error rates of individual (dotted) and collective (solid) errors for dephasing (blue) and decay/pumping (red) process.  The ensemble contains $N = 10^3$ spins and encoded states occupy levels $M_1 = 5$ and $M_2 = 2$, decoherence rates in the ensemble are $\gamma_m = \Gamma_m = 1$ for all $m$.}
\end{figure}

Table~\ref{ApTab:ErrRate} shows the error rates for a logical qubit encoded with our encoding in Eq.~\eqref{Eq:AssyCode}. The error rate is independent of the logical information as guaranteed by KL conditions. It can be seen that the collective error rates depend on the encoding levels and $J$, while the individual error rates are determined by only the total number of spins $N$.
For a fixed choice of $M_1$ and $M_2$, the collective error rate generally increases as $J$ due to supperradiance, and is approximately proportional to $J^2$ when $J$ is large.  On the other hand, because individual errors originate from spins interacting with local environment, so its rate naturally does not depend on the global properties of the ensemble, such as $J$. Figure~\ref{Apfig:ErrRate} shows the error rates for several error processes in the ensemble for different $J$. 

%%%-----------------------
\section{Solution to Knill-Laflamme conditions\label{App:B}}
Under logical mapping, the state after an error can be described as,
\begin{equation}
    \hat{\rho}_L(t+dt) = \mathcal{E}(\hat{\rho}_L(t)) = \sum_{j,m \in \{\varnothing,0,\pm1\}} \hat{E}_{j,m}\hat{\rho}_L(t) \hat{E}_{j,m}^\dagger,
    \label{AppEq:OSR}
\end{equation}
where $\mathcal{E}(\hat{\rho})$ is the error channel through which the state $\hat{\rho}$ undergoes.
The KL conditions in Eq.~\eqref{Eq:KL} ensure that the decohered states are orthogonal to each other, so that errors can be distinguished. In addition to this, the decoherence rates of $\ket{0_Q}$ and $\ket{1_Q}$ are equal so that the environment doesn't gain any information about the quantum state during decoherence, our encoded information can thus be retained in the system.

The condition for error detection is satisfied if one ensures sufficient level spacing in the encoding states such that the states for decay and pumping errors are mapped to distinct levels. Additionally, we require the following condition to be satisfied 
\begin{equation}
    \bra{0_Q} \hat{E}_{0,0} \ket{0_Q} = \bra{1_Q} \hat{E}_{0,0} \ket{1_Q} = 0,
    \label{Eq:Appdeperr}
\end{equation}
to ensure that the dephased state is distinguishable from the encoded state. Error correction then requires solving the diagonal KL condition,
\begin{equation}
     \bra{0_Q} \hat{E}_{j,m}^\dagger \hat{E}_{j,m} \ket{0_Q} = \bra{1_Q} \hat{E}_{j,m}^\dagger \hat{E}_{j,m} \ket{1_Q}.
    \label{AppEq:KL}
\end{equation}

As discussed earlier, individual decoherence processes could take a particular collective state to one of the three $J$ subspaces i.e, $J$ or $J \pm 1$. Therefore for the code to successfully correct the errors, the KL conditions in Eq.~\eqref{AppEq:KL} must be satisfied irrespective of the subspace of the error state. This problem can be cast into a matrix equation of the form, $K_{m} \vec{c}=0$, where $\vec{c}= \vec{c}^{(0)}-\vec{c}^{(1)}$ is a column vector whose elements are the superposition amplitudes of the logical qubits and $K_{m}$ is a $4 \times (2J+1)$ matrix, with each column $K_{m}^M$ of the matrix corresponding to a $M = \{ -J,\cdots , J\}$ level given by,
\begin{equation}
    K_{m}^M = \left( (\chi_{1,m}^{N,J,M})^2,(\chi_{0,m}^{N,J,M})^2,(\chi_{-1,m}^{N,J,M})^2, 1 \right)^T
\end{equation}
where the last element ensures the orthonormality condition between $\ket{0_Q}$ and $\ket{1_Q}$ and corresponds to no jump $(j,m = \varnothing)$ case when higher order terms in $dt$ are negligible. Since Kraus operators for decay and pumping errors are complex conjugate of each other, they have the same KL conditions. As a result, for the simplest case of 2 levels each in $\ket{0_Q}$ and $\ket{1_Q}$, protection against individual spin decay requires
\begin{equation}
    \begin{pmatrix}
        1 & 0 & 0 & 1\\\addlinespace
        0 & 1 & 0 & \frac{M_1}{ M_2}\\\addlinespace
        0 & 0 & 1 & -\frac{M_1}{ M_2}
    \end{pmatrix}
    \begin{pmatrix}
        |c_{M_1}|^2\\
        -|c_{M_2}|^2\\
        |c_{-M_2}|^2\\
        -|c_{-M_1}|^2
    \end{pmatrix}
    = 0,
\end{equation}
where we have written the matrix equation in row-reduced form and removed all zero rows and columns.
Solution to above equation is then given by
\begin{subequations}
    \begin{equation}
        |c_{M_1}|^2 = |c_{-M_1}|^2 = \frac{M_2 }{ M_1+M_2}
    \end{equation}
    \begin{equation}
        |c_{M_2}|^2 = |c_{-M_2}|^2 = \frac{M_1 }{ M_1+M_2}
    \end{equation}
    \label{AppEq:SolKL}
\end{subequations}

%-------------------------------------===========================
\section{Recovery process}\label{App:C}
{Consider the encoded qubit  $\hat{\rho}_L = \sum_{M,M'} \rho_{M,M'} \overline{\ket{J,M}\bra{J,M'}}$, then under an error process parameterized by indices $(j,m)$, the state is transformed as 
\begin{multline}
    \hat{\rho}_E = {1\over r_{j,m} d_N^J}\sum_{i_J=1}^{d_N^J} \sum_{M,M'} \rho_{M,M'} X_{j,m}^{J,M} X_{j,m}^{J,M'} 
    \\ \ket{J+j,M+m,i_J}\bra{J+j,M'+m,i_J}
\end{multline}
Note that we get this state after measurement of $\hat{J}^2$, which projects the state to $J+j$ subspace and $r_{j,m}$ is the appropriate normalization. In practice there could be several methods to perform recovery based on different universal control techniques. Here, we focus on one such process using specific unitaries described in Sec.~\ref{Sec:IVB} of main text.
\begin{figure*}
    \centering
    \includegraphics[width=0.65\linewidth]{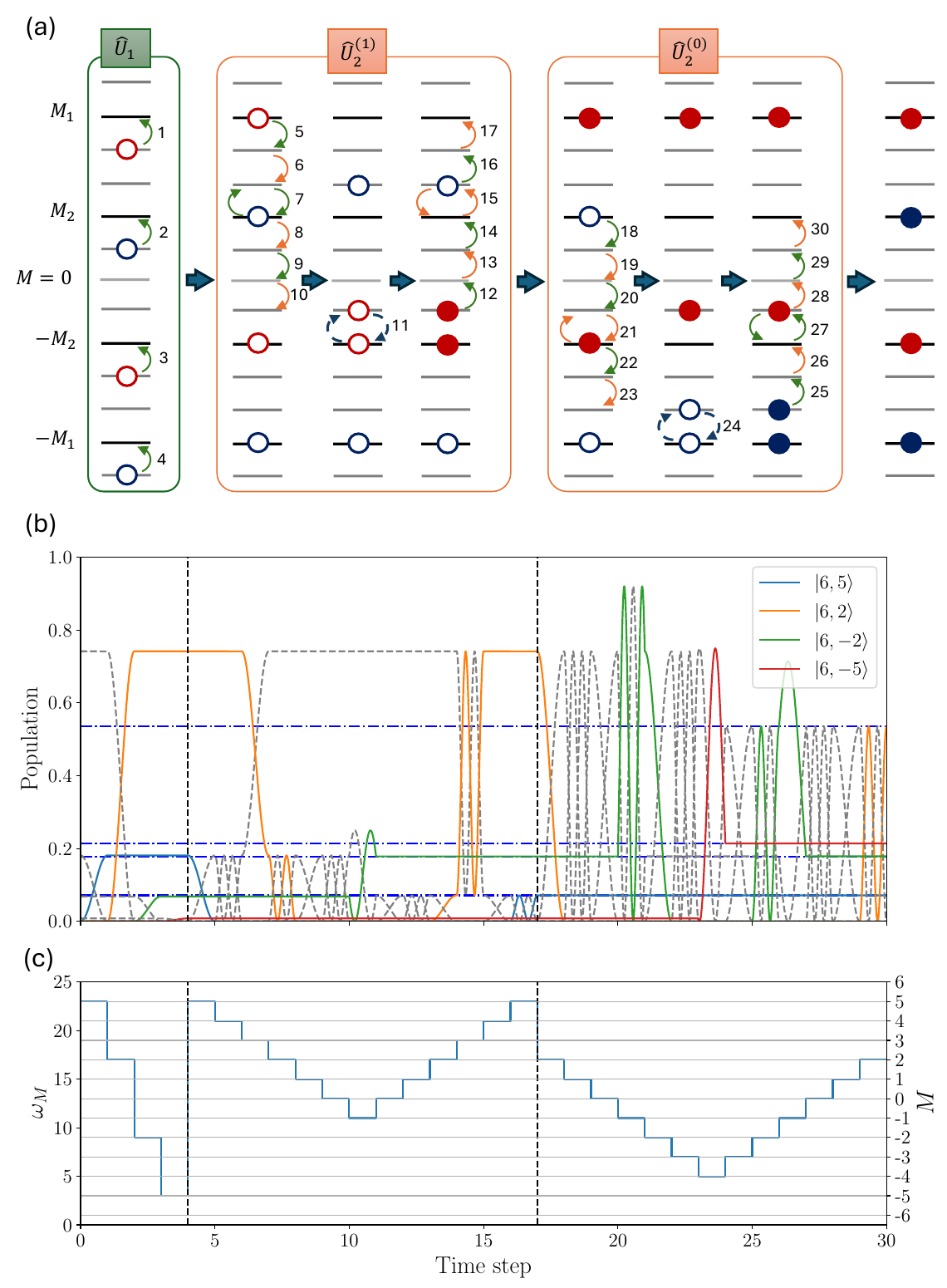}
    \caption{{\textbf{Schematic of recovery process using collective control.} (a) Outline of the process to recover from individual decay error to lower $J$ subspace, i.e., $(j,m) = (-1,-1)$. (b) Population dynamics of the encoding levels during the entire process. Population of encoded state is marked by blue dot-dashed lines for reference and the populations for levels other than the encoding space is shown using gray dashed lines. (c) Frequency pulse for the linear drive applied during the entire process.}}
    \label{Apfig:collcont}
\end{figure*}

The first step in recovery is to bring the state back to the encoding levels by using unitaries $\hat{U}_1 = \prod_M \hat{U}_{M+m}(\pi/2g^J_{M+(m+1)/2},0)$, where product runs over all levels $\{M\}$ in the codespace, which gives the state
\begin{widetext}
\begin{align}
    \hat{\sigma}_E &= \hat{U}_1~ \hat{\rho}_E ~ \hat{U}^\dagger_1 
    = {1\over r_{j,m} d_N^J}\sum_{i_J=1}^{d_N^J} \sum_{M,M'} \rho_{M,M'} X_{j,m}^{J,M} X_{j,m}^{J,M'} 
     \ket{J+j,M,i_J}\bra{J+j,M',i_J}
\end{align}
This step is however required only if $m\neq 0$. 
Next, we adjust the superposition amplitude and phase to the encoding states in the $J+j$ subspace.  This can be implemented by the unitary $\hat{U}_2 = \hat{U}_2^{(0)}(\tau_0)~ \hat{U}_2^{(1)}(\tau_1)$, where

    \begin{subequations}
        \begin{multline}
            \hat{U}_2^{(0)}(\tau_0) = \prod_{i=0}^{M_2 - (-M_1+2)} \left( \hat{U}_{M_2-i}((2i+1){\pi \over 2g^J_{M_2-i}},0 )\right) \hat{U}_{-M_1+1}(\tau_0,\pi/2 ) \\
            \prod_{i=0}^{M_2 - (-M_1+2)} \left( \hat{U}_{-M_1+2+i}((2i+3){\pi \over 2g^J_{-M_1+2+i}},0 )\right),
        \end{multline}
        \begin{multline}
            \hat{U}_2^{(1)}(\tau_1) = \prod_{i=0}^{M_1 - (-M_2+2)} \left( \hat{U}_{M_1-i}((2i+1){\pi \over 2g^J_{M_1-i}},0 )\right) \hat{U}_{-M_2+1}(\tau_1,\pi/2 ) \\
            \prod_{i=0}^{M_1 - (-M_2+2)} \left( \hat{U}_{-M_2+2+i}((2i+3){\pi \over 2g^J_{-M_2+2+i}},0 )\right).
        \end{multline}
        \label{AppEq:Redist_unit}
    \end{subequations}
\end{widetext}

As evident the entire unitary $\hat{U}_2$ is a multiple step process and is completed in $4(M_1+M_2)-2$ steps. The state after implementing $\hat{U}_2$ is,
\begin{align}
    \hat{\sigma}_Q &= \hat{U}_2~ \hat{\sigma}_E ~ \hat{U}^\dagger_2 \nonumber \\
    &= {1\over d_N^J}\sum_{i_J=1}^{d_N^J}\sum_{M,M'} \sigma_{M,M'} \ket{J+j,M,i_J}\bra{J+j,M',i_J}
\end{align}
where $\sigma_{M,M'} = \rho_{M,M'}$. We implement $\hat{U}_{-M_1+1}$ and $\hat{U}_{-M_2+1}$ with $\phi = \pi/2$ to avoid any unwanted phase in the recovered state of logical qubits. The time $\tau_0$ for unitary evolution can be determined by demanding that after action of unitary on state, population of state $ \ket{J+j,M_1,i_J}\bra{J+j,M_1,i_J}$ should become $\left|\alpha_0 C_{M_1} \right|^2$, this gives
\begin{equation}
    \cos^2{(g_{12}\tau_0)} = \dfrac{ \left| c_{M_1} \right|^2 X_{j,m}^{J,M_1} + \left| c_{-M_2} \right|^2  X_{j,m}^{J,-M_2} }{ \left| c_{M_1} \right|^2 \left( X_{j,m}^{J,M_1} \right)^2 + \left| c_{-M_2} \right|^2  \left( X_{j,m}^{J,-M_2} \right)^2}
\end{equation}
Similarly, $\tau_1$ is found by changing $M_1 \rightarrow -M_1$ and $-M_2 \rightarrow M_2$ in the equation above. Note that the expression for time $\tau_0$ here is independent of $(\alpha_0, \alpha_1)$ because both levels $M_1$ and $-M_2$ belong to $\ket{0_Q}$ state. 
Similar procedure can be adopted to correct for errors introduced during the no--jump evolution. Figure~\ref{Apfig:collcont} outlines the entire unitary process described above to recover a state which had undergone an individual decay process to a lower $J$ subspace. The unitaries in Eq.~\eqref{AppEq:Redist_unit} have been outlined in Fig.~\ref{Apfig:collcont}(a). 

%========================================================================
{\section{Example QEC protocol}\label{App:D0}
Suppose a protocol requires storing the state $\ket{+}=(\ket{0}+\ket{1})/\sqrt{2}$ for a finite duration. Physically, this information is encoded in the collective state of an ensemble of spins, which undergo spontaneous emission (i.e., individual spin decay). We now outline how this information can be protected over an extended period using the quantum error correction (QEC) scheme described in the main text. This forms the basis of the simulation presented in Fig.~\ref{fig:Encoding}(c) of the main text.
\begin{enumerate}
\item[1.] \textbf{State preparation}.-- In spin-ensemble systems (e.g., with $N=20$ and frequency $\omega_0$), the state can be initialized in the Dicke subspace with $J=10$ by cooling the ensemble to its collective ground state $\ket{\psi}=\ket{J=10,M=-10,i_J=1}$. The required superposition is then generated by applying the control unitary with drive frequency $\omega_{-9}=\omega_0-19\chi$ for a duration $t=\pi/4g_{-9}^{-10}$ with phase $\phi=0$, yielding
\begin{align}
\ket{\psi_I}
&=
\hat{U}_{-9}(\frac{\pi}{4g_{-9}^{10}},0)\,\ket{10,-10,1} \nonumber\\
&=
\frac{\ket{10,-9,1}+\ket{10,-10,1}}{\sqrt{2}} .
\end{align}

\item[2.] \textbf{Encoding information}.-- To preserve the information in the initial state, it is then encoded into levels with $M = \{\pm2, \pm5\}$ using the sequence of control unitaries as follows,
\begin{subequations}
    \begin{multline}
        \hat{U}_{L}^1 = \prod_{M=5}^1 \left(\hat{U}_M(\frac{\pi}{2g_M^{10}},0)\right)~\hat{U}_0(\frac{\arccos{\sqrt{5/7}}}{g_0^{10}},\frac{3\pi}{2}) \\ 
        \prod_{M=-1}^{-8} \left(\hat{U}_M(\frac{\pi}{2g_M^{10}},0)\right),
    \end{multline}
    \begin{multline}
        \hat{U}_{L}^0 = \prod_{M=2}^{-3} \left(\hat{U}_M(\frac{\pi}{2g_M^{10}},0)\right)~\hat{U}_{-4}(\frac{\arccos{\sqrt{2/7}}}{g_{-4}^{10}},\frac{\pi}{2}) \\
        \prod_{M=-5}^{-9} \left(\hat{U}_M(\frac{\pi}{2g_M^{10}},0)\right).
    \end{multline}
\end{subequations}
After applying these unitaries on the initial state, we get the encoded physical state as,
\begin{align}
    \ket{\psi_c(0)} &= \hat{U}_L^0\hat{U}_L^1\ket{\psi_I} \nonumber \\
    &=  \left. \frac{-i}{\sqrt{2}} \left(\sqrt{\frac{2}{7}}\ket{10,-5,1} + \sqrt{\frac{5}{7}}\ket{10,2,1} \right) \right. \nonumber \\
    &\left.+ \frac{-i}{\sqrt{2}}\left(\sqrt{\frac{5}{7}}\ket{10,-2,1} + \sqrt{\frac{2}{7}}\ket{10,5,1} \right) \right.,
    \label{Eq:PhysSt}
\end{align}
where the global phase $-i$ can be neglected and by using the logical map (in main text), this physical state can be written in logical basis as,
\begin{multline}
    \ket{\psi_L} = \frac{1}{\sqrt{2}} \left(\sqrt{\frac{2}{7}}\ket{-5_L} + \sqrt{\frac{5}{7}}\ket{2_L} \right) \\+ \frac{1}{\sqrt{2}}\left(\sqrt{\frac{5}{7}}\ket{-2_L} + \sqrt{\frac{2}{7}}\ket{5_L} \right).
\end{multline}
Note that the physical state in this case is pure as the information was encoded in the Dicke subspace ($J=10$), which has the maximum degeneracy $d_{20}^{10}=1$.

\item[3.] \textbf{Error process}.-- As time elapses, the physical state in Eq.~\eqref{Eq:PhysSt} undergoes decoherences and after a small time $dt$ it is a mixture of no jump and error states. Assuming that time is small enough such that only single error occurs, then as per Eq.~\eqref{AppEq:StDyn}, the state at time $dt$ is,
\begin{widetext}
   \begin{multline}
        \hat{\rho}_c(dt) = \left(1-\frac{\gamma_{-1}dt}{4}N \right) \hat{\rho}_c(0) - \frac{\gamma_{-1}dt}{2} \left(\hat{\rho}_c(0)\hat{J}_z + \hat{J}_z \hat{\rho}_c(0)\right)\\
        + \frac{\gamma_{-1}dt}{N}\sum_{k,k'=0}^1\sum_{l,l'=1}^2 \alpha_k^{} \alpha_{k'}^* c^{}_{M_{l^{}}}c^*_{M_{l'}}~{X^{10,M_l}_{0,-1}X^{10,M_{l'}}_{0,-1}}~\big|10,(-1)^{k+l}M^{}_l-1,1\big\rangle \big\langle 10,(-1)^{k'+l'}M_{l'}-1,1 \big|\\
        + \frac{\gamma_{-1}dt}{N}\frac{1}{d_{20}^{9}}\sum_{i_{9} = 1}^{d_{20}^{9}}\sum_{k,k'=0}^1\sum_{l,l'=1}^2 \alpha_k^{} \alpha_{k'}^* c^{}_{M_{l^{}}}c^*_{M_{l'}}~{X^{10,M_l}_{-1,-1}X^{10,M_{l'}}_{-1,-1}}~\big|9,(-1)^{k+l}M^{}_l-1,i_{9}\big\rangle \big\langle 9,(-1)^{k'+l'}M_{l'}-1,i_{9} \big| \\+ \mathcal{O}(\gamma_{-1}^2dt^2),
    \end{multline} 
\end{widetext}
where $\hat{\rho}_c(0)=\ket{\psi_c(0)}\bra{\psi_c(0)}$, $i_9$ is the degeneracy index in $J=9$ subspace with the maximum value $d_{N}^J=19$, $M_1=5$, $M_2=2$, $c_{2} = \sqrt{5/7}$, $c_{5} = \sqrt{2/7}$ and the coefficients $X^{10,M}_{-1,-1}$ can be calculated from Table~\ref{tab:decoherences}. Here, we have neglected the cases where multiple errors occur within time $dt$, i.e., the terms of order $\gamma_{-1}^2dt^2$ and above, which are not correctable for the code state in Eq.~\eqref{Eq:PhysSt}.

\item[4.] \textbf{Syndrome extraction}.-- After a short time interval $dt$, syndrome measurements are performed to determine whether an error has occurred. During this process, the cross terms $\hat{\rho}_c(0)\hat{J}_z$ and $\hat{J}_z \hat{\rho}_c(0)$ vanish~\cite{nielsen_chuang_2019}, and the remaining components of the mixture can be distinguished via the generalized parity measurements described in the main text. 

First, the change in magnetization $m$ is measured using generalized parity measurements~\cite{Fong2026} to distinguish the no jump case ($m = 0$) from states corresponding to individual spin decay ($m = -1$). Conditioned on detecting $m = -1$, a subsequent measurement of the total angular momentum projects the state onto a specific $J$ subspace. For instance, if the generalized parity measurement yields $m = -1$ and the total angular momentum measurement yields $J = 9$ (corresponding to $j = -1$), then the post-measurement state of the ensemble becomes
\begin{multline}
    \hat{\rho}_{E}^{\text{syn}} = \frac{1}{\mathcal{N}_{9}} \sum_{i_{9} = 1}^{d_{20}^{9}}\sum_{k,k'=0}^1\sum_{l,l'=1}^2 \alpha_k^{} \alpha_{k'}^* c^{}_{M_{l^{}}}c^*_{M_{l'}}~{X^{10,M_l}_{-1,-1}X^{10,M_{l'}}_{-1,-1}}~\\
    \big|9,(-1)^{k+l}M^{}_l-1,i_{9}\big\rangle \big\langle 9,(-1)^{k'+l'}M_{l'}-1,i_{9} \big|,
\end{multline}
where $\mathcal{N}_J$ is the appropriate normalization to keep $\text{Tr}\left[\hat{\rho}_{E}^{\text{syn}}\right]=1$.

\item[5.] \textbf{Recovery process}.-- This error can then be corrected by applying the appropriate universal control unitaries, as described in the previous section or in the main text. After applying these unitaries, the encoded state in the new $J=9$ subspace becomes
\begin{multline}
    \hat{\rho}_R = \frac{1}{d_{20}^9} \sum_{i_{9} = 1}^{d_{20}^{9}}\sum_{k,k'=0}^1\sum_{l,l'=1}^2 \alpha_k^{} \alpha_{k'}^* c^{}_{M_{l^{}}}c^*_{M_{l'}}\\
    \big|9,(-1)^{k+l}M^{}_l,i_{9}\big\rangle \big\langle 9,(-1)^{k'+l'}M_{l'},i_{9} \big|.
\end{multline}
\end{enumerate}
Steps 3--5 are repeated at short time intervals $dt$ until the information is either extracted or the syndrome measurement yields $J = M_1$. The only difference in subsequent rounds is that the state may also decohere into $J+1$ subspaces. However, the syndrome extraction and recovery procedure remain essentially the same, with the addition of one extra subspace that must be distinguished and corrected. If the syndrome measurement yields $J = M_1$, the (error-corrected) information is deliberately extracted and subsequently re-encoded (or reinitialized) into the $J = N/2$ subspace of the ensemble using the teleportation protocol described in the main text.
}

%-----------------------------------------------------------------------
\section{Spin loss errors}\label{App:D}
Spin loss errors in ensemble can be represented by taking a partial trace over the spins which are lost. Moreover since all the spins are identical, without loss of generality it suffices to trace out the last spin if a spin loss occurs. The error map can then be found by deconstructing the permutationally invariant (PI) basis to a tensor product of PI basis of $N-1$ spin and 1 spin as \cite{Chase2008},
\begin{widetext}
    \begin{align}
      \overline{\ket{J,M,N}\bra{J,M',N}} &= {d_{N-1}^{J+{1\over2}}\over d_N^J} \sum_{m_1,m_1^{'}} {}^{J,M}\!G^{{1\over2},m_1}_{J+{1\over2}, M-m_1}\ket{{1\over2},m_1}\bra{{1\over2},m^{'}_1} \nonumber \\ & ~~~~ \otimes \overline{\ket{J+{1\over2},M-m_1,N-1}\bra{J+{1\over2},M'-m^{'}_1,N-1}} ~ {}^{J,M}\!G^{{1\over2},m^{'}_1}_{J+{1\over2}, M'-m^{'}_1} \nonumber \\
      & ~~ + {d_{N-1}^{J-{1\over2}}\over d_N^J} \sum_{m_2,m_2^{'}} {}^{J,M}\!G^{{1\over2},m_2}_{J-{1\over2}, M-m_2}\ket{{1\over2},m_2}\bra{{1\over2},m^{'}_2} \nonumber \\ & ~~~~ \otimes \overline{\ket{J-{1\over2},M-m_2,N-1}\bra{J-{1\over2},M'-m^{'}_2,N-1}} ~ {}^{J,M}\!G^{{1\over2},m^{'}_2}_{J-{1\over2}, M'-m^{'}_2},
      \label{Eq:DeletionDeriv}
  \end{align} 
\end{widetext}
where ${}^{J,M}\!G^{j_1,m_1}_{j_2,m_2}$ are the Clebsch-Gordan coefficients, $d_N^J$ is the degeneracy of the state with total angular momentum $J$ and $m_i,m_i{'}$ sums over single spin values $\pm {1\over2}$. Taking the partial trace over the single spin term in Eq.~\eqref{Eq:DeletionDeriv} and inserting the values of the Clebsch-Gordan coefficients using relations in \cite{Varshalovich1988}, we get
\begin{widetext}
    \begin{align}
        \text{Tr}_1 \left[ \overline{\ket{J,M,N}\bra{J,M',N}} \right] &= \sum_{j,m}{d_{N-1}^{J+j}\over d_N^J} {}^{J,M}\!G^{{1\over2},-m}_{J+j, M+m}    {}^{J,M'}\!G^{{1\over2},-m}_{J+j, M'+m}\nonumber\\ &\hspace{3.8cm} \times \overline{\ket{J+{j},M+{m},N-1}\bra{J+{j},M'+{m},N-1}}\nonumber\\ 
        &= \sum_{j,m} D_{j}^{N,J} Y_{j,m}^{J,M} Y_{j,m}^{J,M'} \times \overline{\ket{J+{j},M+{m},N-1}\bra{J+{j},M'+{m},N-1}} \label{AppEq:SpinLoss}
    \end{align}
\end{widetext}
where indices $j,m$ takes values $\{-{1/2},{1/2}\}$ and the coefficients $D_{j}^{N,J} ,Y_{j,m}^{J,M}$ are mentioned in Table.~\ref{tab:deletions}. This is the required error map for single spin loss error. By comparing this with the error map in Eq.~\eqref{Eq:DeletionErr}, one can see that $\xi_{j,m}^{N,J,M} = \sqrt{D_{j}^{N,J}} Y_{j,m}^{J,M}$. 
One can learn that a spin loss has occurred by using non-demolition measurements~\cite{Wang2021}.
Additionally, we can identify the polarization of the lost spin by measuring $m$ of the ensemble. If $m$ is measured to be $1/2$, we know the lost spin is $\ket{1/2,-1/2}$, and if $m=-1/2$, the lost spin is $\ket{1/2,1/2}$.
\begin{table}
    \caption{\textbf{Values of $Y_{j,m}^{J,M}$ and $D_{j}^{N,J}$.}\label{tab:deletions}}
    \centering
    \begin{ruledtabular}
    \begin{tabular}{lccc} 
 &\multicolumn{2}{c}{$Y_{j,m}^{J,M}$}&$D_{j}^{N,J}$\\
         \backslashbox{$j$}{$m$}&-1/2& +1/2&--\\  \hline
         -1/2&$\sqrt{(J+M)}$& {$\sqrt{(J-M)}$}&${1\over N}\left({{N/2}+J+1 \over 2J+1} \right)$\\
         +1/2&{$\sqrt{(J-M+1)}$}& $-\sqrt{(J+M+1)}$&${1\over N}\left({{N/2}-J \over 2J+1} \right)$\\
    \end{tabular}
    \end{ruledtabular}
\end{table}

Though the process is different from the spin number  preserving errors discussed in previous sections, some similarities can be drawn between spin loss and dephasing errors. Suppose after determining $m$, a new spin with the same polarization as the lost one is added back into the ensemble (we will call this spin gain). This transforms the states as
\begin{widetext}
    \begin{multline}
        {d_{N-1}^{J+j}\over d_N^J}\overline{\ket{J+{j},M+{m},N-1}\bra{J+{j},M'+{m},N-1}} \xrightarrow[gain]{\ket{{1\over2},-m}\bra{{1\over2},-m}} \\
        {d_{N}^{J+j_+}\over d_N^J} {}^{J+j_+,M}\!G^{{1\over2},-m}_{J+j, M+m} \overline{\ket{J+{j_+},M,N}\bra{J+{j_+},M',N}}  {}^{J+j_+,M'}\!G^{{1\over2},-m}_{J+j, M'+m}\\
        +{d_{N}^{J+j_-}\over d_N^J} {}^{J+j_-,M}\!G^{{1\over2},-m}_{J+j, M+m} \overline{\ket{J+{j_-},M,N}\bra{J+{j_-},M',N}}  {}^{J+j_-,M'}\!G^{{1\over2},-m}_{J+j, M'+m}
    \end{multline}
with $j_\pm = j \pm 1/2$. For each of the four measurement outcomes (of $j$ and $m$) in Eq.\eqref{AppEq:SpinLoss}, we find that the overall transformation of the ensemble state after spin loss and gain is given by
    \begin{multline}
        \overline{\ket{J,M,N}\bra{J,M',N}} \xrightarrow[]{loss + gain} \overline{\ket{J,M,N}\bra{J,M',N}} \\
        + {1\over 2J^2} \left(1+{J^2\over(J+1)^2} \right) X_{0,0}^{J,M} \overline{\ket{J,M,N}\bra{J,M',N}} X_{0,0}^{J,M'}\\
        + {1\over 2J^2} {d_N^{J-1}\over d_N^J} X_{-1,0}^{J,M} \overline{\ket{J-1,M,N}\bra{J-1,M',N}} X_{-1,0}^{J,M'} \\
        + {1\over 2(J+1)^2} {d_N^{J+1}\over d_N^J} X_{+1,0}^{J,M} \overline{\ket{J+1,M,N}\bra{J+1,M',N}} X_{+1,0}^{J,M'}.
    \end{multline}
\end{widetext}
Interestingly, the transformation in last three terms is equivalent (upto some global phase) to the individual spin dephasing errors ($m=0$) defined in Eq.~\eqref{Eq:ChaseIdentity} and the first term represents the case with no errors. 
The result is not surprising  as the entire procedure described above only affects the spin that is lost and independent of other spins and is therefore an individual spin process.

%-------------------------------------===========================
\section{KL conditions for sensing}\label{App:E}
For a code to protect against individual decay errors, the KL conditions require, $$\bra{k'_Q} \hat{E}_{j',m'}^\dagger \hat{E}_{j,m} \ket{k_Q}= K_{j,m;j',m'}  \delta_{kk'},$$ where $m = \varnothing, -1$. Since the PI states are orthogonal to each other, for sufficient level spacing, we can reduce the conditions to cases with $j'=j$ and $m'=m$. Also, these conditions are satisfied for each $j$ separately, it implies that the following statement has to hold as well
\begin{equation}
   \sum_j \bra{0_Q} \hat{E}_{j,-1}^\dagger \hat{E}_{j,-1} \ket{0_Q} = \sum_j \bra{1_Q} \hat{E}_{j,-1}^\dagger \hat{E}_{j,-1} \ket{1_Q}.
   \label{AppEq:KLSensing}
\end{equation}

From Eqs.~\eqref{AppEq:StDyn} and~\eqref{AppEq:OSR}, we can relate the action of Kraus operators on logical states $\ket{k_Q}$ and the physical operators on the collective states $\ket{k_c}$, which then allows us to evaluate the above expression as follows:
\begin{align}
   \sum_j \bra{k_Q} \hat{E}_{j,-1}^\dagger \hat{E}_{j,-1} \ket{k_Q} &= \text{Tr}\left[ \sum_j \hat{E}_{j,-1} \ket{k_Q}\bra{k_Q} \hat{E}_{j,-1}^\dagger\right]\nonumber\\
   &= \text{Tr}\left[\sum_{n=1}^N \hat{\sigma}_{-1,n} \ket{k_c}\bra{k_c} \hat{\sigma}_{-1,n}^{\dagger}\right]\nonumber\\
   &= \text{Tr}\left[\sum_{n=1}^N {\mathbb{I} + \hat{\sigma}_{0,n}\over2} \ket{k_c}\bra{k_c} \right]\nonumber\\
   &= {N\over 2} + \text{Tr}\left[ \hat{J}_z \ket{k_c}\bra{k_c}\right].
   \label{AppEq:KLEval}
\end{align}
This result then implies that $\bra{0_c} \hat{J}_z \ket{0_c}=\bra{1_c} \hat{J}_z \ket{1_c}${, which means that the accumulated phase in Eq.~\eqref{Eq:QSPhase} vanishes, so no QEC code can protect against individual decay errors}. Similar analysis can be performed for individual pumping errors, which also leads to the same result.
% \end{widetext}

\bibliography{references}

 \newcommand{\noop}[1]{}
\begin{thebibliography}{87}%
\makeatletter
\providecommand \@ifxundefined [1]{%
 \@ifx{#1\undefined}
}%
\providecommand \@ifnum [1]{%
 \ifnum #1\expandafter \@firstoftwo
 \else \expandafter \@secondoftwo
 \fi
}%
\providecommand \@ifx [1]{%
 \ifx #1\expandafter \@firstoftwo
 \else \expandafter \@secondoftwo
 \fi
}%
\providecommand \natexlab [1]{#1}%
\providecommand \enquote  [1]{``#1''}%
\providecommand \bibnamefont  [1]{#1}%
\providecommand \bibfnamefont [1]{#1}%
\providecommand \citenamefont [1]{#1}%
\providecommand \href@noop [0]{\@secondoftwo}%
\providecommand \href [0]{\begingroup \@sanitize@url \@href}%
\providecommand \@href[1]{\@@startlink{#1}\@@href}%
\providecommand \@@href[1]{\endgroup#1\@@endlink}%
\providecommand \@sanitize@url [0]{\catcode `\\12\catcode `\$12\catcode `\&12\catcode `\#12\catcode `\^12\catcode `\_12\catcode `\%12\relax}%
\providecommand \@@startlink[1]{}%
\providecommand \@@endlink[0]{}%
\providecommand \url  [0]{\begingroup\@sanitize@url \@url }%
\providecommand \@url [1]{\endgroup\@href {#1}{\urlprefix }}%
\providecommand \urlprefix  [0]{URL }%
\providecommand \Eprint [0]{\href }%
\providecommand \doibase [0]{https://doi.org/}%
\providecommand \selectlanguage [0]{\@gobble}%
\providecommand \bibinfo  [0]{\@secondoftwo}%
\providecommand \bibfield  [0]{\@secondoftwo}%
\providecommand \translation [1]{[#1]}%
\providecommand \BibitemOpen [0]{}%
\providecommand \bibitemStop [0]{}%
\providecommand \bibitemNoStop [0]{.\EOS\space}%
\providecommand \EOS [0]{\spacefactor3000\relax}%
\providecommand \BibitemShut  [1]{\csname bibitem#1\endcsname}%
\let\auto@bib@innerbib\@empty
%</preamble>
\bibitem [{\citenamefont {Nielsen}\ and\ \citenamefont {Chuang}(2019)}]{nielsen_chuang_2019}%
  \BibitemOpen
  \bibfield  {author} {\bibinfo {author} {\bibfnamefont {M.~A.}\ \bibnamefont {Nielsen}}\ and\ \bibinfo {author} {\bibfnamefont {I.~L.}\ \bibnamefont {Chuang}},\ }\href@noop {} {\emph {\bibinfo {title} {Quantum computation and quantum information}}}\ (\bibinfo  {publisher} {Cambridge Cambridge University Press},\ \bibinfo {year} {2019})\ p.\ \bibinfo {pages} {425–499}\BibitemShut {NoStop}%
\bibitem [{\citenamefont {Tordrup}\ \emph {et~al.}(2008)\citenamefont {Tordrup}, \citenamefont {Negretti},\ and\ \citenamefont {M\o{}lmer}}]{Tordrup2008}%
  \BibitemOpen
  \bibfield  {author} {\bibinfo {author} {\bibfnamefont {K.}~\bibnamefont {Tordrup}}, \bibinfo {author} {\bibfnamefont {A.}~\bibnamefont {Negretti}},\ and\ \bibinfo {author} {\bibfnamefont {K.}~\bibnamefont {M\o{}lmer}},\ }\bibfield  {title} {\bibinfo {title} {Holographic quantum computing},\ }\href {https://doi.org/10.1103/PhysRevLett.101.040501} {\bibfield  {journal} {\bibinfo  {journal} {Phys. Rev. Lett.}\ }\textbf {\bibinfo {volume} {101}},\ \bibinfo {pages} {040501} (\bibinfo {year} {2008})}\BibitemShut {NoStop}%
\bibitem [{\citenamefont {Wesenberg}\ \emph {et~al.}(2009)\citenamefont {Wesenberg}, \citenamefont {Ardavan}, \citenamefont {Briggs}, \citenamefont {Morton}, \citenamefont {Schoelkopf}, \citenamefont {Schuster},\ and\ \citenamefont {M\o{}lmer}}]{Wesenberg2009}%
  \BibitemOpen
  \bibfield  {author} {\bibinfo {author} {\bibfnamefont {J.~H.}\ \bibnamefont {Wesenberg}}, \bibinfo {author} {\bibfnamefont {A.}~\bibnamefont {Ardavan}}, \bibinfo {author} {\bibfnamefont {G.~A.~D.}\ \bibnamefont {Briggs}}, \bibinfo {author} {\bibfnamefont {J.~J.~L.}\ \bibnamefont {Morton}}, \bibinfo {author} {\bibfnamefont {R.~J.}\ \bibnamefont {Schoelkopf}}, \bibinfo {author} {\bibfnamefont {D.~I.}\ \bibnamefont {Schuster}},\ and\ \bibinfo {author} {\bibfnamefont {K.}~\bibnamefont {M\o{}lmer}},\ }\bibfield  {title} {\bibinfo {title} {Quantum computing with an electron spin ensemble},\ }\href {https://doi.org/10.1103/PhysRevLett.103.070502} {\bibfield  {journal} {\bibinfo  {journal} {Phys. Rev. Lett.}\ }\textbf {\bibinfo {volume} {103}},\ \bibinfo {pages} {070502} (\bibinfo {year} {2009})}\BibitemShut {NoStop}%
\bibitem [{\citenamefont {Cox}\ \emph {et~al.}(2022)\citenamefont {Cox}, \citenamefont {Bienias}, \citenamefont {Meyer}, \citenamefont {Fahey}, \citenamefont {Kunz},\ and\ \citenamefont {Gorshkov}}]{Cox2022}%
  \BibitemOpen
  \bibfield  {author} {\bibinfo {author} {\bibfnamefont {K.~C.}\ \bibnamefont {Cox}}, \bibinfo {author} {\bibfnamefont {P.}~\bibnamefont {Bienias}}, \bibinfo {author} {\bibfnamefont {D.~H.}\ \bibnamefont {Meyer}}, \bibinfo {author} {\bibfnamefont {D.~P.}\ \bibnamefont {Fahey}}, \bibinfo {author} {\bibfnamefont {P.~D.}\ \bibnamefont {Kunz}},\ and\ \bibinfo {author} {\bibfnamefont {A.~V.}\ \bibnamefont {Gorshkov}},\ }\bibfield  {title} {\bibinfo {title} {Spin-wave quantum computing with atoms in a single-mode cavity},\ }\href {https://doi.org/10.1103/PhysRevResearch.4.033149} {\bibfield  {journal} {\bibinfo  {journal} {Phys. Rev. Res.}\ }\textbf {\bibinfo {volume} {4}},\ \bibinfo {pages} {033149} (\bibinfo {year} {2022})}\BibitemShut {NoStop}%
\bibitem [{\citenamefont {Afzelius}\ \emph {et~al.}(2009)\citenamefont {Afzelius}, \citenamefont {Simon}, \citenamefont {de~Riedmatten},\ and\ \citenamefont {Gisin}}]{Afzelius2009}%
  \BibitemOpen
  \bibfield  {author} {\bibinfo {author} {\bibfnamefont {M.}~\bibnamefont {Afzelius}}, \bibinfo {author} {\bibfnamefont {C.}~\bibnamefont {Simon}}, \bibinfo {author} {\bibfnamefont {H.}~\bibnamefont {de~Riedmatten}},\ and\ \bibinfo {author} {\bibfnamefont {N.}~\bibnamefont {Gisin}},\ }\bibfield  {title} {\bibinfo {title} {Multimode quantum memory based on atomic frequency combs},\ }\href {https://doi.org/10.1103/PhysRevA.79.052329} {\bibfield  {journal} {\bibinfo  {journal} {Phys. Rev. A}\ }\textbf {\bibinfo {volume} {79}},\ \bibinfo {pages} {052329} (\bibinfo {year} {2009})}\BibitemShut {NoStop}%
\bibitem [{\citenamefont {Lvovsky}\ \emph {et~al.}(2009)\citenamefont {Lvovsky}, \citenamefont {Sanders},\ and\ \citenamefont {Tittel}}]{lvovsky2009}%
  \BibitemOpen
  \bibfield  {author} {\bibinfo {author} {\bibfnamefont {A.~I.}\ \bibnamefont {Lvovsky}}, \bibinfo {author} {\bibfnamefont {B.~C.}\ \bibnamefont {Sanders}},\ and\ \bibinfo {author} {\bibfnamefont {W.}~\bibnamefont {Tittel}},\ }\bibfield  {title} {\bibinfo {title} {Optical quantum memory},\ }\href {https://doi.org/https://doi.org/10.1038/nphoton.2009.231} {\bibfield  {journal} {\bibinfo  {journal} {Nature Photonics}\ }\textbf {\bibinfo {volume} {3}},\ \bibinfo {pages} {706} (\bibinfo {year} {2009})}\BibitemShut {NoStop}%
\bibitem [{\citenamefont {Sangouard}\ \emph {et~al.}(2011)\citenamefont {Sangouard}, \citenamefont {Simon}, \citenamefont {de~Riedmatten},\ and\ \citenamefont {Gisin}}]{Sangouard2011}%
  \BibitemOpen
  \bibfield  {author} {\bibinfo {author} {\bibfnamefont {N.}~\bibnamefont {Sangouard}}, \bibinfo {author} {\bibfnamefont {C.}~\bibnamefont {Simon}}, \bibinfo {author} {\bibfnamefont {H.}~\bibnamefont {de~Riedmatten}},\ and\ \bibinfo {author} {\bibfnamefont {N.}~\bibnamefont {Gisin}},\ }\bibfield  {title} {\bibinfo {title} {Quantum repeaters based on atomic ensembles and linear optics},\ }\href {https://doi.org/10.1103/RevModPhys.83.33} {\bibfield  {journal} {\bibinfo  {journal} {Rev. Mod. Phys.}\ }\textbf {\bibinfo {volume} {83}},\ \bibinfo {pages} {33} (\bibinfo {year} {2011})}\BibitemShut {NoStop}%
\bibitem [{\citenamefont {Muschik}\ \emph {et~al.}(2008)\citenamefont {Muschik}, \citenamefont {de~Vega}, \citenamefont {Porras},\ and\ \citenamefont {Cirac}}]{Muschik2008}%
  \BibitemOpen
  \bibfield  {author} {\bibinfo {author} {\bibfnamefont {C.~A.}\ \bibnamefont {Muschik}}, \bibinfo {author} {\bibfnamefont {I.}~\bibnamefont {de~Vega}}, \bibinfo {author} {\bibfnamefont {D.}~\bibnamefont {Porras}},\ and\ \bibinfo {author} {\bibfnamefont {J.~I.}\ \bibnamefont {Cirac}},\ }\bibfield  {title} {\bibinfo {title} {Quantum processing photonic states in optical lattices},\ }\href {https://doi.org/10.1103/PhysRevLett.100.063601} {\bibfield  {journal} {\bibinfo  {journal} {Phys. Rev. Lett.}\ }\textbf {\bibinfo {volume} {100}},\ \bibinfo {pages} {063601} (\bibinfo {year} {2008})}\BibitemShut {NoStop}%
\bibitem [{\citenamefont {Saffman}\ \emph {et~al.}(2010)\citenamefont {Saffman}, \citenamefont {Walker},\ and\ \citenamefont {M\o{}lmer}}]{Saffman2010}%
  \BibitemOpen
  \bibfield  {author} {\bibinfo {author} {\bibfnamefont {M.}~\bibnamefont {Saffman}}, \bibinfo {author} {\bibfnamefont {T.~G.}\ \bibnamefont {Walker}},\ and\ \bibinfo {author} {\bibfnamefont {K.}~\bibnamefont {M\o{}lmer}},\ }\bibfield  {title} {\bibinfo {title} {Quantum information with rydberg atoms},\ }\href {https://doi.org/10.1103/RevModPhys.82.2313} {\bibfield  {journal} {\bibinfo  {journal} {Rev. Mod. Phys.}\ }\textbf {\bibinfo {volume} {82}},\ \bibinfo {pages} {2313} (\bibinfo {year} {2010})}\BibitemShut {NoStop}%
\bibitem [{\citenamefont {Heller}\ \emph {et~al.}(2020)\citenamefont {Heller}, \citenamefont {Farrera}, \citenamefont {Heinze},\ and\ \citenamefont {de~Riedmatten}}]{Heller2020}%
  \BibitemOpen
  \bibfield  {author} {\bibinfo {author} {\bibfnamefont {L.}~\bibnamefont {Heller}}, \bibinfo {author} {\bibfnamefont {P.}~\bibnamefont {Farrera}}, \bibinfo {author} {\bibfnamefont {G.}~\bibnamefont {Heinze}},\ and\ \bibinfo {author} {\bibfnamefont {H.}~\bibnamefont {de~Riedmatten}},\ }\bibfield  {title} {\bibinfo {title} {Cold-atom temporally multiplexed quantum memory with cavity-enhanced noise suppression},\ }\href {https://doi.org/10.1103/PhysRevLett.124.210504} {\bibfield  {journal} {\bibinfo  {journal} {Phys. Rev. Lett.}\ }\textbf {\bibinfo {volume} {124}},\ \bibinfo {pages} {210504} (\bibinfo {year} {2020})}\BibitemShut {NoStop}%
\bibitem [{\citenamefont {Morello}\ \emph {et~al.}(2020)\citenamefont {Morello}, \citenamefont {Pla}, \citenamefont {Bertet},\ and\ \citenamefont {Jamieson}}]{Morello2020}%
  \BibitemOpen
  \bibfield  {author} {\bibinfo {author} {\bibfnamefont {A.}~\bibnamefont {Morello}}, \bibinfo {author} {\bibfnamefont {J.~J.}\ \bibnamefont {Pla}}, \bibinfo {author} {\bibfnamefont {P.}~\bibnamefont {Bertet}},\ and\ \bibinfo {author} {\bibfnamefont {D.~N.}\ \bibnamefont {Jamieson}},\ }\bibfield  {title} {\bibinfo {title} {Donor spins in silicon for quantum technologies},\ }\href {https://doi.org/https://doi.org/10.1002/qute.202000005} {\bibfield  {journal} {\bibinfo  {journal} {Adv. Quantum Technol.}\ }\textbf {\bibinfo {volume} {3}},\ \bibinfo {pages} {2000005} (\bibinfo {year} {2020})}\BibitemShut {NoStop}%
\bibitem [{\citenamefont {Nemoto}\ \emph {et~al.}(2014)\citenamefont {Nemoto}, \citenamefont {Trupke}, \citenamefont {Devitt}, \citenamefont {Stephens}, \citenamefont {Scharfenberger}, \citenamefont {Buczak}, \citenamefont {N\"obauer}, \citenamefont {Everitt}, \citenamefont {Schmiedmayer},\ and\ \citenamefont {Munro}}]{Nemoto2014}%
  \BibitemOpen
  \bibfield  {author} {\bibinfo {author} {\bibfnamefont {K.}~\bibnamefont {Nemoto}}, \bibinfo {author} {\bibfnamefont {M.}~\bibnamefont {Trupke}}, \bibinfo {author} {\bibfnamefont {S.~J.}\ \bibnamefont {Devitt}}, \bibinfo {author} {\bibfnamefont {A.~M.}\ \bibnamefont {Stephens}}, \bibinfo {author} {\bibfnamefont {B.}~\bibnamefont {Scharfenberger}}, \bibinfo {author} {\bibfnamefont {K.}~\bibnamefont {Buczak}}, \bibinfo {author} {\bibfnamefont {T.}~\bibnamefont {N\"obauer}}, \bibinfo {author} {\bibfnamefont {M.~S.}\ \bibnamefont {Everitt}}, \bibinfo {author} {\bibfnamefont {J.}~\bibnamefont {Schmiedmayer}},\ and\ \bibinfo {author} {\bibfnamefont {W.~J.}\ \bibnamefont {Munro}},\ }\bibfield  {title} {\bibinfo {title} {Photonic architecture for scalable quantum information processing in diamond},\ }\href {https://doi.org/10.1103/PhysRevX.4.031022} {\bibfield  {journal} {\bibinfo  {journal} {Phys. Rev. X}\ }\textbf {\bibinfo {volume} {4}},\ \bibinfo {pages} {031022} (\bibinfo {year} {2014})}\BibitemShut {NoStop}%
\bibitem [{\citenamefont {Pezzagna}\ and\ \citenamefont {Meijer}(2021)}]{Pezzagna2021}%
  \BibitemOpen
  \bibfield  {author} {\bibinfo {author} {\bibfnamefont {S.}~\bibnamefont {Pezzagna}}\ and\ \bibinfo {author} {\bibfnamefont {J.}~\bibnamefont {Meijer}},\ }\bibfield  {title} {\bibinfo {title} {{Quantum computer based on color centers in diamond}},\ }\href {https://doi.org/10.1063/5.0007444} {\bibfield  {journal} {\bibinfo  {journal} {App. Phys. Rev.}\ }\textbf {\bibinfo {volume} {8}},\ \bibinfo {pages} {011308} (\bibinfo {year} {2021})}\BibitemShut {NoStop}%
\bibitem [{\citenamefont {Gilmore}\ \emph {et~al.}(2021)\citenamefont {Gilmore}, \citenamefont {Affolter}, \citenamefont {Lewis-Swan}, \citenamefont {Barberena}, \citenamefont {Jordan}, \citenamefont {Rey},\ and\ \citenamefont {Bollinger}}]{Gilmore2021}%
  \BibitemOpen
  \bibfield  {author} {\bibinfo {author} {\bibfnamefont {K.~A.}\ \bibnamefont {Gilmore}}, \bibinfo {author} {\bibfnamefont {M.}~\bibnamefont {Affolter}}, \bibinfo {author} {\bibfnamefont {R.~J.}\ \bibnamefont {Lewis-Swan}}, \bibinfo {author} {\bibfnamefont {D.}~\bibnamefont {Barberena}}, \bibinfo {author} {\bibfnamefont {E.}~\bibnamefont {Jordan}}, \bibinfo {author} {\bibfnamefont {A.~M.}\ \bibnamefont {Rey}},\ and\ \bibinfo {author} {\bibfnamefont {J.~J.}\ \bibnamefont {Bollinger}},\ }\bibfield  {title} {\bibinfo {title} {{Quantum-enhanced sensing of displacements and electric fields with two-dimensional trapped-ion crystals}},\ }\href {https://doi.org/10.1126/science.abi5226} {\bibfield  {journal} {\bibinfo  {journal} {Science}\ }\textbf {\bibinfo {volume} {373}},\ \bibinfo {pages} {673} (\bibinfo {year} {2021})}\BibitemShut {NoStop}%
\bibitem [{\citenamefont {Polloreno}\ \emph {et~al.}(2022)\citenamefont {Polloreno}, \citenamefont {Rey},\ and\ \citenamefont {Bollinger}}]{Polloreno2022}%
  \BibitemOpen
  \bibfield  {author} {\bibinfo {author} {\bibfnamefont {A.~M.}\ \bibnamefont {Polloreno}}, \bibinfo {author} {\bibfnamefont {A.~M.}\ \bibnamefont {Rey}},\ and\ \bibinfo {author} {\bibfnamefont {J.~J.}\ \bibnamefont {Bollinger}},\ }\bibfield  {title} {\bibinfo {title} {Individual qubit addressing of rotating ion crystals in a penning trap},\ }\href {https://doi.org/10.1103/PhysRevResearch.4.033076} {\bibfield  {journal} {\bibinfo  {journal} {Phys. Rev. Res.}\ }\textbf {\bibinfo {volume} {4}},\ \bibinfo {pages} {033076} (\bibinfo {year} {2022})}\BibitemShut {NoStop}%
\bibitem [{\citenamefont {Longdell}\ \emph {et~al.}(2005)\citenamefont {Longdell}, \citenamefont {Fraval}, \citenamefont {Sellars},\ and\ \citenamefont {Manson}}]{Longdell2005}%
  \BibitemOpen
  \bibfield  {author} {\bibinfo {author} {\bibfnamefont {J.~J.}\ \bibnamefont {Longdell}}, \bibinfo {author} {\bibfnamefont {E.}~\bibnamefont {Fraval}}, \bibinfo {author} {\bibfnamefont {M.~J.}\ \bibnamefont {Sellars}},\ and\ \bibinfo {author} {\bibfnamefont {N.~B.}\ \bibnamefont {Manson}},\ }\bibfield  {title} {\bibinfo {title} {Stopped light with storage times greater than one second using electromagnetically induced transparency in a solid},\ }\href {https://doi.org/10.1103/PhysRevLett.95.063601} {\bibfield  {journal} {\bibinfo  {journal} {Phys. Rev. Lett.}\ }\textbf {\bibinfo {volume} {95}},\ \bibinfo {pages} {063601} (\bibinfo {year} {2005})}\BibitemShut {NoStop}%
\bibitem [{\citenamefont {Dudin}\ \emph {et~al.}(2013)\citenamefont {Dudin}, \citenamefont {Li},\ and\ \citenamefont {Kuzmich}}]{Dudin2013}%
  \BibitemOpen
  \bibfield  {author} {\bibinfo {author} {\bibfnamefont {Y.~O.}\ \bibnamefont {Dudin}}, \bibinfo {author} {\bibfnamefont {L.}~\bibnamefont {Li}},\ and\ \bibinfo {author} {\bibfnamefont {A.}~\bibnamefont {Kuzmich}},\ }\bibfield  {title} {\bibinfo {title} {Light storage on the time scale of a minute},\ }\href {https://doi.org/10.1103/PhysRevA.87.031801} {\bibfield  {journal} {\bibinfo  {journal} {Phys. Rev. A}\ }\textbf {\bibinfo {volume} {87}},\ \bibinfo {pages} {031801(R)} (\bibinfo {year} {2013})}\BibitemShut {NoStop}%
\bibitem [{\citenamefont {Bar-Gill}\ \emph {et~al.}(2013)\citenamefont {Bar-Gill}, \citenamefont {Pham}, \citenamefont {Jarmola}, \citenamefont {Budker},\ and\ \citenamefont {Walsworth}}]{Bar-Gill2013}%
  \BibitemOpen
  \bibfield  {author} {\bibinfo {author} {\bibfnamefont {N.}~\bibnamefont {Bar-Gill}}, \bibinfo {author} {\bibfnamefont {L.~M.}\ \bibnamefont {Pham}}, \bibinfo {author} {\bibfnamefont {A.}~\bibnamefont {Jarmola}}, \bibinfo {author} {\bibfnamefont {D.}~\bibnamefont {Budker}},\ and\ \bibinfo {author} {\bibfnamefont {R.~L.}\ \bibnamefont {Walsworth}},\ }\bibfield  {title} {\bibinfo {title} {Solid-state electronic spin coherence time approaching one second},\ }\href {https://doi.org/10.1038/ncomms2771} {\bibfield  {journal} {\bibinfo  {journal} {Nature Communications}\ }\textbf {\bibinfo {volume} {4}},\ \bibinfo {pages} {1743} (\bibinfo {year} {2013})}\BibitemShut {NoStop}%
\bibitem [{\citenamefont {Kubo}\ \emph {et~al.}(2010)\citenamefont {Kubo}, \citenamefont {Ong}, \citenamefont {Bertet}, \citenamefont {Vion}, \citenamefont {Jacques}, \citenamefont {Zheng}, \citenamefont {Dr\'eau}, \citenamefont {Roch}, \citenamefont {Auffeves}, \citenamefont {Jelezko}, \citenamefont {Wrachtrup}, \citenamefont {Barthe}, \citenamefont {Bergonzo},\ and\ \citenamefont {Esteve}}]{Kubo2010}%
  \BibitemOpen
  \bibfield  {author} {\bibinfo {author} {\bibfnamefont {Y.}~\bibnamefont {Kubo}}, \bibinfo {author} {\bibfnamefont {F.~R.}\ \bibnamefont {Ong}}, \bibinfo {author} {\bibfnamefont {P.}~\bibnamefont {Bertet}}, \bibinfo {author} {\bibfnamefont {D.}~\bibnamefont {Vion}}, \bibinfo {author} {\bibfnamefont {V.}~\bibnamefont {Jacques}}, \bibinfo {author} {\bibfnamefont {D.}~\bibnamefont {Zheng}}, \bibinfo {author} {\bibfnamefont {A.}~\bibnamefont {Dr\'eau}}, \bibinfo {author} {\bibfnamefont {J.-F.}\ \bibnamefont {Roch}}, \bibinfo {author} {\bibfnamefont {A.}~\bibnamefont {Auffeves}}, \bibinfo {author} {\bibfnamefont {F.}~\bibnamefont {Jelezko}}, \bibinfo {author} {\bibfnamefont {J.}~\bibnamefont {Wrachtrup}}, \bibinfo {author} {\bibfnamefont {M.~F.}\ \bibnamefont {Barthe}}, \bibinfo {author} {\bibfnamefont {P.}~\bibnamefont {Bergonzo}},\ and\ \bibinfo {author} {\bibfnamefont {D.}~\bibnamefont {Esteve}},\ }\bibfield  {title} {\bibinfo {title} {Strong coupling of a spin ensemble to a superconducting resonator},\ }\href
  {https://doi.org/10.1103/PhysRevLett.105.140502} {\bibfield  {journal} {\bibinfo  {journal} {Phys. Rev. Lett.}\ }\textbf {\bibinfo {volume} {105}},\ \bibinfo {pages} {140502} (\bibinfo {year} {2010})}\BibitemShut {NoStop}%
\bibitem [{\citenamefont {Schuster}\ \emph {et~al.}(2010)\citenamefont {Schuster}, \citenamefont {Sears}, \citenamefont {Ginossar}, \citenamefont {DiCarlo}, \citenamefont {Frunzio}, \citenamefont {Morton}, \citenamefont {Wu}, \citenamefont {Briggs}, \citenamefont {Buckley}, \citenamefont {Awschalom},\ and\ \citenamefont {Schoelkopf}}]{Schuster2010}%
  \BibitemOpen
  \bibfield  {author} {\bibinfo {author} {\bibfnamefont {D.~I.}\ \bibnamefont {Schuster}}, \bibinfo {author} {\bibfnamefont {A.~P.}\ \bibnamefont {Sears}}, \bibinfo {author} {\bibfnamefont {E.}~\bibnamefont {Ginossar}}, \bibinfo {author} {\bibfnamefont {L.}~\bibnamefont {DiCarlo}}, \bibinfo {author} {\bibfnamefont {L.}~\bibnamefont {Frunzio}}, \bibinfo {author} {\bibfnamefont {J.~J.~L.}\ \bibnamefont {Morton}}, \bibinfo {author} {\bibfnamefont {H.}~\bibnamefont {Wu}}, \bibinfo {author} {\bibfnamefont {G.~A.~D.}\ \bibnamefont {Briggs}}, \bibinfo {author} {\bibfnamefont {B.~B.}\ \bibnamefont {Buckley}}, \bibinfo {author} {\bibfnamefont {D.~D.}\ \bibnamefont {Awschalom}},\ and\ \bibinfo {author} {\bibfnamefont {R.~J.}\ \bibnamefont {Schoelkopf}},\ }\bibfield  {title} {\bibinfo {title} {High-cooperativity coupling of electron-spin ensembles to superconducting cavities},\ }\href {https://doi.org/10.1103/PhysRevLett.105.140501} {\bibfield  {journal} {\bibinfo  {journal} {Phys. Rev. Lett.}\ }\textbf {\bibinfo
  {volume} {105}},\ \bibinfo {pages} {140501} (\bibinfo {year} {2010})}\BibitemShut {NoStop}%
\bibitem [{\citenamefont {Xiang}\ \emph {et~al.}(2013)\citenamefont {Xiang}, \citenamefont {Ashhab}, \citenamefont {You},\ and\ \citenamefont {Nori}}]{Xiang2013}%
  \BibitemOpen
  \bibfield  {author} {\bibinfo {author} {\bibfnamefont {Z.-L.}\ \bibnamefont {Xiang}}, \bibinfo {author} {\bibfnamefont {S.}~\bibnamefont {Ashhab}}, \bibinfo {author} {\bibfnamefont {J.~Q.}\ \bibnamefont {You}},\ and\ \bibinfo {author} {\bibfnamefont {F.}~\bibnamefont {Nori}},\ }\bibfield  {title} {\bibinfo {title} {Hybrid quantum circuits: Superconducting circuits interacting with other quantum systems},\ }\href {https://doi.org/10.1103/RevModPhys.85.623} {\bibfield  {journal} {\bibinfo  {journal} {Rev. Mod. Phys.}\ }\textbf {\bibinfo {volume} {85}},\ \bibinfo {pages} {623} (\bibinfo {year} {2013})}\BibitemShut {NoStop}%
\bibitem [{\citenamefont {Kurizki}\ \emph {et~al.}(2015)\citenamefont {Kurizki}, \citenamefont {Bertet}, \citenamefont {Kubo}, \citenamefont {M\o{}lmer}, \citenamefont {Petrosyan}, \citenamefont {Rabl},\ and\ \citenamefont {Schmiedmayer}}]{Kurizki2015}%
  \BibitemOpen
  \bibfield  {author} {\bibinfo {author} {\bibfnamefont {G.}~\bibnamefont {Kurizki}}, \bibinfo {author} {\bibfnamefont {P.}~\bibnamefont {Bertet}}, \bibinfo {author} {\bibfnamefont {Y.}~\bibnamefont {Kubo}}, \bibinfo {author} {\bibfnamefont {K.}~\bibnamefont {M\o{}lmer}}, \bibinfo {author} {\bibfnamefont {D.}~\bibnamefont {Petrosyan}}, \bibinfo {author} {\bibfnamefont {P.}~\bibnamefont {Rabl}},\ and\ \bibinfo {author} {\bibfnamefont {J.}~\bibnamefont {Schmiedmayer}},\ }\bibfield  {title} {\bibinfo {title} {Quantum technologies with hybrid systems},\ }\href {https://doi.org/https://doi.org/10.1073/pnas.1419326112} {\bibfield  {journal} {\bibinfo  {journal} {Proceedings of the National Academy of Sciences of the United States of America}\ }\textbf {\bibinfo {volume} {112}},\ \bibinfo {pages} {3866} (\bibinfo {year} {2015})}\BibitemShut {NoStop}%
\bibitem [{\citenamefont {Bloch}\ \emph {et~al.}(2012)\citenamefont {Bloch}, \citenamefont {Dalibard},\ and\ \citenamefont {Nascimbène}}]{bloch2012}%
  \BibitemOpen
  \bibfield  {author} {\bibinfo {author} {\bibfnamefont {I.}~\bibnamefont {Bloch}}, \bibinfo {author} {\bibfnamefont {J.}~\bibnamefont {Dalibard}},\ and\ \bibinfo {author} {\bibfnamefont {S.}~\bibnamefont {Nascimbène}},\ }\bibfield  {title} {\bibinfo {title} {Quantum simulations with ultracold quantum gases},\ }\href {https://doi.org/https://doi.org/10.1038/nphys2259} {\bibfield  {journal} {\bibinfo  {journal} {Nature Physics}\ }\textbf {\bibinfo {volume} {8}},\ \bibinfo {pages} {267} (\bibinfo {year} {2012})}\BibitemShut {NoStop}%
\bibitem [{\citenamefont {Endres}\ \emph {et~al.}(2016)\citenamefont {Endres}, \citenamefont {Bernien}, \citenamefont {Keesling}, \citenamefont {Levine}, \citenamefont {Anschuetz}, \citenamefont {Krajenbrink}, \citenamefont {Senko}, \citenamefont {Vuletic}, \citenamefont {Greiner},\ and\ \citenamefont {Lukin}}]{Endres2016}%
  \BibitemOpen
  \bibfield  {author} {\bibinfo {author} {\bibfnamefont {M.}~\bibnamefont {Endres}}, \bibinfo {author} {\bibfnamefont {H.}~\bibnamefont {Bernien}}, \bibinfo {author} {\bibfnamefont {A.}~\bibnamefont {Keesling}}, \bibinfo {author} {\bibfnamefont {H.}~\bibnamefont {Levine}}, \bibinfo {author} {\bibfnamefont {E.~R.}\ \bibnamefont {Anschuetz}}, \bibinfo {author} {\bibfnamefont {A.}~\bibnamefont {Krajenbrink}}, \bibinfo {author} {\bibfnamefont {C.}~\bibnamefont {Senko}}, \bibinfo {author} {\bibfnamefont {V.}~\bibnamefont {Vuletic}}, \bibinfo {author} {\bibfnamefont {M.}~\bibnamefont {Greiner}},\ and\ \bibinfo {author} {\bibfnamefont {M.~D.}\ \bibnamefont {Lukin}},\ }\bibfield  {title} {\bibinfo {title} {Atom-by-atom assembly of defect-free one-dimensional cold atom arrays},\ }\href {https://doi.org/10.1126/science.aah3752} {\bibfield  {journal} {\bibinfo  {journal} {Science}\ }\textbf {\bibinfo {volume} {354}},\ \bibinfo {pages} {1024} (\bibinfo {year} {2016})}\BibitemShut {NoStop}%
\bibitem [{\citenamefont {Pezz\`e}\ \emph {et~al.}(2018)\citenamefont {Pezz\`e}, \citenamefont {Smerzi}, \citenamefont {Oberthaler}, \citenamefont {Schmied},\ and\ \citenamefont {Treutlein}}]{Pezze2018}%
  \BibitemOpen
  \bibfield  {author} {\bibinfo {author} {\bibfnamefont {L.}~\bibnamefont {Pezz\`e}}, \bibinfo {author} {\bibfnamefont {A.}~\bibnamefont {Smerzi}}, \bibinfo {author} {\bibfnamefont {M.~K.}\ \bibnamefont {Oberthaler}}, \bibinfo {author} {\bibfnamefont {R.}~\bibnamefont {Schmied}},\ and\ \bibinfo {author} {\bibfnamefont {P.}~\bibnamefont {Treutlein}},\ }\bibfield  {title} {\bibinfo {title} {Quantum metrology with nonclassical states of atomic ensembles},\ }\href {https://doi.org/10.1103/RevModPhys.90.035005} {\bibfield  {journal} {\bibinfo  {journal} {Rev. Mod. Phys.}\ }\textbf {\bibinfo {volume} {90}},\ \bibinfo {pages} {035005} (\bibinfo {year} {2018})}\BibitemShut {NoStop}%
\bibitem [{\citenamefont {{Mu\~noz}-Arias}\ \emph {et~al.}(2023)\citenamefont {{Mu\~noz}-Arias}, \citenamefont {Deutsch},\ and\ \citenamefont {Poggi}}]{Mu2023}%
  \BibitemOpen
  \bibfield  {author} {\bibinfo {author} {\bibfnamefont {M.~H.}\ \bibnamefont {{Mu\~noz}-Arias}}, \bibinfo {author} {\bibfnamefont {I.~H.}\ \bibnamefont {Deutsch}},\ and\ \bibinfo {author} {\bibfnamefont {P.~M.}\ \bibnamefont {Poggi}},\ }\bibfield  {title} {\bibinfo {title} {Phase-space geometry and optimal state preparation in quantum metrology with collective spins},\ }\href {https://doi.org/10.1103/PRXQuantum.4.020314} {\bibfield  {journal} {\bibinfo  {journal} {PRX Quantum}\ }\textbf {\bibinfo {volume} {4}},\ \bibinfo {pages} {020314} (\bibinfo {year} {2023})}\BibitemShut {NoStop}%
\bibitem [{\citenamefont {Ouyang}\ and\ \citenamefont {Brennen}(2026{\natexlab{a}})}]{Ouyang2026a}%
  \BibitemOpen
  \bibfield  {author} {\bibinfo {author} {\bibfnamefont {Y.}~\bibnamefont {Ouyang}}\ and\ \bibinfo {author} {\bibfnamefont {G.~K.}\ \bibnamefont {Brennen}},\ }\href@noop {} {\bibinfo {title} {Finite-round quantum error correction on symmetric quantum sensors}} (\bibinfo {year} {2026}{\natexlab{a}}),\ \Eprint {https://arxiv.org/abs/2212.06285v5} {arXiv:2212.06285v5 [quant-ph]} \BibitemShut {NoStop}%
\bibitem [{\citenamefont {Wineland}\ \emph {et~al.}(1998)\citenamefont {Wineland}, \citenamefont {Monroe}, \citenamefont {Itano}, \citenamefont {Leibfried}, \citenamefont {King},\ and\ \citenamefont {Meekhof}}]{Wineland1998}%
  \BibitemOpen
  \bibfield  {author} {\bibinfo {author} {\bibfnamefont {D.~J.}\ \bibnamefont {Wineland}}, \bibinfo {author} {\bibfnamefont {C.}~\bibnamefont {Monroe}}, \bibinfo {author} {\bibfnamefont {W.~M.}\ \bibnamefont {Itano}}, \bibinfo {author} {\bibfnamefont {D.}~\bibnamefont {Leibfried}}, \bibinfo {author} {\bibfnamefont {B.~E.}\ \bibnamefont {King}},\ and\ \bibinfo {author} {\bibfnamefont {D.~M.}\ \bibnamefont {Meekhof}},\ }\bibfield  {title} {\bibinfo {title} {Experimental issues in coherent quantum-state manipulation of trapped atomic ions},\ }\href {https://tf.nist.gov/general/pdf/1275.pdf} {\bibfield  {journal} {\bibinfo  {journal} {J. Res. Natl. Inst. Stand. Technol}\ }\textbf {\bibinfo {volume} {103}} (\bibinfo {year} {1998})}\BibitemShut {NoStop}%
\bibitem [{\citenamefont {Chase}\ and\ \citenamefont {Geremia}(2008)}]{Chase2008}%
  \BibitemOpen
  \bibfield  {author} {\bibinfo {author} {\bibfnamefont {B.~A.}\ \bibnamefont {Chase}}\ and\ \bibinfo {author} {\bibfnamefont {J.~M.}\ \bibnamefont {Geremia}},\ }\bibfield  {title} {\bibinfo {title} {Collective processes of an ensemble of spin-$1/2$ particles},\ }\href {https://doi.org/10.1103/PhysRevA.78.052101} {\bibfield  {journal} {\bibinfo  {journal} {Phys. Rev. A}\ }\textbf {\bibinfo {volume} {78}},\ \bibinfo {pages} {052101} (\bibinfo {year} {2008})}\BibitemShut {NoStop}%
\bibitem [{\citenamefont {Biercuk}\ \emph {et~al.}(2009)\citenamefont {Biercuk}, \citenamefont {Uys}, \citenamefont {Vandevender}, \citenamefont {Shiga}, \citenamefont {Itano},\ and\ \citenamefont {Bollinger}}]{Biercuk2009}%
  \BibitemOpen
  \bibfield  {author} {\bibinfo {author} {\bibfnamefont {M.~J.}\ \bibnamefont {Biercuk}}, \bibinfo {author} {\bibfnamefont {H.}~\bibnamefont {Uys}}, \bibinfo {author} {\bibfnamefont {A.~P.}\ \bibnamefont {Vandevender}}, \bibinfo {author} {\bibfnamefont {N.}~\bibnamefont {Shiga}}, \bibinfo {author} {\bibfnamefont {W.~M.}\ \bibnamefont {Itano}},\ and\ \bibinfo {author} {\bibfnamefont {J.~J.}\ \bibnamefont {Bollinger}},\ }\bibfield  {title} {\bibinfo {title} {High-fidelity quantum control using ion crystals in a penning trap},\ }\href {https://doi.org/10.26421/QIC9.11-12-2} {\bibfield  {journal} {\bibinfo  {journal} {Quantum Information and Computation}\ }\textbf {\bibinfo {volume} {9}},\ \bibinfo {pages} {0920} (\bibinfo {year} {2009})}\BibitemShut {NoStop}%
\bibitem [{\citenamefont {Choi}\ \emph {et~al.}(2017)\citenamefont {Choi}, \citenamefont {Choi}, \citenamefont {Kucsko}, \citenamefont {Maurer}, \citenamefont {Shields}, \citenamefont {Sumiya}, \citenamefont {Onoda}, \citenamefont {Isoya}, \citenamefont {Demler}, \citenamefont {Jelezko}, \citenamefont {Yao},\ and\ \citenamefont {Lukin}}]{Choi2017}%
  \BibitemOpen
  \bibfield  {author} {\bibinfo {author} {\bibfnamefont {J.}~\bibnamefont {Choi}}, \bibinfo {author} {\bibfnamefont {S.}~\bibnamefont {Choi}}, \bibinfo {author} {\bibfnamefont {G.}~\bibnamefont {Kucsko}}, \bibinfo {author} {\bibfnamefont {P.~C.}\ \bibnamefont {Maurer}}, \bibinfo {author} {\bibfnamefont {B.~J.}\ \bibnamefont {Shields}}, \bibinfo {author} {\bibfnamefont {H.}~\bibnamefont {Sumiya}}, \bibinfo {author} {\bibfnamefont {S.}~\bibnamefont {Onoda}}, \bibinfo {author} {\bibfnamefont {J.}~\bibnamefont {Isoya}}, \bibinfo {author} {\bibfnamefont {E.}~\bibnamefont {Demler}}, \bibinfo {author} {\bibfnamefont {F.}~\bibnamefont {Jelezko}}, \bibinfo {author} {\bibfnamefont {N.~Y.}\ \bibnamefont {Yao}},\ and\ \bibinfo {author} {\bibfnamefont {M.~D.}\ \bibnamefont {Lukin}},\ }\bibfield  {title} {\bibinfo {title} {Depolarization dynamics in a strongly interacting solid-state spin ensemble},\ }\href {https://doi.org/10.1103/PhysRevLett.118.093601} {\bibfield  {journal} {\bibinfo  {journal} {Phys. Rev. Lett.}\
  }\textbf {\bibinfo {volume} {118}},\ \bibinfo {pages} {093601} (\bibinfo {year} {2017})}\BibitemShut {NoStop}%
\bibitem [{\citenamefont {Kirton}\ and\ \citenamefont {Keeling}(2017)}]{Kirton2017}%
  \BibitemOpen
  \bibfield  {author} {\bibinfo {author} {\bibfnamefont {P.}~\bibnamefont {Kirton}}\ and\ \bibinfo {author} {\bibfnamefont {J.}~\bibnamefont {Keeling}},\ }\bibfield  {title} {\bibinfo {title} {Suppressing and restoring the dicke superradiance transition by dephasing and decay},\ }\href {https://doi.org/10.1103/PhysRevLett.118.123602} {\bibfield  {journal} {\bibinfo  {journal} {Phys. Rev. Lett.}\ }\textbf {\bibinfo {volume} {118}},\ \bibinfo {pages} {123602} (\bibinfo {year} {2017})}\BibitemShut {NoStop}%
\bibitem [{\citenamefont {Shammah}\ \emph {et~al.}(2018)\citenamefont {Shammah}, \citenamefont {Ahmed}, \citenamefont {Lambert}, \citenamefont {De~Liberato},\ and\ \citenamefont {Nori}}]{Shammah2018}%
  \BibitemOpen
  \bibfield  {author} {\bibinfo {author} {\bibfnamefont {N.}~\bibnamefont {Shammah}}, \bibinfo {author} {\bibfnamefont {S.}~\bibnamefont {Ahmed}}, \bibinfo {author} {\bibfnamefont {N.}~\bibnamefont {Lambert}}, \bibinfo {author} {\bibfnamefont {S.}~\bibnamefont {De~Liberato}},\ and\ \bibinfo {author} {\bibfnamefont {F.}~\bibnamefont {Nori}},\ }\bibfield  {title} {\bibinfo {title} {Open quantum systems with local and collective incoherent processes: Efficient numerical simulations using permutational invariance},\ }\href {https://doi.org/10.1103/PhysRevA.98.063815} {\bibfield  {journal} {\bibinfo  {journal} {Phys. Rev. A}\ }\textbf {\bibinfo {volume} {98}},\ \bibinfo {pages} {063815} (\bibinfo {year} {2018})}\BibitemShut {NoStop}%
\bibitem [{\citenamefont {Ruskai}(2000)}]{Ruskai2000}%
  \BibitemOpen
  \bibfield  {author} {\bibinfo {author} {\bibfnamefont {M.~B.}\ \bibnamefont {Ruskai}},\ }\bibfield  {title} {\bibinfo {title} {Pauli exchange errors in quantum computation},\ }\href {https://doi.org/10.1103/PhysRevLett.85.194} {\bibfield  {journal} {\bibinfo  {journal} {Phys. Rev. Lett.}\ }\textbf {\bibinfo {volume} {85}},\ \bibinfo {pages} {194} (\bibinfo {year} {2000})}\BibitemShut {NoStop}%
\bibitem [{\citenamefont {Pollatsek}\ and\ \citenamefont {Ruskai}(2004)}]{Pollatsek2004}%
  \BibitemOpen
  \bibfield  {author} {\bibinfo {author} {\bibfnamefont {H.}~\bibnamefont {Pollatsek}}\ and\ \bibinfo {author} {\bibfnamefont {M.~B.}\ \bibnamefont {Ruskai}},\ }\bibfield  {title} {\bibinfo {title} {Permutationally invariant codes for quantum error correction},\ }\href {https://doi.org/https://doi.org/10.1016/j.laa.2004.06.014} {\bibfield  {journal} {\bibinfo  {journal} {Linear Algebra and its Applications}\ }\textbf {\bibinfo {volume} {392}},\ \bibinfo {pages} {255} (\bibinfo {year} {2004})}\BibitemShut {NoStop}%
\bibitem [{\citenamefont {Brion}\ \emph {et~al.}(2008)\citenamefont {Brion}, \citenamefont {Pedersen}, \citenamefont {Saffman},\ and\ \citenamefont {M\o{}lmer}}]{Brion2008}%
  \BibitemOpen
  \bibfield  {author} {\bibinfo {author} {\bibfnamefont {E.}~\bibnamefont {Brion}}, \bibinfo {author} {\bibfnamefont {L.~H.}\ \bibnamefont {Pedersen}}, \bibinfo {author} {\bibfnamefont {M.}~\bibnamefont {Saffman}},\ and\ \bibinfo {author} {\bibfnamefont {K.}~\bibnamefont {M\o{}lmer}},\ }\bibfield  {title} {\bibinfo {title} {Error correction in ensemble registers for quantum repeaters and quantum computers},\ }\href {https://doi.org/10.1103/PhysRevLett.100.110506} {\bibfield  {journal} {\bibinfo  {journal} {Phys. Rev. Lett.}\ }\textbf {\bibinfo {volume} {100}},\ \bibinfo {pages} {110506} (\bibinfo {year} {2008})}\BibitemShut {NoStop}%
\bibitem [{\citenamefont {Ouyang}(2014)}]{Ouyang2014}%
  \BibitemOpen
  \bibfield  {author} {\bibinfo {author} {\bibfnamefont {Y.}~\bibnamefont {Ouyang}},\ }\bibfield  {title} {\bibinfo {title} {Permutation-invariant quantum codes},\ }\href {https://doi.org/10.1103/PhysRevA.90.062317} {\bibfield  {journal} {\bibinfo  {journal} {Phys. Rev. A}\ }\textbf {\bibinfo {volume} {90}},\ \bibinfo {pages} {062317} (\bibinfo {year} {2014})}\BibitemShut {NoStop}%
\bibitem [{\citenamefont {Ouyang}\ and\ \citenamefont {Brennen}(2026{\natexlab{b}})}]{Ouyang2026b}%
  \BibitemOpen
  \bibfield  {author} {\bibinfo {author} {\bibfnamefont {Y.}~\bibnamefont {Ouyang}}\ and\ \bibinfo {author} {\bibfnamefont {G.~K.}\ \bibnamefont {Brennen}},\ }\href@noop {} {\bibinfo {title} {A theory of quantum error correction for permutation-invariant codes}} (\bibinfo {year} {2026}{\natexlab{b}}),\ \Eprint {https://arxiv.org/abs/2602.13638} {arXiv:2602.13638 [quant-ph]} \BibitemShut {NoStop}%
\bibitem [{\citenamefont {Omanakuttan}\ and\ \citenamefont {Gross}(2023)}]{Omanakuttan2023a}%
  \BibitemOpen
  \bibfield  {author} {\bibinfo {author} {\bibfnamefont {S.}~\bibnamefont {Omanakuttan}}\ and\ \bibinfo {author} {\bibfnamefont {J.~A.}\ \bibnamefont {Gross}},\ }\bibfield  {title} {\bibinfo {title} {Multispin clifford codes for angular momentum errors in spin systems},\ }\href {https://doi.org/10.1103/PhysRevA.108.022424} {\bibfield  {journal} {\bibinfo  {journal} {Phys. Rev. A}\ }\textbf {\bibinfo {volume} {108}},\ \bibinfo {pages} {022424} (\bibinfo {year} {2023})}\BibitemShut {NoStop}%
\bibitem [{\citenamefont {Omanakuttan}\ and\ \citenamefont {Volkoff}(2023)}]{Omanakuttan2023b}%
  \BibitemOpen
  \bibfield  {author} {\bibinfo {author} {\bibfnamefont {S.}~\bibnamefont {Omanakuttan}}\ and\ \bibinfo {author} {\bibfnamefont {T.~J.}\ \bibnamefont {Volkoff}},\ }\bibfield  {title} {\bibinfo {title} {Spin-squeezed gottesman-kitaev-preskill codes for quantum error correction in atomic ensembles},\ }\href {https://doi.org/10.1103/PhysRevA.108.022428} {\bibfield  {journal} {\bibinfo  {journal} {Phys. Rev. A}\ }\textbf {\bibinfo {volume} {108}},\ \bibinfo {pages} {022428} (\bibinfo {year} {2023})}\BibitemShut {NoStop}%
\bibitem [{\citenamefont {Brion}\ \emph {et~al.}(2007)\citenamefont {Brion}, \citenamefont {M\o{}lmer},\ and\ \citenamefont {Saffman}}]{Biron2007}%
  \BibitemOpen
  \bibfield  {author} {\bibinfo {author} {\bibfnamefont {E.}~\bibnamefont {Brion}}, \bibinfo {author} {\bibfnamefont {K.}~\bibnamefont {M\o{}lmer}},\ and\ \bibinfo {author} {\bibfnamefont {M.}~\bibnamefont {Saffman}},\ }\bibfield  {title} {\bibinfo {title} {Quantum computing with collective ensembles of multilevel systems},\ }\href {https://doi.org/10.1103/PhysRevLett.99.260501} {\bibfield  {journal} {\bibinfo  {journal} {Phys. Rev. Lett.}\ }\textbf {\bibinfo {volume} {99}},\ \bibinfo {pages} {260501} (\bibinfo {year} {2007})}\BibitemShut {NoStop}%
\bibitem [{\citenamefont {McMahon}\ \emph {et~al.}(2024)\citenamefont {McMahon}, \citenamefont {Brown}, \citenamefont {Herold},\ and\ \citenamefont {Sawyer}}]{McMahon2024}%
  \BibitemOpen
  \bibfield  {author} {\bibinfo {author} {\bibfnamefont {B.~J.}\ \bibnamefont {McMahon}}, \bibinfo {author} {\bibfnamefont {K.~R.}\ \bibnamefont {Brown}}, \bibinfo {author} {\bibfnamefont {C.~D.}\ \bibnamefont {Herold}},\ and\ \bibinfo {author} {\bibfnamefont {B.~C.}\ \bibnamefont {Sawyer}},\ }\bibfield  {title} {\bibinfo {title} {Individual-ion addressing and readout in a penning trap},\ }\href {https://doi.org/10.1103/PhysRevLett.133.173201} {\bibfield  {journal} {\bibinfo  {journal} {Phys. Rev. Lett.}\ }\textbf {\bibinfo {volume} {133}},\ \bibinfo {pages} {173201} (\bibinfo {year} {2024})}\BibitemShut {NoStop}%
\bibitem [{\citenamefont {Bohnet}\ \emph {et~al.}(2016)\citenamefont {Bohnet}, \citenamefont {Sawyer}, \citenamefont {Britton}, \citenamefont {Wall}, \citenamefont {Rey}, \citenamefont {Foss-Feig},\ and\ \citenamefont {Bollinger}}]{Bohnet2016}%
  \BibitemOpen
  \bibfield  {author} {\bibinfo {author} {\bibfnamefont {J.~G.}\ \bibnamefont {Bohnet}}, \bibinfo {author} {\bibfnamefont {B.~C.}\ \bibnamefont {Sawyer}}, \bibinfo {author} {\bibfnamefont {J.~W.}\ \bibnamefont {Britton}}, \bibinfo {author} {\bibfnamefont {M.~L.}\ \bibnamefont {Wall}}, \bibinfo {author} {\bibfnamefont {A.~M.}\ \bibnamefont {Rey}}, \bibinfo {author} {\bibfnamefont {M.}~\bibnamefont {Foss-Feig}},\ and\ \bibinfo {author} {\bibfnamefont {J.~J.}\ \bibnamefont {Bollinger}},\ }\bibfield  {title} {\bibinfo {title} {Quantum spin dynamics and entanglement generation with hundreds of trapped ions},\ }\href {https://doi.org/10.1126/science.aad9958} {\bibfield  {journal} {\bibinfo  {journal} {Science}\ }\textbf {\bibinfo {volume} {352}},\ \bibinfo {pages} {1297} (\bibinfo {year} {2016})}\BibitemShut {NoStop}%
\bibitem [{\citenamefont {Albert}\ \emph {et~al.}(2018)\citenamefont {Albert}, \citenamefont {Noh}, \citenamefont {Duivenvoorden}, \citenamefont {Young}, \citenamefont {Brierley}, \citenamefont {Reinhold}, \citenamefont {Vuillot}, \citenamefont {Li}, \citenamefont {Shen}, \citenamefont {Girvin}, \citenamefont {Terhal},\ and\ \citenamefont {Jiang}}]{Albert2018}%
  \BibitemOpen
  \bibfield  {author} {\bibinfo {author} {\bibfnamefont {V.~V.}\ \bibnamefont {Albert}}, \bibinfo {author} {\bibfnamefont {K.}~\bibnamefont {Noh}}, \bibinfo {author} {\bibfnamefont {K.}~\bibnamefont {Duivenvoorden}}, \bibinfo {author} {\bibfnamefont {D.~J.}\ \bibnamefont {Young}}, \bibinfo {author} {\bibfnamefont {R.~T.}\ \bibnamefont {Brierley}}, \bibinfo {author} {\bibfnamefont {P.}~\bibnamefont {Reinhold}}, \bibinfo {author} {\bibfnamefont {C.}~\bibnamefont {Vuillot}}, \bibinfo {author} {\bibfnamefont {L.}~\bibnamefont {Li}}, \bibinfo {author} {\bibfnamefont {C.}~\bibnamefont {Shen}}, \bibinfo {author} {\bibfnamefont {S.~M.}\ \bibnamefont {Girvin}}, \bibinfo {author} {\bibfnamefont {B.~M.}\ \bibnamefont {Terhal}},\ and\ \bibinfo {author} {\bibfnamefont {L.}~\bibnamefont {Jiang}},\ }\bibfield  {title} {\bibinfo {title} {Performance and structure of single-mode bosonic codes},\ }\href {https://doi.org/10.1103/PhysRevA.97.032346} {\bibfield  {journal} {\bibinfo  {journal} {Phys. Rev. A}\ }\textbf {\bibinfo
  {volume} {97}},\ \bibinfo {pages} {032346} (\bibinfo {year} {2018})}\BibitemShut {NoStop}%
\bibitem [{\citenamefont {Joshi}\ \emph {et~al.}(2021)\citenamefont {Joshi}, \citenamefont {Noh},\ and\ \citenamefont {Gao}}]{Joshi_2021}%
  \BibitemOpen
  \bibfield  {author} {\bibinfo {author} {\bibfnamefont {A.}~\bibnamefont {Joshi}}, \bibinfo {author} {\bibfnamefont {K.}~\bibnamefont {Noh}},\ and\ \bibinfo {author} {\bibfnamefont {Y.~Y.}\ \bibnamefont {Gao}},\ }\bibfield  {title} {\bibinfo {title} {Quantum information processing with bosonic qubits in circuit qed},\ }\href {https://doi.org/10.1088/2058-9565/abe989} {\bibfield  {journal} {\bibinfo  {journal} {Quantum Science and Technology}\ }\textbf {\bibinfo {volume} {6}},\ \bibinfo {pages} {033001} (\bibinfo {year} {2021})}\BibitemShut {NoStop}%
\bibitem [{\citenamefont {Gross}(2021)}]{Gross2021}%
  \BibitemOpen
  \bibfield  {author} {\bibinfo {author} {\bibfnamefont {J.~A.}\ \bibnamefont {Gross}},\ }\bibfield  {title} {\bibinfo {title} {Designing codes around interactions: The case of a spin},\ }\href {https://doi.org/10.1103/PhysRevLett.127.010504} {\bibfield  {journal} {\bibinfo  {journal} {Phys. Rev. Lett.}\ }\textbf {\bibinfo {volume} {127}},\ \bibinfo {pages} {010504} (\bibinfo {year} {2021})}\BibitemShut {NoStop}%
\bibitem [{\citenamefont {Albert}\ \emph {et~al.}(2020)\citenamefont {Albert}, \citenamefont {Covey},\ and\ \citenamefont {Preskill}}]{Albert2020}%
  \BibitemOpen
  \bibfield  {author} {\bibinfo {author} {\bibfnamefont {V.~V.}\ \bibnamefont {Albert}}, \bibinfo {author} {\bibfnamefont {J.~P.}\ \bibnamefont {Covey}},\ and\ \bibinfo {author} {\bibfnamefont {J.}~\bibnamefont {Preskill}},\ }\bibfield  {title} {\bibinfo {title} {Robust encoding of a qubit in a molecule},\ }\href {https://doi.org/10.1103/PhysRevX.10.031050} {\bibfield  {journal} {\bibinfo  {journal} {Phys. Rev. X}\ }\textbf {\bibinfo {volume} {10}},\ \bibinfo {pages} {031050} (\bibinfo {year} {2020})}\BibitemShut {NoStop}%
\bibitem [{\citenamefont {Jain}\ \emph {et~al.}(2024)\citenamefont {Jain}, \citenamefont {Hudson}, \citenamefont {Campbell},\ and\ \citenamefont {Albert}}]{Jain2024}%
  \BibitemOpen
  \bibfield  {author} {\bibinfo {author} {\bibfnamefont {S.~P.}\ \bibnamefont {Jain}}, \bibinfo {author} {\bibfnamefont {E.~R.}\ \bibnamefont {Hudson}}, \bibinfo {author} {\bibfnamefont {W.~C.}\ \bibnamefont {Campbell}},\ and\ \bibinfo {author} {\bibfnamefont {V.~V.}\ \bibnamefont {Albert}},\ }\bibfield  {title} {\bibinfo {title} {Absorption-emission codes for atomic and molecular quantum information platforms},\ }\href {https://doi.org/10.1103/PhysRevLett.133.260601} {\bibfield  {journal} {\bibinfo  {journal} {Phys. Rev. Lett.}\ }\textbf {\bibinfo {volume} {133}},\ \bibinfo {pages} {260601} (\bibinfo {year} {2024})}\BibitemShut {NoStop}%
\bibitem [{\citenamefont {Furey}\ \emph {et~al.}(2024)\citenamefont {Furey}, \citenamefont {Wu}, \citenamefont {Isaza-Monsalve}, \citenamefont {Walser}, \citenamefont {Mattivi}, \citenamefont {Nardi},\ and\ \citenamefont {Schindler}}]{Furey2024}%
  \BibitemOpen
  \bibfield  {author} {\bibinfo {author} {\bibfnamefont {B.~J.}\ \bibnamefont {Furey}}, \bibinfo {author} {\bibfnamefont {Z.}~\bibnamefont {Wu}}, \bibinfo {author} {\bibfnamefont {M.}~\bibnamefont {Isaza-Monsalve}}, \bibinfo {author} {\bibfnamefont {S.}~\bibnamefont {Walser}}, \bibinfo {author} {\bibfnamefont {E.}~\bibnamefont {Mattivi}}, \bibinfo {author} {\bibfnamefont {R.}~\bibnamefont {Nardi}},\ and\ \bibinfo {author} {\bibfnamefont {P.}~\bibnamefont {Schindler}},\ }\bibfield  {title} {\bibinfo {title} {Strategies for implementing quantum error correction in molecular rotation},\ }\href {https://doi.org/10.22331/q-2024-12-27-1578} {\bibfield  {journal} {\bibinfo  {journal} {{Quantum}}\ }\textbf {\bibinfo {volume} {8}},\ \bibinfo {pages} {1578} (\bibinfo {year} {2024})}\BibitemShut {NoStop}%
\bibitem [{\citenamefont {Varshalovich}\ \emph {et~al.}(1988)\citenamefont {Varshalovich}, \citenamefont {Moskalev},\ and\ \citenamefont {Khersonskii}}]{Varshalovich1988}%
  \BibitemOpen
  \bibfield  {author} {\bibinfo {author} {\bibfnamefont {D.~A.}\ \bibnamefont {Varshalovich}}, \bibinfo {author} {\bibfnamefont {A.~N.}\ \bibnamefont {Moskalev}},\ and\ \bibinfo {author} {\bibfnamefont {V.~K.}\ \bibnamefont {Khersonskii}},\ }\href {https://doi.org/10.1142/0270} {\emph {\bibinfo {title} {Quantum Theory of Angular Momentum}}}\ (\bibinfo  {publisher} {World Scientific},\ \bibinfo {year} {1988})\ pp.\ \bibinfo {pages} {235--289}\BibitemShut {NoStop}%
\bibitem [{\citenamefont {Dicke}(1954)}]{Dicke1954}%
  \BibitemOpen
  \bibfield  {author} {\bibinfo {author} {\bibfnamefont {R.~H.}\ \bibnamefont {Dicke}},\ }\bibfield  {title} {\bibinfo {title} {Coherence in spontaneous radiation processes},\ }\href {https://doi.org/10.1103/PhysRev.93.99} {\bibfield  {journal} {\bibinfo  {journal} {Phys. Rev.}\ }\textbf {\bibinfo {volume} {93}},\ \bibinfo {pages} {99} (\bibinfo {year} {1954})}\BibitemShut {NoStop}%
\bibitem [{\citenamefont {Breuer}\ and\ \citenamefont {Petruccione}(2007)}]{Breuer2007}%
  \BibitemOpen
  \bibfield  {author} {\bibinfo {author} {\bibfnamefont {H.-P.}\ \bibnamefont {Breuer}}\ and\ \bibinfo {author} {\bibfnamefont {F.}~\bibnamefont {Petruccione}},\ }\href {https://doi.org/10.1093/acprof:oso/9780199213900.001.0001} {\emph {\bibinfo {title} {{The Theory of Open Quantum Systems}}}}\ (\bibinfo  {publisher} {Oxford University Press},\ \bibinfo {year} {2007})\BibitemShut {NoStop}%
\bibitem [{Note1()}]{Note1}%
  \BibitemOpen
  \bibinfo {note} {The individual decoherence rates of all spins are assumed to be same, which preserves permutational invariance of the system. This condition is naturally satisfied when the spins are identical and couple identically to the environment.}\BibitemShut {Stop}%
\bibitem [{\citenamefont {Dalibard}\ \emph {et~al.}(1992)\citenamefont {Dalibard}, \citenamefont {Castin},\ and\ \citenamefont {M\o{}lmer}}]{Dalibard1992}%
  \BibitemOpen
  \bibfield  {author} {\bibinfo {author} {\bibfnamefont {J.}~\bibnamefont {Dalibard}}, \bibinfo {author} {\bibfnamefont {Y.}~\bibnamefont {Castin}},\ and\ \bibinfo {author} {\bibfnamefont {K.}~\bibnamefont {M\o{}lmer}},\ }\bibfield  {title} {\bibinfo {title} {Wave-function approach to dissipative processes in quantum optics},\ }\href {https://doi.org/10.1103/PhysRevLett.68.580} {\bibfield  {journal} {\bibinfo  {journal} {Phys. Rev. Lett.}\ }\textbf {\bibinfo {volume} {68}},\ \bibinfo {pages} {580} (\bibinfo {year} {1992})}\BibitemShut {NoStop}%
\bibitem [{\citenamefont {Grace}\ \emph {et~al.}(2006)\citenamefont {Grace}, \citenamefont {Brif}, \citenamefont {Rabitz}, \citenamefont {Walmsley}, \citenamefont {Kosut},\ and\ \citenamefont {Lidar}}]{Grace2006}%
  \BibitemOpen
  \bibfield  {author} {\bibinfo {author} {\bibfnamefont {M.}~\bibnamefont {Grace}}, \bibinfo {author} {\bibfnamefont {C.}~\bibnamefont {Brif}}, \bibinfo {author} {\bibfnamefont {H.}~\bibnamefont {Rabitz}}, \bibinfo {author} {\bibfnamefont {I.}~\bibnamefont {Walmsley}}, \bibinfo {author} {\bibfnamefont {R.}~\bibnamefont {Kosut}},\ and\ \bibinfo {author} {\bibfnamefont {D.}~\bibnamefont {Lidar}},\ }\bibfield  {title} {\bibinfo {title} {Encoding a qubit into multilevel subspaces},\ }\href {https://doi.org/10.1088/1367-2630/8/3/035} {\bibfield  {journal} {\bibinfo  {journal} {New Journal of Physics}\ }\textbf {\bibinfo {volume} {8}},\ \bibinfo {pages} {35} (\bibinfo {year} {2006})}\BibitemShut {NoStop}%
\bibitem [{\citenamefont {Lau}\ and\ \citenamefont {Plenio}(2017)}]{Lau2017}%
  \BibitemOpen
  \bibfield  {author} {\bibinfo {author} {\bibfnamefont {H.-K.}\ \bibnamefont {Lau}}\ and\ \bibinfo {author} {\bibfnamefont {M.~B.}\ \bibnamefont {Plenio}},\ }\bibfield  {title} {\bibinfo {title} {Universal continuous-variable quantum computation without cooling},\ }\href {https://doi.org/10.1103/PhysRevA.95.022303} {\bibfield  {journal} {\bibinfo  {journal} {Phys. Rev. A}\ }\textbf {\bibinfo {volume} {95}},\ \bibinfo {pages} {022303} (\bibinfo {year} {2017})}\BibitemShut {NoStop}%
\bibitem [{\citenamefont {Zhou}\ \emph {et~al.}(2000)\citenamefont {Zhou}, \citenamefont {Leung},\ and\ \citenamefont {Chuang}}]{Zhou2000}%
  \BibitemOpen
  \bibfield  {author} {\bibinfo {author} {\bibfnamefont {X.}~\bibnamefont {Zhou}}, \bibinfo {author} {\bibfnamefont {D.~W.}\ \bibnamefont {Leung}},\ and\ \bibinfo {author} {\bibfnamefont {I.~L.}\ \bibnamefont {Chuang}},\ }\bibfield  {title} {\bibinfo {title} {Methodology for quantum logic gate construction},\ }\href {https://doi.org/10.1103/PhysRevA.62.052316} {\bibfield  {journal} {\bibinfo  {journal} {Phys. Rev. A}\ }\textbf {\bibinfo {volume} {62}},\ \bibinfo {pages} {052316} (\bibinfo {year} {2000})}\BibitemShut {NoStop}%
\bibitem [{\citenamefont {Leung}\ \emph {et~al.}(1997)\citenamefont {Leung}, \citenamefont {Nielsen}, \citenamefont {Chuang},\ and\ \citenamefont {Yamamoto}}]{Leung1997}%
  \BibitemOpen
  \bibfield  {author} {\bibinfo {author} {\bibfnamefont {D.~W.}\ \bibnamefont {Leung}}, \bibinfo {author} {\bibfnamefont {M.~A.}\ \bibnamefont {Nielsen}}, \bibinfo {author} {\bibfnamefont {I.~L.}\ \bibnamefont {Chuang}},\ and\ \bibinfo {author} {\bibfnamefont {Y.}~\bibnamefont {Yamamoto}},\ }\bibfield  {title} {\bibinfo {title} {Approximate quantum error correction can lead to better codes},\ }\href {https://doi.org/10.1103/PhysRevA.56.2567} {\bibfield  {journal} {\bibinfo  {journal} {Phys. Rev. A}\ }\textbf {\bibinfo {volume} {56}},\ \bibinfo {pages} {2567} (\bibinfo {year} {1997})}\BibitemShut {NoStop}%
\bibitem [{\citenamefont {Wang}\ and\ \citenamefont {Terhal}(2021)}]{Wang2021}%
  \BibitemOpen
  \bibfield  {author} {\bibinfo {author} {\bibfnamefont {Y.}~\bibnamefont {Wang}}\ and\ \bibinfo {author} {\bibfnamefont {B.~M.}\ \bibnamefont {Terhal}},\ }\bibfield  {title} {\bibinfo {title} {Preparing dicke states in a spin ensemble using phase estimation},\ }\href {https://doi.org/10.1103/PhysRevA.104.032407} {\bibfield  {journal} {\bibinfo  {journal} {Phys. Rev. A}\ }\textbf {\bibinfo {volume} {104}},\ \bibinfo {pages} {032407} (\bibinfo {year} {2021})}\BibitemShut {NoStop}%
\bibitem [{\citenamefont {Knill}\ and\ \citenamefont {Laflamme}(1997)}]{Knill1997}%
  \BibitemOpen
  \bibfield  {author} {\bibinfo {author} {\bibfnamefont {E.}~\bibnamefont {Knill}}\ and\ \bibinfo {author} {\bibfnamefont {R.}~\bibnamefont {Laflamme}},\ }\bibfield  {title} {\bibinfo {title} {Theory of quantum error-correcting codes},\ }\href {https://doi.org/10.1103/PhysRevA.55.900} {\bibfield  {journal} {\bibinfo  {journal} {Phys. Rev. A}\ }\textbf {\bibinfo {volume} {55}},\ \bibinfo {pages} {900} (\bibinfo {year} {1997})}\BibitemShut {NoStop}%
\bibitem [{\citenamefont {Motlagh}\ and\ \citenamefont {Wiebe}(2024)}]{Motlagh2024}%
  \BibitemOpen
  \bibfield  {author} {\bibinfo {author} {\bibfnamefont {D.}~\bibnamefont {Motlagh}}\ and\ \bibinfo {author} {\bibfnamefont {N.}~\bibnamefont {Wiebe}},\ }\bibfield  {title} {\bibinfo {title} {Generalized quantum signal processing},\ }\href {https://doi.org/10.1103/PRXQuantum.5.020368} {\bibfield  {journal} {\bibinfo  {journal} {PRX Quantum}\ }\textbf {\bibinfo {volume} {5}},\ \bibinfo {pages} {020368} (\bibinfo {year} {2024})}\BibitemShut {NoStop}%
\bibitem [{\citenamefont {Fong}\ and\ \citenamefont {Lau}(2026)}]{Fong2026}%
  \BibitemOpen
  \bibfield  {author} {\bibinfo {author} {\bibfnamefont {P.-T.}\ \bibnamefont {Fong}}\ and\ \bibinfo {author} {\bibfnamefont {H.-K.}\ \bibnamefont {Lau}},\ }\bibfield  {title} {\bibinfo {title} {Engineering non-gaussian bosonic gates through quantum signal processing},\ }\href {https://doi.org/10.1103/gdbf-cjz4} {\bibfield  {journal} {\bibinfo  {journal} {Phys. Rev. A}\ } (\bibinfo {year} {2026})},\ \bibinfo {note} {(to be published)}\BibitemShut {NoStop}%
\bibitem [{\citenamefont {Gutman}\ \emph {et~al.}(2024)\citenamefont {Gutman}, \citenamefont {Gorlach}, \citenamefont {Tziperman}, \citenamefont {Ruimy},\ and\ \citenamefont {Kaminer}}]{Gutman2023}%
  \BibitemOpen
  \bibfield  {author} {\bibinfo {author} {\bibfnamefont {N.}~\bibnamefont {Gutman}}, \bibinfo {author} {\bibfnamefont {A.}~\bibnamefont {Gorlach}}, \bibinfo {author} {\bibfnamefont {O.}~\bibnamefont {Tziperman}}, \bibinfo {author} {\bibfnamefont {R.}~\bibnamefont {Ruimy}},\ and\ \bibinfo {author} {\bibfnamefont {I.}~\bibnamefont {Kaminer}},\ }\bibfield  {title} {\bibinfo {title} {Universal control of symmetric states using spin squeezing},\ }\href {https://doi.org/10.1103/PhysRevLett.132.153601} {\bibfield  {journal} {\bibinfo  {journal} {Phys. Rev. Lett.}\ }\textbf {\bibinfo {volume} {132}},\ \bibinfo {pages} {153601} (\bibinfo {year} {2024})}\BibitemShut {NoStop}%
\bibitem [{\citenamefont {Kitagawa}\ and\ \citenamefont {Ueda}(1993)}]{Kitagawa1993}%
  \BibitemOpen
  \bibfield  {author} {\bibinfo {author} {\bibfnamefont {M.}~\bibnamefont {Kitagawa}}\ and\ \bibinfo {author} {\bibfnamefont {M.}~\bibnamefont {Ueda}},\ }\bibfield  {title} {\bibinfo {title} {Squeezed spin states},\ }\href {https://doi.org/10.1103/PhysRevA.47.5138} {\bibfield  {journal} {\bibinfo  {journal} {Phys. Rev. A}\ }\textbf {\bibinfo {volume} {47}},\ \bibinfo {pages} {5138} (\bibinfo {year} {1993})}\BibitemShut {NoStop}%
\bibitem [{\citenamefont {Keating}\ \emph {et~al.}(2016)\citenamefont {Keating}, \citenamefont {Baldwin}, \citenamefont {Jau}, \citenamefont {Lee}, \citenamefont {Biedermann},\ and\ \citenamefont {Deutsch}}]{Keating2016}%
  \BibitemOpen
  \bibfield  {author} {\bibinfo {author} {\bibfnamefont {T.}~\bibnamefont {Keating}}, \bibinfo {author} {\bibfnamefont {C.~H.}\ \bibnamefont {Baldwin}}, \bibinfo {author} {\bibfnamefont {Y.-Y.}\ \bibnamefont {Jau}}, \bibinfo {author} {\bibfnamefont {J.}~\bibnamefont {Lee}}, \bibinfo {author} {\bibfnamefont {G.~W.}\ \bibnamefont {Biedermann}},\ and\ \bibinfo {author} {\bibfnamefont {I.~H.}\ \bibnamefont {Deutsch}},\ }\bibfield  {title} {\bibinfo {title} {Arbitrary dicke-state control of symmetric rydberg ensembles},\ }\href {https://doi.org/10.1103/PhysRevLett.117.213601} {\bibfield  {journal} {\bibinfo  {journal} {Phys. Rev. Lett.}\ }\textbf {\bibinfo {volume} {117}},\ \bibinfo {pages} {213601} (\bibinfo {year} {2016})}\BibitemShut {NoStop}%
\bibitem [{\citenamefont {Bennett}\ \emph {et~al.}(2013)\citenamefont {Bennett}, \citenamefont {Yao}, \citenamefont {Otterbach}, \citenamefont {Zoller}, \citenamefont {Rabl},\ and\ \citenamefont {Lukin}}]{Bennett2013}%
  \BibitemOpen
  \bibfield  {author} {\bibinfo {author} {\bibfnamefont {S.~D.}\ \bibnamefont {Bennett}}, \bibinfo {author} {\bibfnamefont {N.~Y.}\ \bibnamefont {Yao}}, \bibinfo {author} {\bibfnamefont {J.}~\bibnamefont {Otterbach}}, \bibinfo {author} {\bibfnamefont {P.}~\bibnamefont {Zoller}}, \bibinfo {author} {\bibfnamefont {P.}~\bibnamefont {Rabl}},\ and\ \bibinfo {author} {\bibfnamefont {M.~D.}\ \bibnamefont {Lukin}},\ }\bibfield  {title} {\bibinfo {title} {Phonon-induced spin-spin interactions in diamond nanostructures: Application to spin squeezing},\ }\href {https://doi.org/10.1103/PhysRevLett.110.156402} {\bibfield  {journal} {\bibinfo  {journal} {Phys. Rev. Lett.}\ }\textbf {\bibinfo {volume} {110}},\ \bibinfo {pages} {156402} (\bibinfo {year} {2013})}\BibitemShut {NoStop}%
\bibitem [{\citenamefont {Teissier}\ \emph {et~al.}(2014)\citenamefont {Teissier}, \citenamefont {Barfuss}, \citenamefont {Appel}, \citenamefont {Neu},\ and\ \citenamefont {Maletinsky}}]{Teissier2014}%
  \BibitemOpen
  \bibfield  {author} {\bibinfo {author} {\bibfnamefont {J.}~\bibnamefont {Teissier}}, \bibinfo {author} {\bibfnamefont {A.}~\bibnamefont {Barfuss}}, \bibinfo {author} {\bibfnamefont {P.}~\bibnamefont {Appel}}, \bibinfo {author} {\bibfnamefont {E.}~\bibnamefont {Neu}},\ and\ \bibinfo {author} {\bibfnamefont {P.}~\bibnamefont {Maletinsky}},\ }\bibfield  {title} {\bibinfo {title} {Strain coupling of a nitrogen-vacancy center spin to a diamond mechanical oscillator},\ }\href {https://doi.org/10.1103/PhysRevLett.113.020503} {\bibfield  {journal} {\bibinfo  {journal} {Phys. Rev. Lett.}\ }\textbf {\bibinfo {volume} {113}},\ \bibinfo {pages} {020503} (\bibinfo {year} {2014})}\BibitemShut {NoStop}%
\bibitem [{\citenamefont {Pradana}\ and\ \citenamefont {Chew}(2021)}]{Pradana2021}%
  \BibitemOpen
  \bibfield  {author} {\bibinfo {author} {\bibfnamefont {A.}~\bibnamefont {Pradana}}\ and\ \bibinfo {author} {\bibfnamefont {L.~Y.}\ \bibnamefont {Chew}},\ }\bibfield  {title} {\bibinfo {title} {Entanglement of nitrogen-vacancy-center ensembles with initial squeezing},\ }\href {https://doi.org/10.1103/PhysRevA.104.022435} {\bibfield  {journal} {\bibinfo  {journal} {Phys. Rev. A}\ }\textbf {\bibinfo {volume} {104}},\ \bibinfo {pages} {022435} (\bibinfo {year} {2021})}\BibitemShut {NoStop}%
\bibitem [{\citenamefont {Tsyganok}\ \emph {et~al.}(2023)\citenamefont {Tsyganok}, \citenamefont {Pershin}, \citenamefont {Khlebnikov}, \citenamefont {Kumpilov}, \citenamefont {Pyrkh}, \citenamefont {Rudnev}, \citenamefont {Fedotova}, \citenamefont {Gaifudinov}, \citenamefont {Cojocaru}, \citenamefont {Khoruzhii}, \citenamefont {Aksentsev}, \citenamefont {Zykova},\ and\ \citenamefont {Akimov}}]{Tsyganok2023}%
  \BibitemOpen
  \bibfield  {author} {\bibinfo {author} {\bibfnamefont {V.~V.}\ \bibnamefont {Tsyganok}}, \bibinfo {author} {\bibfnamefont {D.~A.}\ \bibnamefont {Pershin}}, \bibinfo {author} {\bibfnamefont {V.~A.}\ \bibnamefont {Khlebnikov}}, \bibinfo {author} {\bibfnamefont {D.~A.}\ \bibnamefont {Kumpilov}}, \bibinfo {author} {\bibfnamefont {I.~A.}\ \bibnamefont {Pyrkh}}, \bibinfo {author} {\bibfnamefont {A.~E.}\ \bibnamefont {Rudnev}}, \bibinfo {author} {\bibfnamefont {E.~A.}\ \bibnamefont {Fedotova}}, \bibinfo {author} {\bibfnamefont {D.~V.}\ \bibnamefont {Gaifudinov}}, \bibinfo {author} {\bibfnamefont {I.~S.}\ \bibnamefont {Cojocaru}}, \bibinfo {author} {\bibfnamefont {K.~A.}\ \bibnamefont {Khoruzhii}}, \bibinfo {author} {\bibfnamefont {P.~A.}\ \bibnamefont {Aksentsev}}, \bibinfo {author} {\bibfnamefont {A.~K.}\ \bibnamefont {Zykova}},\ and\ \bibinfo {author} {\bibfnamefont {A.~V.}\ \bibnamefont {Akimov}},\ }\bibfield  {title} {\bibinfo {title} {Losses of thulium atoms from optical dipole traps operating at 532 and 1064
  nm},\ }\href {https://doi.org/10.1103/PhysRevA.107.023315} {\bibfield  {journal} {\bibinfo  {journal} {Phys. Rev. A}\ }\textbf {\bibinfo {volume} {107}},\ \bibinfo {pages} {023315} (\bibinfo {year} {2023})}\BibitemShut {NoStop}%
\bibitem [{\citenamefont {Guerin}\ \emph {et~al.}(2016)\citenamefont {Guerin}, \citenamefont {Ara\'ujo},\ and\ \citenamefont {Kaiser}}]{Guerin2016}%
  \BibitemOpen
  \bibfield  {author} {\bibinfo {author} {\bibfnamefont {W.}~\bibnamefont {Guerin}}, \bibinfo {author} {\bibfnamefont {M.~O.}\ \bibnamefont {Ara\'ujo}},\ and\ \bibinfo {author} {\bibfnamefont {R.}~\bibnamefont {Kaiser}},\ }\bibfield  {title} {\bibinfo {title} {Subradiance in a large cloud of cold atoms},\ }\href {https://doi.org/10.1103/PhysRevLett.116.083601} {\bibfield  {journal} {\bibinfo  {journal} {Phys. Rev. Lett.}\ }\textbf {\bibinfo {volume} {116}},\ \bibinfo {pages} {083601} (\bibinfo {year} {2016})}\BibitemShut {NoStop}%
\bibitem [{\citenamefont {Plenio}\ \emph {et~al.}(1999)\citenamefont {Plenio}, \citenamefont {Huelga}, \citenamefont {Beige},\ and\ \citenamefont {Knight}}]{Plenio1999}%
  \BibitemOpen
  \bibfield  {author} {\bibinfo {author} {\bibfnamefont {M.~B.}\ \bibnamefont {Plenio}}, \bibinfo {author} {\bibfnamefont {S.~F.}\ \bibnamefont {Huelga}}, \bibinfo {author} {\bibfnamefont {A.}~\bibnamefont {Beige}},\ and\ \bibinfo {author} {\bibfnamefont {P.~L.}\ \bibnamefont {Knight}},\ }\bibfield  {title} {\bibinfo {title} {Cavity-loss-induced generation of entangled atoms},\ }\href {https://doi.org/10.1103/PhysRevA.59.2468} {\bibfield  {journal} {\bibinfo  {journal} {Phys. Rev. A}\ }\textbf {\bibinfo {volume} {59}},\ \bibinfo {pages} {2468} (\bibinfo {year} {1999})}\BibitemShut {NoStop}%
\bibitem [{\citenamefont {Reiter}\ \emph {et~al.}(2016)\citenamefont {Reiter}, \citenamefont {Reeb},\ and\ \citenamefont {S\o{}rensen}}]{Reiter2016}%
  \BibitemOpen
  \bibfield  {author} {\bibinfo {author} {\bibfnamefont {F.}~\bibnamefont {Reiter}}, \bibinfo {author} {\bibfnamefont {D.}~\bibnamefont {Reeb}},\ and\ \bibinfo {author} {\bibfnamefont {A.~S.}\ \bibnamefont {S\o{}rensen}},\ }\bibfield  {title} {\bibinfo {title} {Scalable dissipative preparation of many-body entanglement},\ }\href {https://doi.org/10.1103/PhysRevLett.117.040501} {\bibfield  {journal} {\bibinfo  {journal} {Phys. Rev. Lett.}\ }\textbf {\bibinfo {volume} {117}},\ \bibinfo {pages} {040501} (\bibinfo {year} {2016})}\BibitemShut {NoStop}%
\bibitem [{\citenamefont {Kastoryano}\ \emph {et~al.}(2011)\citenamefont {Kastoryano}, \citenamefont {Reiter},\ and\ \citenamefont {S\o{}rensen}}]{Kastoryano2011}%
  \BibitemOpen
  \bibfield  {author} {\bibinfo {author} {\bibfnamefont {M.~J.}\ \bibnamefont {Kastoryano}}, \bibinfo {author} {\bibfnamefont {F.}~\bibnamefont {Reiter}},\ and\ \bibinfo {author} {\bibfnamefont {A.~S.}\ \bibnamefont {S\o{}rensen}},\ }\bibfield  {title} {\bibinfo {title} {Dissipative preparation of entanglement in optical cavities},\ }\href {https://doi.org/10.1103/PhysRevLett.106.090502} {\bibfield  {journal} {\bibinfo  {journal} {Phys. Rev. Lett.}\ }\textbf {\bibinfo {volume} {106}},\ \bibinfo {pages} {090502} (\bibinfo {year} {2011})}\BibitemShut {NoStop}%
\bibitem [{\citenamefont {Lin}\ \emph {et~al.}(2013)\citenamefont {Lin}, \citenamefont {Gaebler}, \citenamefont {Reiter}, \citenamefont {Tan}, \citenamefont {Bowler}, \citenamefont {S{\o}rensen}, \citenamefont {Leibfried},\ and\ \citenamefont {Wineland}}]{Lin2013}%
  \BibitemOpen
  \bibfield  {author} {\bibinfo {author} {\bibfnamefont {Y.}~\bibnamefont {Lin}}, \bibinfo {author} {\bibfnamefont {J.}~\bibnamefont {Gaebler}}, \bibinfo {author} {\bibfnamefont {F.}~\bibnamefont {Reiter}}, \bibinfo {author} {\bibfnamefont {T.~R.}\ \bibnamefont {Tan}}, \bibinfo {author} {\bibfnamefont {R.}~\bibnamefont {Bowler}}, \bibinfo {author} {\bibfnamefont {A.}~\bibnamefont {S{\o}rensen}}, \bibinfo {author} {\bibfnamefont {D.}~\bibnamefont {Leibfried}},\ and\ \bibinfo {author} {\bibfnamefont {D.~J.}\ \bibnamefont {Wineland}},\ }\bibfield  {title} {\bibinfo {title} {Dissipative production of a maximally entangled steady state of two quantum bits},\ }\href {https://doi.org/10.1038/nature12801} {\bibfield  {journal} {\bibinfo  {journal} {Nature}\ }\textbf {\bibinfo {volume} {504}},\ \bibinfo {pages} {415} (\bibinfo {year} {2013})}\BibitemShut {NoStop}%
\bibitem [{\citenamefont {Behbood}\ \emph {et~al.}(2014)\citenamefont {Behbood}, \citenamefont {Martin~Ciurana}, \citenamefont {Colangelo}, \citenamefont {Napolitano}, \citenamefont {T\'oth}, \citenamefont {Sewell},\ and\ \citenamefont {Mitchell}}]{Behbood2014}%
  \BibitemOpen
  \bibfield  {author} {\bibinfo {author} {\bibfnamefont {N.}~\bibnamefont {Behbood}}, \bibinfo {author} {\bibfnamefont {F.}~\bibnamefont {Martin~Ciurana}}, \bibinfo {author} {\bibfnamefont {G.}~\bibnamefont {Colangelo}}, \bibinfo {author} {\bibfnamefont {M.}~\bibnamefont {Napolitano}}, \bibinfo {author} {\bibfnamefont {G.}~\bibnamefont {T\'oth}}, \bibinfo {author} {\bibfnamefont {R.~J.}\ \bibnamefont {Sewell}},\ and\ \bibinfo {author} {\bibfnamefont {M.~W.}\ \bibnamefont {Mitchell}},\ }\bibfield  {title} {\bibinfo {title} {Generation of macroscopic singlet states in a cold atomic ensemble},\ }\href {https://doi.org/10.1103/PhysRevLett.113.093601} {\bibfield  {journal} {\bibinfo  {journal} {Phys. Rev. Lett.}\ }\textbf {\bibinfo {volume} {113}},\ \bibinfo {pages} {093601} (\bibinfo {year} {2014})}\BibitemShut {NoStop}%
\bibitem [{\citenamefont {Ilo-Okeke}\ \emph {et~al.}(2022)\citenamefont {Ilo-Okeke}, \citenamefont {Ji}, \citenamefont {Chen}, \citenamefont {Mao}, \citenamefont {Kondappan}, \citenamefont {Ivannikov}, \citenamefont {Xiao},\ and\ \citenamefont {Byrnes}}]{Ilo2022}%
  \BibitemOpen
  \bibfield  {author} {\bibinfo {author} {\bibfnamefont {E.~O.}\ \bibnamefont {Ilo-Okeke}}, \bibinfo {author} {\bibfnamefont {Y.}~\bibnamefont {Ji}}, \bibinfo {author} {\bibfnamefont {P.}~\bibnamefont {Chen}}, \bibinfo {author} {\bibfnamefont {Y.}~\bibnamefont {Mao}}, \bibinfo {author} {\bibfnamefont {M.}~\bibnamefont {Kondappan}}, \bibinfo {author} {\bibfnamefont {V.}~\bibnamefont {Ivannikov}}, \bibinfo {author} {\bibfnamefont {Y.}~\bibnamefont {Xiao}},\ and\ \bibinfo {author} {\bibfnamefont {T.}~\bibnamefont {Byrnes}},\ }\bibfield  {title} {\bibinfo {title} {Deterministic preparation of supersinglets with collective spin projections},\ }\href {https://doi.org/10.1103/PhysRevA.106.033314} {\bibfield  {journal} {\bibinfo  {journal} {Phys. Rev. A}\ }\textbf {\bibinfo {volume} {106}},\ \bibinfo {pages} {033314} (\bibinfo {year} {2022})}\BibitemShut {NoStop}%
\bibitem [{\citenamefont {Cole}\ \emph {et~al.}(2021)\citenamefont {Cole}, \citenamefont {Wu}, \citenamefont {Erickson}, \citenamefont {Hou}, \citenamefont {Wilson}, \citenamefont {Leibfried},\ and\ \citenamefont {Reiter}}]{Cole2021}%
  \BibitemOpen
  \bibfield  {author} {\bibinfo {author} {\bibfnamefont {D.~C.}\ \bibnamefont {Cole}}, \bibinfo {author} {\bibfnamefont {J.~J.}\ \bibnamefont {Wu}}, \bibinfo {author} {\bibfnamefont {S.~D.}\ \bibnamefont {Erickson}}, \bibinfo {author} {\bibfnamefont {P.}~\bibnamefont {Hou}}, \bibinfo {author} {\bibfnamefont {A.~C.}\ \bibnamefont {Wilson}}, \bibinfo {author} {\bibfnamefont {D.}~\bibnamefont {Leibfried}},\ and\ \bibinfo {author} {\bibfnamefont {F.}~\bibnamefont {Reiter}},\ }\bibfield  {title} {\bibinfo {title} {Dissipative preparation of w states in trapped ion systems},\ }\href {https://doi.org/10.1088/1367-2630/ac09c8} {\bibfield  {journal} {\bibinfo  {journal} {New Journal of Physics}\ }\textbf {\bibinfo {volume} {23}},\ \bibinfo {pages} {073011} (\bibinfo {year} {2021})}\BibitemShut {NoStop}%
\bibitem [{\citenamefont {Tullu}\ \emph {et~al.}(2023)\citenamefont {Tullu}, \citenamefont {Dhar},\ and\ \citenamefont {Lau}}]{Tullu2023}%
  \BibitemOpen
  \bibfield  {author} {\bibinfo {author} {\bibfnamefont {R.~K.}\ \bibnamefont {Tullu}}, \bibinfo {author} {\bibfnamefont {H.~S.}\ \bibnamefont {Dhar}},\ and\ \bibinfo {author} {\bibfnamefont {H.-K.}\ \bibnamefont {Lau}},\ }\bibfield  {title} {\bibinfo {title} {Deterministic construction of arbitrary states in permutationally-invariant spin ensemble},\ }\href {https://meetings.aps.org/Meeting/MAR23/Session/N67.11} {\bibfield  {journal} {\bibinfo  {journal} {Bulletin of the American Physical Society}\ } (\bibinfo {year} {2023})}\BibitemShut {NoStop}%
\bibitem [{\citenamefont {Giovannetti}\ \emph {et~al.}(2006)\citenamefont {Giovannetti}, \citenamefont {Lloyd},\ and\ \citenamefont {Maccone}}]{Giovannetti2006}%
  \BibitemOpen
  \bibfield  {author} {\bibinfo {author} {\bibfnamefont {V.}~\bibnamefont {Giovannetti}}, \bibinfo {author} {\bibfnamefont {S.}~\bibnamefont {Lloyd}},\ and\ \bibinfo {author} {\bibfnamefont {L.}~\bibnamefont {Maccone}},\ }\bibfield  {title} {\bibinfo {title} {Quantum metrology},\ }\href {https://doi.org/10.1103/PhysRevLett.96.010401} {\bibfield  {journal} {\bibinfo  {journal} {Phys. Rev. Lett.}\ }\textbf {\bibinfo {volume} {96}},\ \bibinfo {pages} {010401} (\bibinfo {year} {2006})}\BibitemShut {NoStop}%
\bibitem [{\citenamefont {Degen}\ \emph {et~al.}(2017)\citenamefont {Degen}, \citenamefont {Reinhard},\ and\ \citenamefont {Cappellaro}}]{Degen2017}%
  \BibitemOpen
  \bibfield  {author} {\bibinfo {author} {\bibfnamefont {C.~L.}\ \bibnamefont {Degen}}, \bibinfo {author} {\bibfnamefont {F.}~\bibnamefont {Reinhard}},\ and\ \bibinfo {author} {\bibfnamefont {P.}~\bibnamefont {Cappellaro}},\ }\bibfield  {title} {\bibinfo {title} {Quantum sensing},\ }\href {https://doi.org/10.1103/RevModPhys.89.035002} {\bibfield  {journal} {\bibinfo  {journal} {Rev. Mod. Phys.}\ }\textbf {\bibinfo {volume} {89}},\ \bibinfo {pages} {035002} (\bibinfo {year} {2017})}\BibitemShut {NoStop}%
\bibitem [{\citenamefont {Huelga}\ \emph {et~al.}(1997)\citenamefont {Huelga}, \citenamefont {Macchiavello}, \citenamefont {Pellizzari}, \citenamefont {Ekert}, \citenamefont {Plenio},\ and\ \citenamefont {Cirac}}]{Huelga1997}%
  \BibitemOpen
  \bibfield  {author} {\bibinfo {author} {\bibfnamefont {S.~F.}\ \bibnamefont {Huelga}}, \bibinfo {author} {\bibfnamefont {C.}~\bibnamefont {Macchiavello}}, \bibinfo {author} {\bibfnamefont {T.}~\bibnamefont {Pellizzari}}, \bibinfo {author} {\bibfnamefont {A.~K.}\ \bibnamefont {Ekert}}, \bibinfo {author} {\bibfnamefont {M.~B.}\ \bibnamefont {Plenio}},\ and\ \bibinfo {author} {\bibfnamefont {J.~I.}\ \bibnamefont {Cirac}},\ }\bibfield  {title} {\bibinfo {title} {Improvement of frequency standards with quantum entanglement},\ }\href {https://doi.org/10.1103/PhysRevLett.79.3865} {\bibfield  {journal} {\bibinfo  {journal} {Phys. Rev. Lett.}\ }\textbf {\bibinfo {volume} {79}},\ \bibinfo {pages} {3865} (\bibinfo {year} {1997})}\BibitemShut {NoStop}%
\bibitem [{\citenamefont {Zhou}\ \emph {et~al.}(2018)\citenamefont {Zhou}, \citenamefont {Zhang}, \citenamefont {Preskill},\ and\ \citenamefont {Jiang}}]{Zhou2018}%
  \BibitemOpen
  \bibfield  {author} {\bibinfo {author} {\bibfnamefont {S.}~\bibnamefont {Zhou}}, \bibinfo {author} {\bibfnamefont {M.}~\bibnamefont {Zhang}}, \bibinfo {author} {\bibfnamefont {J.}~\bibnamefont {Preskill}},\ and\ \bibinfo {author} {\bibfnamefont {L.}~\bibnamefont {Jiang}},\ }\bibfield  {title} {\bibinfo {title} {Achieving the heisenberg limit in quantum metrology using quantum error correction},\ }\href {https://doi.org/10.1038/s41467-017-02510-3} {\bibfield  {journal} {\bibinfo  {journal} {Nature Communications}\ }\textbf {\bibinfo {volume} {9}},\ \bibinfo {pages} {78} (\bibinfo {year} {2018})}\BibitemShut {NoStop}%
\bibitem [{\citenamefont {Gross}\ and\ \citenamefont {Haroche}(1982)}]{Gross1982}%
  \BibitemOpen
  \bibfield  {author} {\bibinfo {author} {\bibfnamefont {M.}~\bibnamefont {Gross}}\ and\ \bibinfo {author} {\bibfnamefont {S.}~\bibnamefont {Haroche}},\ }\bibfield  {title} {\bibinfo {title} {Superradiance: An essay on the theory of collective spontaneous emission},\ }\href {https://doi.org/https://doi.org/10.1016/0370-1573(82)90102-8} {\bibfield  {journal} {\bibinfo  {journal} {Physics Reports}\ }\textbf {\bibinfo {volume} {93}},\ \bibinfo {pages} {301} (\bibinfo {year} {1982})}\BibitemShut {NoStop}%
\bibitem [{\citenamefont {Braunstein}\ and\ \citenamefont {Caves}(1994)}]{Braunstein1994}%
  \BibitemOpen
  \bibfield  {author} {\bibinfo {author} {\bibfnamefont {S.~L.}\ \bibnamefont {Braunstein}}\ and\ \bibinfo {author} {\bibfnamefont {C.~M.}\ \bibnamefont {Caves}},\ }\bibfield  {title} {\bibinfo {title} {Statistical distance and the geometry of quantum states},\ }\href {https://doi.org/10.1103/PhysRevLett.72.3439} {\bibfield  {journal} {\bibinfo  {journal} {Phys. Rev. Lett.}\ }\textbf {\bibinfo {volume} {72}},\ \bibinfo {pages} {3439} (\bibinfo {year} {1994})}\BibitemShut {NoStop}%
\bibitem [{\citenamefont {Sawyer}\ \emph {et~al.}(2012)\citenamefont {Sawyer}, \citenamefont {Britton}, \citenamefont {Keith}, \citenamefont {Wang}, \citenamefont {Freericks}, \citenamefont {Uys}, \citenamefont {Biercuk},\ and\ \citenamefont {Bollinger}}]{Sawyer2012}%
  \BibitemOpen
  \bibfield  {author} {\bibinfo {author} {\bibfnamefont {B.~C.}\ \bibnamefont {Sawyer}}, \bibinfo {author} {\bibfnamefont {J.~W.}\ \bibnamefont {Britton}}, \bibinfo {author} {\bibfnamefont {A.~C.}\ \bibnamefont {Keith}}, \bibinfo {author} {\bibfnamefont {C.-C.~J.}\ \bibnamefont {Wang}}, \bibinfo {author} {\bibfnamefont {J.~K.}\ \bibnamefont {Freericks}}, \bibinfo {author} {\bibfnamefont {H.}~\bibnamefont {Uys}}, \bibinfo {author} {\bibfnamefont {M.~J.}\ \bibnamefont {Biercuk}},\ and\ \bibinfo {author} {\bibfnamefont {J.~J.}\ \bibnamefont {Bollinger}},\ }\bibfield  {title} {\bibinfo {title} {Spectroscopy and thermometry of drumhead modes in a mesoscopic trapped-ion crystal using entanglement},\ }\href {https://doi.org/10.1103/PhysRevLett.108.213003} {\bibfield  {journal} {\bibinfo  {journal} {Phys. Rev. Lett.}\ }\textbf {\bibinfo {volume} {108}},\ \bibinfo {pages} {213003} (\bibinfo {year} {2012})}\BibitemShut {NoStop}%
\bibitem [{\citenamefont {Safavi-Naini}\ \emph {et~al.}(2018)\citenamefont {Safavi-Naini}, \citenamefont {Lewis-Swan}, \citenamefont {Bohnet}, \citenamefont {G\"arttner}, \citenamefont {Gilmore}, \citenamefont {Jordan}, \citenamefont {Cohn}, \citenamefont {Freericks}, \citenamefont {Rey},\ and\ \citenamefont {Bollinger}}]{Safvi2018}%
  \BibitemOpen
  \bibfield  {author} {\bibinfo {author} {\bibfnamefont {A.}~\bibnamefont {Safavi-Naini}}, \bibinfo {author} {\bibfnamefont {R.~J.}\ \bibnamefont {Lewis-Swan}}, \bibinfo {author} {\bibfnamefont {J.~G.}\ \bibnamefont {Bohnet}}, \bibinfo {author} {\bibfnamefont {M.}~\bibnamefont {G\"arttner}}, \bibinfo {author} {\bibfnamefont {K.~A.}\ \bibnamefont {Gilmore}}, \bibinfo {author} {\bibfnamefont {J.~E.}\ \bibnamefont {Jordan}}, \bibinfo {author} {\bibfnamefont {J.}~\bibnamefont {Cohn}}, \bibinfo {author} {\bibfnamefont {J.~K.}\ \bibnamefont {Freericks}}, \bibinfo {author} {\bibfnamefont {A.~M.}\ \bibnamefont {Rey}},\ and\ \bibinfo {author} {\bibfnamefont {J.~J.}\ \bibnamefont {Bollinger}},\ }\bibfield  {title} {\bibinfo {title} {Verification of a many-ion simulator of the dicke model through slow quenches across a phase transition},\ }\href {https://doi.org/10.1103/PhysRevLett.121.040503} {\bibfield  {journal} {\bibinfo  {journal} {Phys. Rev. Lett.}\ }\textbf {\bibinfo {volume} {121}},\ \bibinfo {pages} {040503}
  (\bibinfo {year} {2018})}\BibitemShut {NoStop}%
\bibitem [{\citenamefont {Zhu}\ \emph {et~al.}(2011)\citenamefont {Zhu}, \citenamefont {Saito}, \citenamefont {Kemp}, \citenamefont {Kakuyanagi}, \citenamefont {Karimoto}, \citenamefont {Nakano}, \citenamefont {Munro}, \citenamefont {Tokura}, \citenamefont {Everitt}, \citenamefont {Nemoto}, \citenamefont {Kasu}, \citenamefont {Mizuochi},\ and\ \citenamefont {Semba}}]{Zhu2011}%
  \BibitemOpen
  \bibfield  {author} {\bibinfo {author} {\bibfnamefont {X.}~\bibnamefont {Zhu}}, \bibinfo {author} {\bibfnamefont {S.}~\bibnamefont {Saito}}, \bibinfo {author} {\bibfnamefont {A.}~\bibnamefont {Kemp}}, \bibinfo {author} {\bibfnamefont {K.}~\bibnamefont {Kakuyanagi}}, \bibinfo {author} {\bibfnamefont {S.-I.}\ \bibnamefont {Karimoto}}, \bibinfo {author} {\bibfnamefont {H.}~\bibnamefont {Nakano}}, \bibinfo {author} {\bibfnamefont {W.~J.}\ \bibnamefont {Munro}}, \bibinfo {author} {\bibfnamefont {Y.}~\bibnamefont {Tokura}}, \bibinfo {author} {\bibfnamefont {M.~S.}\ \bibnamefont {Everitt}}, \bibinfo {author} {\bibfnamefont {K.}~\bibnamefont {Nemoto}}, \bibinfo {author} {\bibfnamefont {M.}~\bibnamefont {Kasu}}, \bibinfo {author} {\bibfnamefont {N.}~\bibnamefont {Mizuochi}},\ and\ \bibinfo {author} {\bibfnamefont {K.}~\bibnamefont {Semba}},\ }\bibfield  {title} {\bibinfo {title} {Coherent coupling of a superconducting flux qubit to an electron spin ensemble in diamond},\ }\href {https://doi.org/10.1038/nature10462}
  {\bibfield  {journal} {\bibinfo  {journal} {Nature}\ }\textbf {\bibinfo {volume} {478}},\ \bibinfo {pages} {221–224} (\bibinfo {year} {2011})}\BibitemShut {NoStop}%
\end{thebibliography}%

\end{document}